\documentclass[prl,twocolumn,amsmath,amssymb,superscriptaddress]{revtex4-1}

\usepackage{epsfig,amsmath}
\usepackage{subfigure}
\usepackage{graphicx}
\usepackage{dcolumn}
\usepackage{stmaryrd}
\usepackage{mathrsfs}
\usepackage{pifont}
\usepackage{amsthm}
\usepackage{amssymb}
\usepackage{bm}
\usepackage{latexsym}
\usepackage{color}
\usepackage{epstopdf}
\usepackage{amsfonts}
\usepackage{units}
\usepackage{times}
\usepackage{upgreek}
\usepackage[colorlinks=true,linkcolor=blue,citecolor=blue,urlcolor=blue]{hyperref}

\begin{document}

\title{General theory of persistent microwave-optical quantum resources in hybrid-system dynamics}

\author{Fan Li}
\affiliation{College of Physics and Hebei Key Laboratory of Photophysics Research and Application, Hebei Normal University, Shijiazhuang, Hebei 050024, China}

\author{Shi-fan Qi}
\email{qishifan@hebtu.edu.cn}
\affiliation{College of Physics and Hebei Key Laboratory of Photophysics Research and Application, Hebei Normal University, Shijiazhuang, Hebei 050024, China}

\author{Z. D. Wang}
\email{zwang@hku.hk}
\affiliation{Frontier Research Institute for Physics, South China Normal University, Guangzhou 510006, China}
\affiliation{HK Institute of Quantum Science \& Technology and Department of Physics, The University of Hong Kong, Pokfulam Road, Hong Kong, China}
\affiliation{Hong Kong Branch for Quantum Science Center of Guangdong-Hong Kong-Macau Greater Bay Area, Shenzhen 518045, China}

\author{Yan-Kui Bai}
\email{ykbai@semi.ac.cn}
\affiliation{College of Physics and Hebei Key Laboratory of Photophysics Research and Application, Hebei Normal University, Shijiazhuang, Hebei 050024, China}
\affiliation{Hong Kong Branch for Quantum Science Center of Guangdong-Hong Kong-Macau Greater Bay Area, Shenzhen 518045, China}

\begin{abstract}
We develop a general theoretical framework for characterizing persistent quantum resources between microwave and optical modes in the dynamics of chain-type hybrid quantum systems with intermediate modes. The effective Hamiltonian for microwave-optical (MO) squeezing is formulated via strong nearest-neighbor interactions in the microwave-intermediate-optical chain, from which rigorous solutions for the dynamics of MO Gaussian entanglement and quantum steering are obtained analytically. Notably, MO quantum resource can survive and approach a finite asymptotic value even in the unsteady regime, and can surpass the steady-state upper bound on the quantum resource. Furthermore, the asymptotic values of MO entanglement as well as one-way and two-way quantum steering are readily controllable by tuning the effective coupling strength. The validity of our theory is demonstrated by applying it to the typical hybrid models of electro-optomechanical and cavity optomagnomechanical systems, and the extension to the nonlinear regime with non-Gaussian MO quantum resources is further studied.

\end{abstract}

\maketitle
\section*{I.~~Introduction}
Microwave and optical modes play distinct roles in quantum technologies ~\cite{cavityoptomechanics14,microwave17,microwave09,microwave19,optomechanics22,photonH24,photonP25}, where the former facilitates quantum control of various physical systems~\cite{microwaveadv_Nori13,microwaveadv_JL14} and the latter is capable of long-distance information propagation~\cite{opticaladv09,opticaladv23}. Efficient conversions between these two modes~\cite{mo_conversionH21,mo_conversionL20,mo_conversion26,mo_conversion07,mo_conversionA14,mo_conversionF20,mo_conversionJ20,microwaveadv_Z22,mo_conversionnpj22,mo_conversionB24} are indispensable in distributed quantum information processing toward a future quantum internet~\cite{quantumnetwork08,quantumnetwork18,quantumnetwork21,quantumcommunication07,quantumcommunication26}. Based on microwave-optical (MO) nonlocal resources such as entanglement~\cite{entanglement} and quantum steering~\cite{steering}, the indirect transduction~\cite{EOM12,mo_entZ20} is a promising approach with high transfer efficiency and strong noise tolerance~\cite{mo_entFink19,mo_entQL23}, where Gaussian quantum resources play a fundamental role owing to their inherent merits~\cite{gaussian07,gaussian12,steeringapp15,steeringapp16,steering22,teleportation98,teleportationHe15,teleportationYam26}. The generation of MO quantum resources typically requires the intermediate modes in hybrid multipartite systems due to the large frequency mismatch between the two target modes~\cite{mo_conversionH21,mo_conversionL20}. Under steady-state conditions, significant progress has been made based on various hybrid multipartite systems, including electro-optomechanical (EOM) systems~\cite{EOM12,EOM13,EOM15,EOM11,EOMLi15,mo_entJ20,EOMJ22,EOM25}, cavity optomagnomechanical (COMM) systems~\cite{COMMLJ23,COMMY23,cavitymagnomechanics,optomechanical07}, magneto-optomechanical (MOM) systems~\cite{microwaveopticalG22,microwaveopticsLJ25}, and so on.

High-precision control in the MO quantum interface hinges on an analytical characterization of MO nonlocal quantum resources. However, the inherent configuration complexity of multipartite hybrid systems~\cite{EOM12,COMMLJ23,microwaveopticalG22} poses a significant challenge to obtaining such a characterization. One may instead resort to the numerical method for full system dynamics which can faithfully capture the evolution of MO quantum resources, yet this approach offers limited insight into the underlying control mechanism. Moreover, because entanglement and quantum steering are monogamous~\cite{CKW00pra,C2N06prl,monogamy06,monogamy07,NegFan07pra,bai09pra,EfMoN14prl,monogamy16,monogamyP17,weili26pra}, the corresponding MO nonlocal quantum resources cannot be freely shared among the constituents of a multipartite hybrid system, and their optimal allocation therefore requires a multi-parameter optimization that analytical characterization for MO entanglement and quantum steering can facilitate.

In multipartite hybrid systems, the generation of MO nonlocal quantum resources is a typical dynamical process~\cite{tingyu04prl,tingyu09science,dynamic09,dynamics14,dynamic18,dynamicW24,Chu25}. Under the steady-state condition, the analysis of MO quantum resources is generally based on the system evolution to a steady state \cite{EOM12,mo_entZ20,mo_entFink19,EOM15,EOM11,EOMLi15,COMMLJ23,microwaveopticsLJ25}. Beyond the steady-state case, it was demonstrated that two-qubit quantum correlations can exhibit the freezing phenomenon under certain dynamical decoherence~\cite{freezing09,freezing10,freezingM10}, where the quantum correlation does not decay while the two-qubit state itself undergoes the unsteady evolution. Recently, a theoretical study also indicates that transient optoacoustic entanglement can be generated in the dynamical process of unsteady states~\cite{dynamicZ24}. However, beyond the steady-state-based schemes, whether MO quantum resource can be persistently generated during unsteady dynamical evolution while approaching a finite asymptotic value remains an open question.

In this work, we propose a general approach to generate MO entanglement and quantum steering of the two non-interacting modes by constructing an effective two-mode squeezing assisted by chain-type intermediate modes with linear interactions. Based on the effective two-mode squeezing Hamiltonian, the system dynamics governed by the quantum Langevin equations within the open-quantum-system framework can be solved analytically. Furthermore, the closed-form formulas of MO Gaussian entanglement and quantum steering are obtained, which facilitate precise control of these resources by modulating the effective coupling strength. It is found that the MO quantum resource can exhibit a persistent dynamical plateau during the unsteady evolution, with its asymptotic value surpassing the steady-state upper bound. Finally, the validity of our theory is analytically demonstrated on two representative hybrid models of EOM~\cite{EOM12,EOM15} and COMM systems~\cite{COMMLJ23}, and the extension of our effective Hamiltonian approach to the nonlinear scenario with non-Gaussian MO quantum resources is further investigated.

\section*{II.~~Results}
\subsection*{A.~~General theory for MO chain-type systems}
\begin{figure}[t] 
\includegraphics[width=0.33\textwidth]{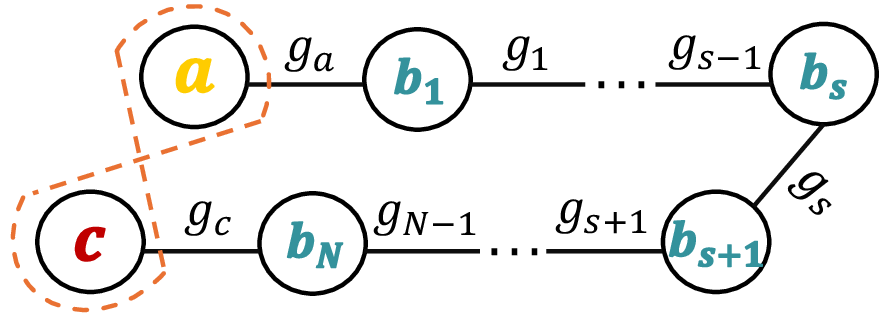}
\caption{\textbf{Chain-coupled multipartite hybrid system.} Diagram of a multipartite hybrid system consisting of two target modes (microwave mode $a$ and optical mode $c$) and chain-type $N$-intermediate modes ($b_1, b_2,\cdots, b_N$). The adjacent couplings are represented by $g_a$, $g_c$, and $g_s$ with $s=1, 2,\dots, N-1$.}
\label{diagram}	
\end{figure}
As shown in Fig. \ref{diagram}, we consider a multipartite hybrid system consisting of microwave mode $a$, optical mode $c$, and chain-type $N$-intermediate modes $b_1, b_2, \dots, b_N$. Without loss of generality, the hybrid system Hamiltonian in the interaction picture may be expressed as ($\hbar\equiv1$)
\begin{equation}\label{generalH} 
\begin{aligned}
H=&~H_0+V,\\
H_0=&~\Delta_aa^\dag a+\Delta_cc^\dag c+\sum^N_{s=1}\omega_sb_s^\dag b_s,\\
V=&~V_a+V_b+V_c,\\
V_a=&~g_a[\cos\theta(a^\dag b_1+ab_1^\dag)+\sin\theta(a^\dag b_1^\dag+ab_1)],\\
V_b=&~\sum^{N-1}_{s=1}g_s(b_s+b_s^\dag)(b_{s+1}+b_{s+1}^\dag),\\
V_c=&~g_c[\cos\phi(c^\dag b_N+cb_N^\dag)+\sin\phi(c^\dag b_N^\dag+cb_N)].\\
\end{aligned}
\end{equation}
Here, $a$ ($a^\dag$), $b_s$ ($b_s^\dag$), and $c$ ($c^\dag$) are the annihilation (creation) operators of the modes $a$, $b_s$, and $c$, respectively. $\Delta_a$ and $\Delta_c$ are the frequency detunings of target modes $a$ and $c$ with respect to the lab frame, and $\omega_s$ is the transition frequency of mode $b_s$. Moreover, $g_a$, $g_s$, and $g_c$ are corresponding linear coupling strengths between the adjacent modes, and the angles $\theta$ and $\phi$ parameterize the magnitude of rotating and counter-rotating couplings between modes $a$ and $b_1$, $c$ and $b_N$, respectively. This MO hybrid model in Eq. \eqref{generalH} is realizable in various physical platforms with different intermediate modes, such as the EOM ~\cite{EOM12,EOM13,EOM15} ($N=1$), COMM~\cite{COMMLJ23,COMMY23} ($N=2$), and MOM systems~\cite{microwaveopticalG22,microwaveopticsLJ25} ($N=3$).

Our theoretical framework for MO quantum resources is derived from engineering an effective two-mode squeezing interaction between microwave mode $a$ and optical mode $c$ assisted by auxiliary modes $b_s$. Using a generalized analysis of virtual transition pathways on the full Hamiltonian in Eq. \eqref{generalH} by the nearly degenerate perturbation theory~\cite{perturbation01,perturbation16,perturbation18,perturbation25}, we can obtain the effective Hamiltonian
\begin{equation}\label{Heffac} 
H_{\rm eff}=g_{\rm eff}(a^\dag c^\dag+ ac),
\end{equation}
where $g_{\rm eff}$ is the effective coupling strength, and we employ the conditions $g_a, g_s, g_c\ll\{|\Delta_a-\omega_s|, |\Delta_c-\omega_s|\}$ and $\Delta_a=-\Delta_c+\delta$ with $\delta$ being the energy shift. The rigorous derivation of $H_{\rm{eff}}$ in Eq. \eqref{Heffac} is presented in Methods, where the analytical formula for $g_{\rm{eff}}$ for an arbitrary number of intermediate modes is given in Eq. \eqref{gn}. The resulting effective Hamiltonian simplifies the complex multipartite hybrid system into the MO subsystem, allowing for an analytical investigation on the dynamics of MO quantum resources.

According to the effective Hamiltonian in Eq. \eqref{Heffac}, the evolution of MO subsystem in the open-quantum-system framework can be analytically derived via the quantum Langevin equations~\cite{input-output}.
Given that the initial state is Gaussian, the dynamics of MO subsystem can be fully characterized by a $4\times 4$ covariance matrix (CM) $v(t)$~\cite{dynamic09,dynamics14}, which satisfies
\begin{equation}\label{hatv1} 
\dot{v}(t)=A_{\rm eff}{v}(t)+{v}(t)A_{\rm eff}^T+D_{\rm eff},
\end{equation}
where $\dot{v}(t)$ denotes the derivation of $v(t)$. The elements of $v(t)$ are defined as $v_{jl}(t)=\langle u_j(t)u_l(t)+ u_l(t)u_j(t)\rangle/2-\langle u_j(t)\rangle\langle u_l(t)\rangle~(j,l=1,2,3,4)$, where $u_j(t)$ is the $j$-term of $u(t)=[X_{a}(t),Y_{a}(t),X_{c}(t),Y_{c}(t)]^T$ with $X_o(t)=(o+o^\dag)/\sqrt{2}$ and $Y_o(t)=i(o^\dag-o)/\sqrt{2}$ for $o=a,c$. Moreover, in Eq. \eqref{hatv1}, $D_{\rm eff}$ is the diffusion matrix corresponding to an initial thermal state, and the drift matrix $A_{\rm eff}$ has the form
\begin{align}\label{Aeff} 
A_{\rm eff}=-\begin{pmatrix}
\kappa_a&0&0&g_{\rm eff}\\
0&\kappa_a&g_{\rm eff}&0\\
0&g_{\rm eff}&\kappa_c&0\\
g_{\rm eff}&0&0&\kappa_c\\
\end{pmatrix},
\end{align}
where $\kappa_a$ and $\kappa_c$ represent the decay rates of the modes $a$ and $c$. After substituting Eq. \eqref{Aeff} into Eq. \eqref{hatv1}, we have the CM
\begin{align}\label{v(t)} 
v(t)=\begin{pmatrix}
v_{11}(t)&0&0&v_{14}(t)\\
0&v_{11}(t)&v_{14}(t)&0\\
0&v_{14}(t)&v_{44}(t)&0\\
v_{14}(t)&0&0&v_{44}(t)\\
\end{pmatrix},
\end{align}
where the analytical expressions of the non-zero matrix elements and the detailed derivation of ${v}(t)$ as well as the validation of the reduced open-system description are provided in Supplementary Notes IA, IIC, and IIIB. According to the Lyapunov stability analysis of the drift matrix~\cite{lyap87}, we derive the steady-state condition $g_{\rm eff}^2<\kappa_a\kappa_c$ and the unsteady-state condition $g_{\rm eff}^2\ge\kappa_a\kappa_c$, respectively. In the former case, the MO system evolves toward a steady state determined by $\dot{v}(t)=0$, whereas, in the latter case, the target system undergoes the unsteady-state evolution characterized by a divergent matrix $v(t)$ (see Supplementary Note IA for details).

\subsection*{B.~~The asymptotic MO quantum resources}
Based on the analytical expression of $v(t)$ in Eq. \eqref{v(t)}, we are able to study the MO dynamics and calculate the Gaussian entanglement and quantum steering for an initial thermal state at any given time. We use the logarithmic negativity~\cite{LNentanglement02,LN2mode04,Mulien04,Plenio05prl,LNentanglement07,entanglementsh24} to quantify the MO entanglement
\begin{equation}\label{Eac} 
E_{ac}(t)={\max[0,-\ln(2\eta^-_{ac})]},
\end{equation}
where $\eta^-_{ac}$ is the minimum symplectic eigenvalue of the partial transpose matrix of the $v(t)$. In the dynamical evolution of MO subsystem, the two-mode entanglement $E_{ac}(t)$ at $t\to\infty$ will converge to a fixed value $E_{ac}$ with its analytical formula being provided in Supplementary Note IB. For the typical initial vacuum state, we have
\begin{equation}\label{eni} 
\begin{aligned}
&E_{ac}=\begin{cases}
\ln\left(\cfrac{\kappa_a\kappa_c-g_{\rm eff}^2}{\kappa_a\kappa_c-g_{\rm eff}^2\chi}\right),~g_{\rm eff}^2<\kappa_a\kappa_c \\[1em]
\ln\left(1+4\cfrac{g^2_{\rm eff}}{\tilde\chi}\right),~~g_{\rm eff}^2\ge\kappa_a\kappa_c\\[1em]
\end{cases}
\end{aligned}
\end{equation}
where the piecewise function corresponds to the MO entanglement for the steady-state and unsteady-state cases, respectively, and the parameters are $\chi=\{1+4\kappa_a\kappa_c(\kappa_a\kappa_c-g_{\rm eff}^2)/[g_{\rm eff}^2(\kappa_a+\kappa_c)^2]\}^{1/2}$ and $\tilde\chi=\Omega(\kappa_a+\kappa_c)+(\kappa_a-\kappa_c)^2$ with $\Omega=[4g_{\rm eff}^2+(\kappa_a-\kappa_c)^2]^{1/2}$. For the given decay rates $\kappa_a$ and $\kappa_c$, the piecewise $E_{ac}$ in Eq. \eqref{eni} is continuous as $g_{\rm{eff}}^2\to (\kappa_a\kappa_c)^-$ and increases monotonically along with the square of effective coupling strength $g_{\rm{eff}}^2$.

\begin{figure} 
\centering
\includegraphics[width=0.48\textwidth]{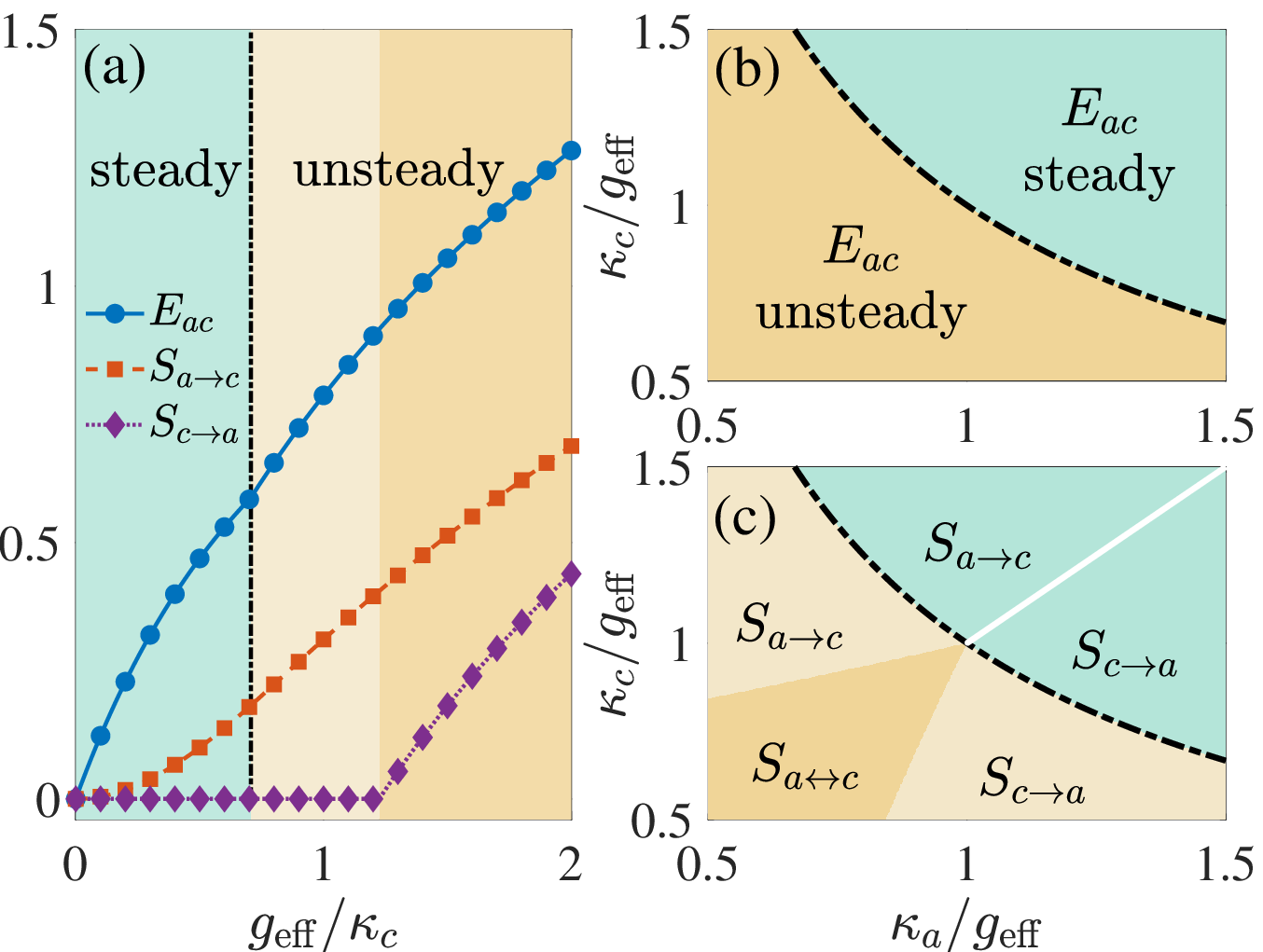}
\caption{\textbf{The control of MO quantum resources via effective parameters.} \textbf{a} The MO entanglement $E_{ac}$ (the blue line), asymmetric quantum steering $S_{a\rightarrow c}$ (the red line) and $S_{c\rightarrow a}$ (the purple line) along with the relative coupling $g_{\rm eff}/\kappa_c$ for the decay rates $\kappa_c=2\kappa_a=1$. \textbf{b} The regional diagram of MO entanglement $E_{ac}$ for the steady-state (the teal area) and the unsteady-state (the golden-yellow area) cases. \textbf{c} The regional diagram of asymmetric quantum steering for the steady-state case (the two teal areas for $S_{a\to c}$ and $S_{c\to a}$ with the white boundary being zero value) and the unsteady-state case (the two beige areas for the asymmetric one-way steering and the golden-yellow area for the two-way steering $S_{a\leftrightarrow c}$). The black dot-dashed line indicates the boundary between the steady-state and unsteady-state dynamics in the three panels, and the effective coupling strength is $g_{\rm{eff}}=1$ in \textbf{b} and \textbf{c}.}
\label{range}
\end{figure}

Quantum steering of continuous variables is a kind of asymmetric quantum resource~\cite{steering}, and enables one-sided device-independent quantum key distribution~\cite{steeringapp15,steeringapp16,steering22} and random number generation~\cite{QRNG17prl,steeringapp25}. Based on the analytical CM $v(t)$ in Eq. \eqref{v(t)}, the two-mode steering from $a$ to $c$ under Gaussian measurements~\cite{LNsteeringK15} can be quantified by
\begin{equation}\label{Sac} 
S_{a\to c}(t)=\max[0,S_{ac}(t)],
\end{equation}
where the quantify $S_{ac}(t)=\ln[\det v_a/(4\det v)]/2$, with $v_a$ being the CM of the microwave mode $a$, evolves to a finite asymptotic value $S_{ac}$ at $t\to\infty$ for an initial thermal state (see the derivation and formula in Supplementary Note IC). For the case of an initial vacuum state, we have
\begin{equation}\label{sact2} 
\begin{aligned}
S_{ac}=\begin{cases}
\ln\left[\cfrac{g_{\rm eff}^2(\kappa_c^2-\kappa_a^2)+\Xi}{g_{\rm eff}^2(\kappa_a-\kappa_c)^2+\Xi}\right],~~g_{\rm eff}^2<\kappa_a\kappa_c\\[1em]
\ln\left(\cfrac{\Omega-\kappa_a+\kappa_c}{2\Omega}\right)+E_{ac},~~g_{\rm eff}^2\ge\kappa_a\kappa_c\\[1em]
\end{cases}
\end{aligned}
\end{equation}	
where the piecewise function corresponds to the steady-state and unsteady-state evolutions, respectively. The parameters are $\Xi=\kappa_a\kappa_c(\kappa_a+\kappa_c)^2$ and $\Omega=[4g_{\rm eff}^2+(\kappa_a-\kappa_c)^2]^{1/2}$, and $E_{ac}$ is the asymptotic entanglement value in the unsteady-state regime in Eq. \eqref{eni}. The nonzero asymptotic steering $S_{a\to c}=S_{a\to c}(\infty)$ arises, when the conditions $g_{\rm eff}^2>0$ for $\kappa_a<\kappa_c$ and $g_{\rm eff}^2>\kappa_a(2\kappa_a-\kappa_c)$ for $\kappa_a\ge\kappa_c$ are satisfied. The asymmetric one-way steering $S_{c\to a}$ from mode $c$ to mode $a$ has a similar formula as those in Eqs. \eqref{Sac} and \eqref{sact2} by interchanging the decay rates $\kappa_a$ and $\kappa_c$. Moreover, the nonzero asymptotic two-way steering $S_{a\leftrightarrow c}$ only exists in the unsteady-state evolution, and the system parameters need to satisfy the condition $g^2_{\rm{eff}}+\kappa_a\kappa_c> 2\kappa_a^2, 2\kappa_c^2$. For the given decay rates $\kappa_a$ and $\kappa_c$, the asymptotic values $S_{a\to c}$ and $S_{c\to a}$ increase monotonically with $g_{\mathrm{eff}}^2$, and both asymmetric steering values are smaller than the value of the MO entanglement $E_{ac}$, as detailed in Supplementary Note IC. Notably, in the unsteady-state regime, despite the divergent system dynamics, the joint action of energy loss and the squeezing drive causes the MO quantum resource to approach a finite asymptotic value, giving rise to persistent entanglement and quantum steering with an asymptotic plateau (see Supplementary Note ID).

These asymptotic quantum-resource values can be quantitatively controlled by modulating the effective coupling strength $g_{\rm{eff}}$ and the decay rates $\kappa_a$ and $\kappa_c$ in terms of the corresponding analytical expressions for MO entanglement and quantum steering. In Fig. \ref{range}a, the asymptotic entanglement $E_{ac}$, quantum steering $S_{a\to c}$ and $S_{c\to a}$ are plotted with increasing relative coupling strength $g_{\rm{eff}}/\kappa_c$, with the decay rates set to $\kappa_c=2\kappa_a=1$. The nonzero quantum resources are increasing along with the effective coupling strength, and the value of entanglement $E_{ac}$ is larger than those of steering $S_{a\to c}$ and $S_{c\to a}$, indicating the higher requirements for generation of one-way and two-way quantum steering. The regional diagram for MO entanglement $E_{ac}$ is illustrated in Fig. \ref{range}b, where the teal area denotes the steady-state case and the golden-yellow area corresponds to the asymptotic entanglement in the unsteady-state evolution. In Fig. \ref{range}c, the regional diagram for asymmetric steering is provided, where the two teal areas represent the nonzero one-way steering $S_{a\to c}$ and $S_{c\to a}$ of the steady-state case (the solid white line between them indicates the zero value for $\kappa_a=\kappa_c$), the two beige areas denote two asymmetric one-way steering in the unsteady-state dynamics, and the golden-yellow area indicates the two-way steering $S_{a\leftrightarrow c}$ for the unsteady-state case.

\subsection*{C.~~Application in MO hybrid systems}
For a concrete MO hybrid system, the effective Hamiltonian corresponding to Eq. \eqref{Heffac} can be constructed by the approach presented in Methods. Then we are able to obtain the rigorous expressions for $g_{\rm{eff}}$ and the persistent Gaussian quantum resources via the analytical CM $v(t)$ in Eq. \eqref{v(t)}. In the dynamics of a realistic MO hybrid system, we introduce a characteristic time $\tau$ to indicate the moment at which the evolution values of entanglement and quantum steering become nearly identical to their analytical asymptotic values at $t\to\infty$, which is expressed as
\begin{equation}\label{tau} 
\tau=\frac{4\pi}{\Omega+\kappa_a+\kappa_c},
\end{equation}
where $\Omega$ is a function of the effective coupling strength $g_{\rm eff}$ and the decay rates $\kappa_a$ and $\kappa_c$. Moreover, in view of the approximations underlying the system Hamiltonian, we further introduce a characteristic time window $t\in[\tau,2\tau]$, over which the quantum resources remain nearly unchanged. Although our general theoretical framework predicts persistent quantum resources in the long-time limit, their persistence in realistic hybrid systems is ultimately constrained by the validity of the approximations employed in deriving the chain-coupled system Hamiltonian in Eq.~\eqref{generalH}. The validity of the characteristic time $\tau$ and the associated time window $t\in[\tau,2\tau]$ for our analytical approach can be further verified numerically against the full system dynamics governed by the original multipartite Hamiltonian.

\begin{figure}[t] 
\centering
\includegraphics[width=0.48\textwidth]{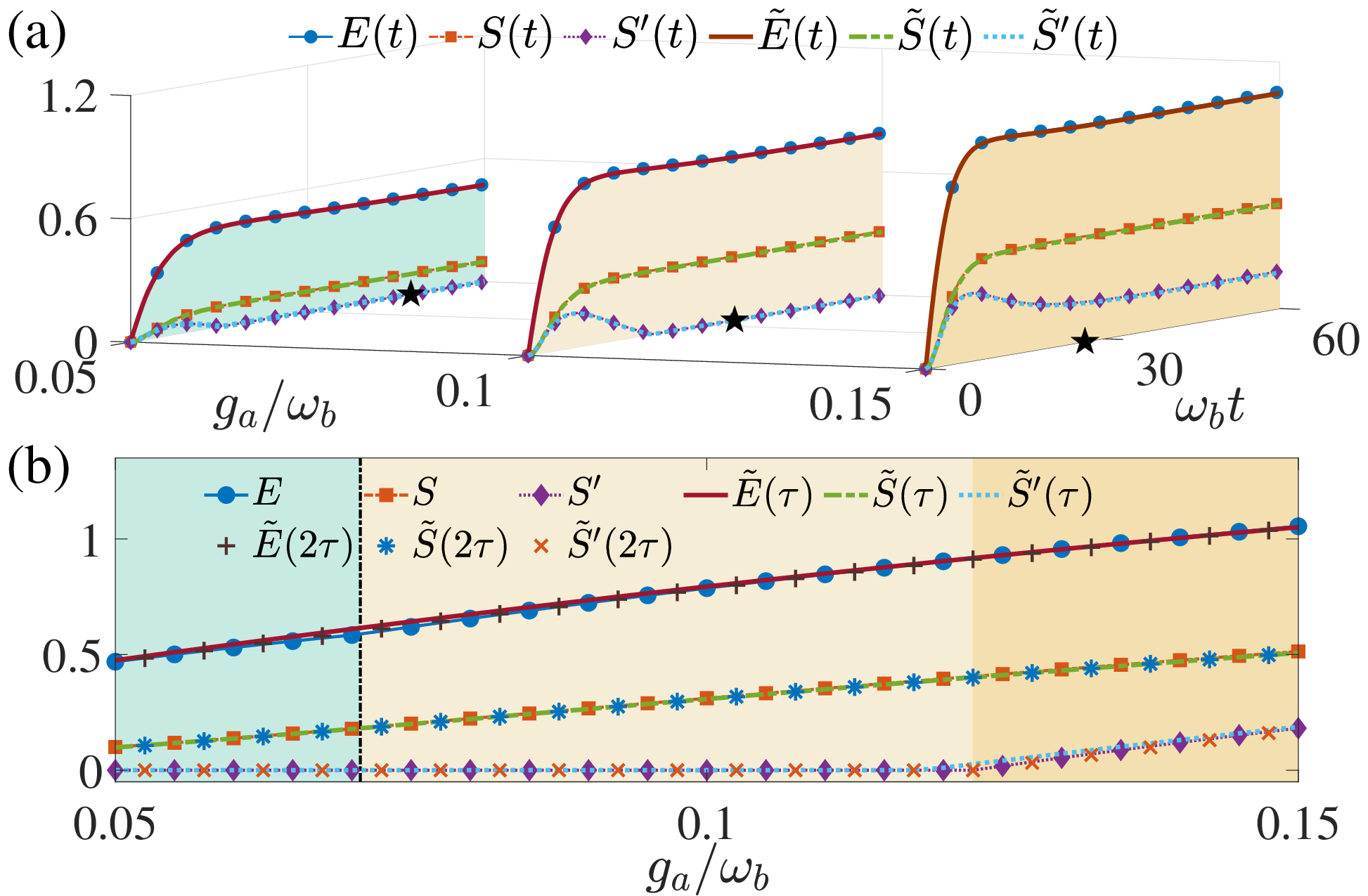}
\caption{\textbf{The persistent MO entanglement and quantum steering in the EOM system.} \textbf{a} The steady-state (the teal region) and unsteady-state (the beige and golden-yellow regions) dynamics of MO quantum resources for three typical values of coupling strength $g_a/\omega_b$ with the star symbols indicating the characteristic time $\tau$. \textbf{b} The asymptotic values of MO quantum resources along with the relative coupling $g_a/\omega_b$, coinciding with the results of the full system dynamics at $t=\tau$ and $t=2\tau$. The MO entanglement is represented by $E$, the steering from mode $a$ to $c$ is denoted $S$, and the one from $c$ to $a$ is represented by $S'$. The quantities with a tilde are the results of full system dynamics. The system parameters are $g_c=0.12\omega_b$, $\Delta_a=5\omega_b$, $\kappa_c=2\kappa_a=10^{-3}\omega_b$, $\kappa_b=10^{-6}\omega_b$, and the thermal occupation numbers are $N_a=N_c=0$, $N_b=10$.}
\label{eom}
\end{figure}

\begin{figure*} 
\centering
\includegraphics[width=0.9\textwidth]{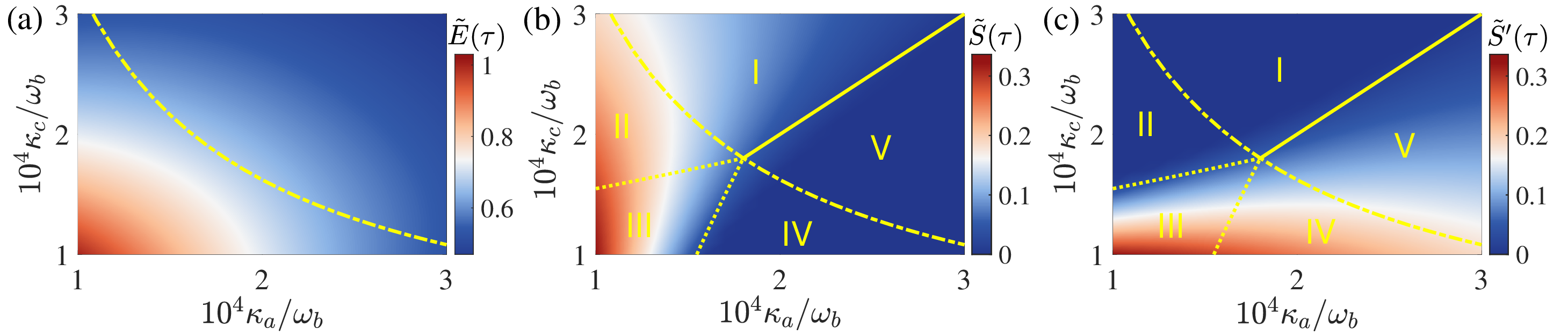}
\caption{\textbf{The regional diagram of persistent MO quantum resources in the COMM system.} \textbf{a} The MO entanglement $\tilde E_{ac}(\tau)$ in the steady-state and unsteady-state evolutions with the yellow dot-dashed line being the boundary. \textbf{b} The MO quantum steering $\tilde S_{a\to c}(\tau)$ in the dynamical process, where the regional map obtained using the effective Hamiltonian method is well reproduced by the numerical results of the full COMM-system dynamics. \textbf{c} The similar case for MO quantum steering $\tilde S_{c\to a}(\tau)$. The parameters are set as $g_a=g_c=0.12\omega_b$, $g_m=0.1\omega_b$, $\Delta_m=\omega_b$, $\Delta_a=3\omega_b$,  $\kappa_m= 10^{-3}\omega_b$, $\kappa_b=10^{-6}\omega_b$, $N_a=N_c=N_m=0$, and $N_b=10$.}
\label{comm}
\end{figure*}

\subsubsection*{Case 1: The electro-optomechanical system}
We first consider the generation of persistent MO entanglement and quantum steering in a multipartite hybrid EOM system~\cite{EOM12,EOM15}, where a mechanical mode $b$ serves as an interface to couple the microwave mode $a$ and the optical mode $c$. After some derivation, we can obtain the linearized Hamiltonian
\begin{equation}\label{HS1} 
H_{S_1}=\omega_bb^\dag b+\sum_{o=a,c}\Delta_oo^\dag o+g_o(o+o^\dag)(b+b^\dag),
\end{equation}
where $\omega_b$ is the transition frequency, $\Delta_o$ denotes the detuning of mode $o$, and $g_o$ represents the coupling strength between the mode $o$ and the mechanical mode $b$ with $o=a, c$. In comparison with general chain-type Hamiltonian in Eq. \eqref{generalH}, the EOM Hamiltonian in Eq. \eqref{HS1} corresponds to $\theta=\phi=\pi/4$ and $N=1$, along with the substitutions $\omega_1\to\omega_b$, $b_1\to b$, and $g_o\to\sqrt{2}g_o$. Then, we construct the effective Hamiltonian $H_{\rm{eff}}$ in Eq. \eqref{Heffac} for the EOM system by the perturbation theory, where the effective coupling strength has the form
\begin{equation}\label{geffeom} 
g_{\rm eff}=\frac{2g_ag_c\omega_b}{\Delta_a^2-\omega_b^2}.
\end{equation}
After substituting this analytical expression into the CM $v(t)$ in Eq. \eqref{v(t)}, we can derive the dynamical MO quantum resources $E_{ac}(t)$, $S_{a\to c}(t)$, $S_{c\to a}(t)$ and the asymptotic values $E_{ac}$, $S_{a\to c}$, $S_{c\to a}$ in Eqs. \eqref{eni} and \eqref{sact2} at $t\to \infty$. Meanwhile, we numerically solve the full EOM dynamics governed by Eq. \eqref{HS1}, which yields the time-dependent quantities $\tilde{E}_{ac}(t)$, $\tilde{S}_{a\to c}(t)$, and $\tilde{S}_{c\to a}(t)$. In Fig. \ref{eom}a, we plot the dynamical process of MO quantum resources for three typical coupling strengths ($g_a/\omega_b=0.05$, $0.1$, and $0.15$), where the results based on our effective Hamiltonian method exhibit good agreement with those obtained by the numerical full EOM-system dynamics. Moreover, the dynamical quantum resources approach their asymptotic values before the characteristic time $\tau$ (the star symbols). In Fig. \ref{eom}b, the asymptotic values of quantum resources in terms of analytical expressions in Eqs. \eqref{eni} and \eqref{sact2} are plotted as $g_a$ increases, which are in good agreement with the numerical results at times $\tau$ and $2\tau$ obtained from the full system dynamics and further validate our theory for persistent quantum-resource generation under both steady-state and unsteady-state evolutions (see Supplementary Note II).

\subsubsection*{Case 2: The cavity optomagnomechanical system}
A yttrium iron garnet crystal is placed inside a microwave cavity (mode $a$) to excite the magnon mode $m$, while simultaneously serving as the vibrating end mirror (mechanical mode $b$) of the optical cavity (mode $c$)~\cite{COMMLJ23}. For this COMM system, we can derive its linearized Hamiltonian
\begin{equation}\label{HS2} 
\begin{aligned}
H_{S_2}=&\sum_{o=a,m,c}\Delta_oo^\dag o+\omega_bb^\dag b+g_a(a^\dag m+am^\dag)\\ &+\sum_{o=m,c}g_o(o+o^\dag)(b+b^\dag),
\end{aligned}
\end{equation}
where $\omega_b$ is the transition frequency, $g_o$s are the coupling strengths, and $\Delta_o$s are the detunings. This COMM system corresponds to the chain-type Hamiltonian in Eq. \eqref{generalH} for $\theta=0$, $\phi=\pi/4$ and $N=2$, with the substitutions $\omega_1\to\Delta_m$, $\omega_2\to\omega_b$, $g_1\to g_m$, $g_c\to\sqrt{2}g_c$, $b_1\to m$, and $b_2\to b$. After constructing the associated $H_{\rm{eff}}$ in Eq. \eqref{Heffac}, we have
\begin{equation}\label{geffcomm} 
g_{\rm eff}
=\frac{2g_ag_mg_c\omega_b}{(\Delta_m-\Delta_a)(\omega_b^2-\Delta_a^2)},
\end{equation}
which can be used to analytically describe the MO dynamics. In Fig. \ref{comm}, we plot the MO entanglement $\tilde E_{ac}(\tau)$ and quantum steering $\tilde S_{a\to c}(\tau)$ and $\tilde{S}_{c\to a}(\tau)$ along with the relative decay rates of microwave and optical modes at the characteristic time $\tau$, where both the asymptotic quantum-resource values and the regional map (the boundaries for the steady-state and unsteady-state evolution as well as different asymmetric areas) derived by the effective Hamiltonian method are well reproduced by the numerical results from full COMM-system dynamics. Moreover, $\tilde E_{ac}(2\tau)$ and $\tilde S_{a\to c}(2\tau)$ are nearly identical to those at the time $\tau$, which coincides with the characteristic time window $t\in[\tau,2\tau]$ over which each corresponding quantum resource remains on its dynamical plateau (see Supplementary Note III).

\section*{III.~~Discussion}
We have developed a general theoretical framework for persistently generating and analytically characterizing MO quantum resources in multipartite chain-type hybrid systems by constructing the effective MO squeezing Hamiltonian $H_{\rm eff}$ given in Eq. \eqref{Heffac}. The effective coupling strength $g_{\rm eff}$ admits a closed-form analytical expression and directly links the system parameters (see Methods). Moreover, the effective-Hamiltonian approach simplifies the multi-mode hybrid-system dynamics within the open-quantum-system framework, and the expressions for the asymptotic values of MO entanglement and quantum steering are analytically accessible, which provides a solid foundation for the high-precision control of MO quantum resources at the quantum interface. Our work provides not only an analytical theory for the generation of persistent MO quantum resources in multipartite hybrid systems but also an effective theoretical tool for the future distributed quantum networks.

\begin{figure}[b] 
\centering
\includegraphics[width=0.48\textwidth]{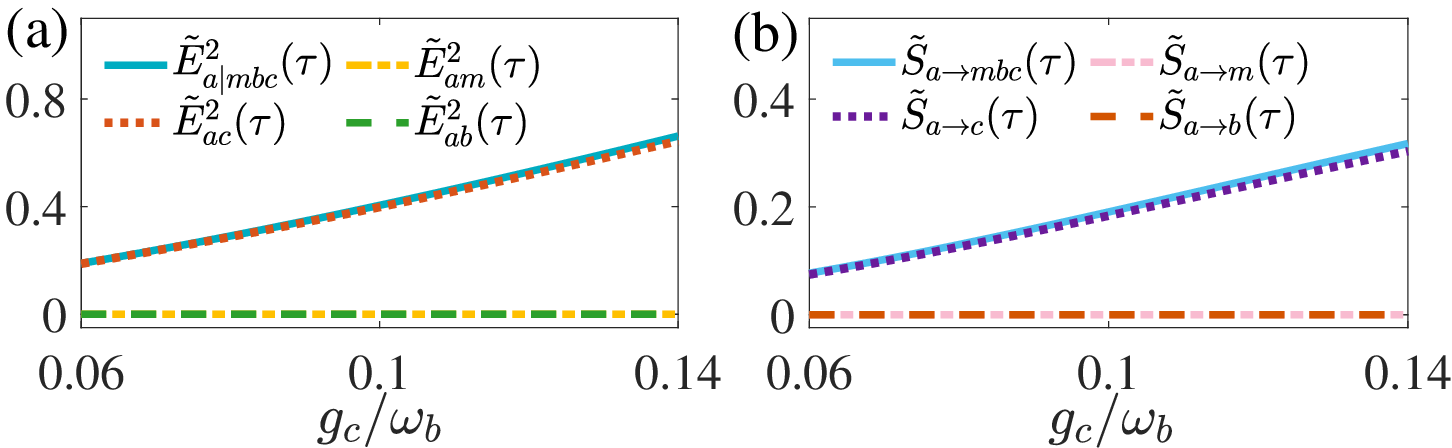}
\caption{\textbf{Multi-parameter optimization-based generation of MO quantum resources.} The distribution properties of entanglement (\textbf{a}) and quantum steering (\textbf{b}) in the COMM system, where the decay rates are $\kappa_a=0.5\kappa_c=10^{-4}\omega_b$, with all other parameters being the same as in Fig.~\ref{comm}.}
\label{monogamy}
\end{figure}

In multipartite hybrid systems, Gaussian quantum entanglement and quantum steering are monogamous~\cite{monogamy06,monogamy07,monogamy16,monogamyP17}, which prevents the quantum resources from being freely shared among the constituents. For example, in the context of the considered COMM system, the entanglement and quantum steering distributions satisfy the following quantitative monogamy inequalities
\begin{eqnarray}
\tilde{E}^2_{a|mbc}&\geq& \tilde{E}^2_{ac}+\tilde{E}^2_{ab}+\tilde{E}^2_{am},\nonumber\\
\tilde{S}_{a\to mbc}&\geq& \tilde{S}_{a\to c}+\tilde{S}_{a\to b}+\tilde{S}_{a\to m},
\end{eqnarray}
where the MO quantum resources $\tilde{E}_{ac}$ and $\tilde{S}_{a\to c}$ are constrained by the allocations of quantum resources in other modes. In Fig. \ref{monogamy}, the quantum resource distributions are plotted along with the increasing of the relative coupling $g_c/\omega_b$ at the characteristic time $\tau$, where the value of MO entanglement $\tilde{E}^2_{ac}(\tau)$ closely approximates that of bipartite multi-mode entanglement $\tilde{E}^2_{a|mbc}(\tau)$ and the values of other two-mode entanglement are negligibly small (the case for MO quantum steering is similar). We also obtain the analogous result for the MO quantum resources in the EOM system (see Supplementary Note IV for details). These distribution properties indicate that our effective Hamiltonian approach enables an efficient multi-parameter optimization strategy for MO quantum-resource generation.

During the unsteady dynamical evolution, the persistent MO quantum resources can survive and approach the finite asymptotic values in Eqs. \eqref{Eac}--\eqref{sact2} surpassing the corresponding steady-state upper bounds (see Supplementary Notes IB and IC for details). Within the framework of open quantum systems, the combined effects of energy loss and the squeezing drive force the MO quantum resource to approach the asymptotic value before the characteristic time $\tau$ given in Eq. \eqref{tau} and subsequently plateau. Moreover, the application to EOM and COMM systems further shows that the persistent MO quantum resource can exhibit the plateau behavior in the characteristic time window $t\in[\tau, 2\tau]$. The validity of our theory is confirmed by the numerical verification of the full dynamics for multi-mode EOM and COMM systems.

Our general theoretical framework can be extended to MO hybrid systems with $N$ chain-coupled intermediate modes. The MOM system~\cite{microwaveopticalG22} corresponds to the case for $N=3$ with the analytical expression of effective coupling strength being available [see Eq. \eqref{gn} in Methods]. It is worth noting that increasing the number of intermediate modes introduces more tunable parameters while imposing additional physical constraints in multipartite MO hybrid systems. Beyond the quantum resource generation in MO subsystem, the general framework can also be applied to arbitrary two target modes in chain-type multipartite systems. As an illustrative case, we apply the analytical theory to cavity magnonic systems~\cite{CM22}, and demonstrate the generation and control of persistent magnon-magnon entanglement and quantum steering (see Supplementary Note V).

\begin{figure} 
\centering
\includegraphics[width=0.48\textwidth]{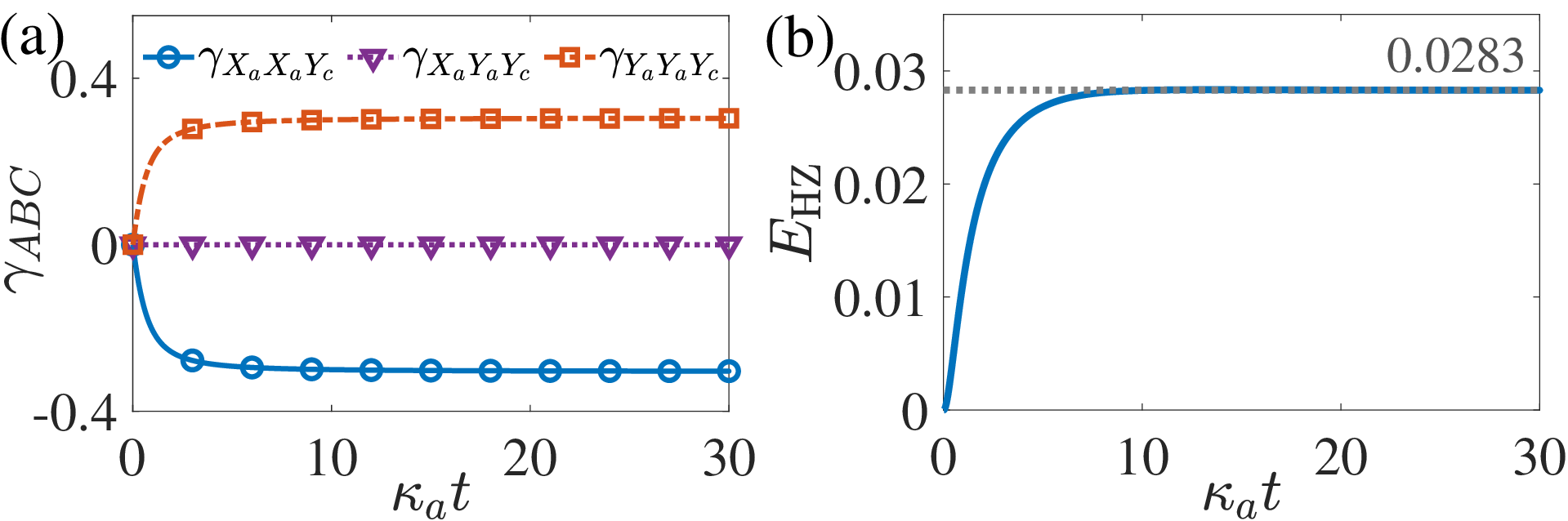}
\caption{\textbf{Persistent high-order correlations and non-Gaussian entanglement in the MO subsystem.} The persistent dynamical behavior of MO third-order correlation $\gamma_{ABC}$ \cite{coskewness} in panel (\textbf{a}) and the indicator of non-Gaussian MO entanglement $E_{\rm HZ}$ \cite{HZ06} in panel (\textbf{b}) with parameter settings given in Supplementary Note VI.}
\label{g013d}
\end{figure}

Non-Gaussian quantum resources play an important role in universal quantum computing. In a three-mode microwave-intermediate-optical hybrid system, the generation of persistent non-Gaussian MO resources can be realized by invoking the nonlinear interactions, where our general effective Hamiltonian approach still holds (see Methods). As an illustrated example, we derive an effective two-mode trisqueezing nonlinear interaction~\cite{coskewness} for MO subsystems 
\begin{equation}\label{nlHeff}
H_{\rm eff}=g_{\rm eff}(a^{\dag 2}c^\dag+a^2c),
\end{equation}
where the analytical expression of the effective coupling $g_{\rm eff}$ is accessible. Under the open-quantum-system framework, the persistent MO third-order correlation quantified by the nonlinear coskewness ~\cite{coskewness} is obtained, and the non-Gaussian MO entanglement~\cite{HZ06} is effectively generated as shown in Fig. \ref{g013d} (see Supplementary Note VI for details).

\section*{IV.~~Methods: effective Hamiltonian approach}

In this work, the effective-Hamiltonian approach enables a multipartite MO hybrid quantum system to be reduced to a description involving only two target modes. Here, we first derive the effective MO Hamiltonian with an analytical coupling strength for the chain-type multipartite hybrid system, and then extend this approach to the case involving nonlinear interactions.

For the linear chain-coupled Hamiltonian in Eq.~\eqref{generalH}, the microwave and optical modes are coupled via $N$ intermediate modes. According to the perturbation theory~\cite{perturbation16}, in the large-detuning regime, $g_a, g_s, g_c \ll \{|\Delta_a-\omega_s|, |\Delta_c-\omega_s|$\}, the Hamiltonian $H_0$ in Eq. \eqref{generalH} can be treated as the unperturbed term, while the interaction term $V$ is regarded as a perturbation. When the detuning of the microwave mode is approximately opposite to that of the optical mode, namely $\Delta_a\approx-\Delta_c$, the energy differences between certain nondegenerate quantum states of the unperturbed Hamiltonian $H_0$ can be comparable to, or even smaller than, the strength of the perturbation $V$. Such states, although formally nondegenerate, can therefore be regarded as nearly degenerate. Under these conditions, the linear interaction described by $V$ can effectively couple the nearly degenerate states, resulting in an effective Hamiltonian defined within the reduced Hilbert subspace spanned by them.

Specifically, we consider two near-degenerate eigenstates $|i\rangle\equiv|n\rangle_a|l_1\rangle_{b_1}\cdot\cdot\cdot|l_N\rangle_{b_N}|k\rangle_c$ and $|j\rangle\equiv|(n+1)\rangle_a|l_1\rangle_{b_1}\cdot\cdot\cdot|l_N\rangle_{b_N}|(k+1)\rangle_c$ of the free Hamiltonian $H_0$, which can be effectively coupled via the perturbation term $V$. The effective Hamiltonian within the subspace spanned by $\{|i\rangle,|j\rangle\}$ can be formally written as
\begin{equation}\label{Heffappa} 
\begin{aligned}
H_{\rm eff}=&\epsilon_i|i\rangle\langle i|+(\Delta_a+\Delta_c+\epsilon_j)|j\rangle\langle j|\\&+\tilde g_{\rm eff}(|i\rangle\langle j|+|j\rangle\langle i|).
\end{aligned}
\end{equation}
Here, $\epsilon_i$ and $\epsilon_j$ represent the energy shifts caused by the coupling for states $|i\rangle$ and $|j\rangle$, respectively, and $\tilde g_{\rm eff}$ is the effective coupling strength between two target states resulting from the interaction term $V$. These are the three coefficients to be determined in this ansatz. Note that we here omitted~\cite{perturbation25} the common unperturbed eigenenergies of the two bases $n\Delta_a+\sum_{s=1}^{N}l_s\omega_s+k\Delta_c$.

According to the standard perturbation theory~\cite{perturbation16}, the effective coupling strength $\tilde{g}_{\rm eff}$ between these two states arising from an $N$-th order perturbed process can be expressed as
\begin{equation}\label{generalgeff} 
\tilde{g}_{\rm eff}=\sum_{m_1m_2\dots m_{N}}\frac{V_{jm_{N}}\dots V_{m_2m_1}V_{m_1i}}{(E_i-E_{m_{N}})\dots(E_i-E_{m_1})}
\end{equation}	
where $V_{m_{k+1}m_k}\equiv\langle m_{k+1}|V|m_k\rangle$ and $|m_k\rangle$ denotes an eigenstate of $H_0$ with eigenenergy $E_{m_k}$. Under the condition that the perturbation strength is much smaller than that of the unperturbed term, i.e., $V_{m_{k+1}m_k}\ll |E_{m_{k+1}}-E_{m_k}|$, the effective coupling strength at $N$-th order is significantly larger than that at $(N+1)$-th order. Consequently, the indirect coupling between eigenstates $|i\rangle$ and $|j\rangle$ is approximated by the dominant leading-order contribution, with higher-order corrections neglected. In particular, when the second-order contribution is included in the effective coupling strength, all third- and higher-order corrections are omitted.

\begin{figure} 
\centering
\includegraphics[width=0.4\textwidth]{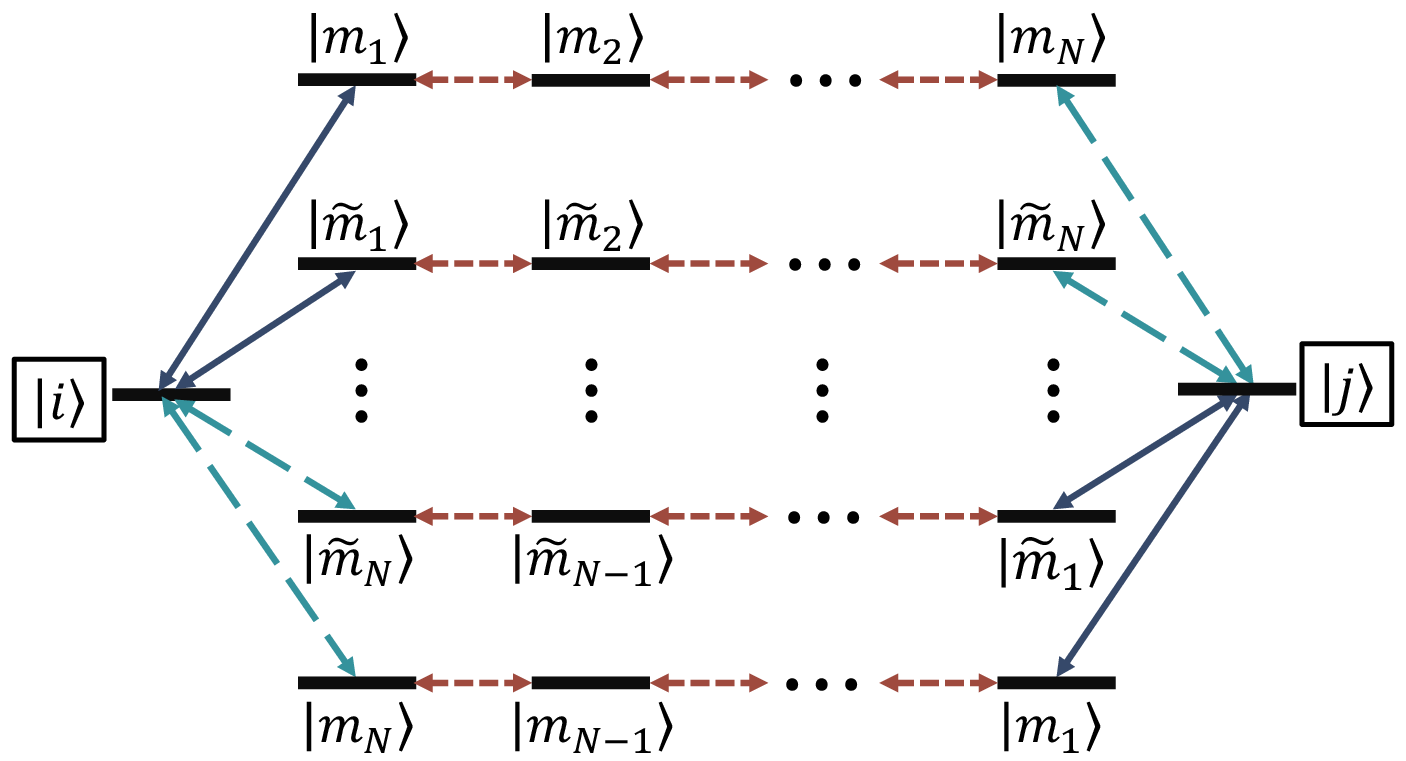}
\caption{\textbf{All leading-order paths connecting $|i\rangle$ and $|j\rangle$ are illustrated.} Here, the target states are $|i\rangle\equiv|n\rangle_a|l_1\rangle_{b_1}\cdot\cdot\cdot|l_N\rangle_{b_N}|k\rangle_c$ and $|j\rangle\equiv|(n+1)\rangle_a|l_1\rangle_{b_1}\cdot\cdot\cdot|l_N\rangle_{b_N}|(k+1)\rangle_c$, and the intermediate states are $|m_1\rangle=|(n+1)(l_1+1)l_2\cdots l_Nk\rangle$, $|m_2\rangle=|(n+1)l_1(l_2+1)\cdots l_Nk\rangle$, $|m_N\rangle=|(n+1)l_1l_2\cdots (l_N+1)k\rangle$, $|\tilde m_1\rangle=|(n+1)(l_1-1)l_2\cdots l_Nk\rangle$, $|\tilde m_2\rangle=|(n+1)l_1(l_2-1)\cdots l_Nk\rangle$, and $|\tilde m_N\rangle=|(n+1)l_1l_2\cdots (l_N-1)k\rangle$. The solid lines represent the coupling between the microwave mode $a$ and the intermediate mode $b_1$, the dashed lines indicate the coupling between the optical mode $c$ and the intermediate mode $b_N$, and the dotted lines denote the coupling between intermediate modes $s$ and $s+1$ ($s=1,2,\dots, N$).}
\label{acpath}
\end{figure}

From the definition of $\tilde g_{\rm eff}$ in Eq. \eqref{generalgeff} and the linear-coupling Hamiltonian in Eq. \eqref{generalH}, the effective coupling strength between two target states $|i\rangle$ and $|j\rangle$ is obtained by summing the leading-order contributions from all paths shown in Fig. \ref{acpath}. Then, we can derive
\begin{equation}\label{geffsub} 
\tilde{g}_{\rm eff}=\sqrt{(n+1)(k+1)}g_{\rm eff},
\end{equation}
where the parameter $g_{\rm eff}$ is given by
\begin{equation}\label{gn} 
\begin{aligned}
g_{\rm eff}=&\begin{cases}& \!\!\!\!\! g_ag_c\left(\cfrac{\cos\theta\sin\phi}{\Delta_a-\omega_1}-\cfrac{\sin\theta\cos\phi}{\Delta_a+\omega_1}\right),~N=1\\[2em]
& \!\!\!\!\! g_ag_1g_c \!\left(\displaystyle\prod_{s=2}^{N-1}\frac{2g_s\omega_s}{\Delta_a^2-\omega_s^2}\!\right)\!\left(\!\cfrac{\sin\theta}{\Delta_a+\omega_1}-\cfrac{\cos\theta}{\Delta_a-\omega_1}\!\right)\\
\end{cases}\\
&\quad\quad \times\left(\!\cfrac{\cos\phi}{\Delta_a+\omega_N}-\cfrac{\sin\phi}{\Delta_a-\omega_N}\!\right),~N\ge2
\end{aligned},
\end{equation}
with the coupling strengths $g_a$, $g_s$, and $g_c$, the detuning $\Delta_a$, the transition frequencies $\omega_1$, $\omega_s$, and $\omega_N$, as well as the angles $\theta$ and $\phi$, following from Eq. \eqref{generalH}.
When $s=2$, the value of $\prod_{s=2}^{N-1}[{2g_s\omega_s}/{(\Delta_a^2-\omega_s^2)}]=1$.
The discrepancy between the cases $N=1$ and $N\ge 2$ is attributed to the line-type structure of the Hamiltonian in Eq. \eqref{generalH}. The effective coupling strength $\tilde g_{\rm eff}$ is proportional to $\sqrt{(n+1)(k+1)}$ and independent of the excitation numbers of intermediate modes. 

Then, we consider the energy shifts of the eigenstates $|i\rangle$ and $|j\rangle$, given by
\begin{equation}\label{generalepsilon} 
\epsilon_i=\sum_m\frac{|V_{mi}|^2}{E_i-E_m},
\epsilon_j=\sum_m\frac{|V_{jm}|^2}{E_j-E_m}.
\end{equation}	
These results are obtained by summing over all virtual paths from $|i\rangle\to |i\rangle$ ($|j\rangle\to |j\rangle$) through intermediate states, corresponding to second-order perturbation processes as shown in Fig. \ref{acpath}. The resulting energy shifts depend only on the target modes and their nearest neighboring modes, and are independent of other intermediate states. When the condition $\epsilon_i=\Delta_a+\Delta_c+\epsilon_j$ is satisfied, the diagonal terms in the first line of Eq. \eqref{Heffappa} form the identity operator within the considered subspace. Defining the difference $\delta\equiv\epsilon_i-\epsilon_j$, we obtain
\begin{equation}\label{generaldelta} 
\delta=\frac{g_a^2[\omega_1+\Delta_a\cos(2\theta)]}{\omega_1^2-\Delta_a^2}+\frac{g_c^2[\omega_N+\Delta_c\cos(2\phi)]}{\omega_N^2-\Delta_c^2},
\end{equation}
which is independent of particular choice of the near-degenerate states.

Under the condition $\Delta_a=-\Delta_c+\delta$, the effective Hamiltonian in Eq. \eqref{Heffappa} can be further reduced to
\begin{equation}\label{Heff1} 
\begin{aligned}
H_{\rm eff}&=\tilde g_{\rm eff}(|i\rangle\langle j|+|j\rangle\langle i|)\\
&=\tilde{g}_{\rm eff}(|nk\rangle \langle (n+1)(k+1)|+{\rm h.c.})\\
&\quad\otimes |l_1l_2\cdots l_N\rangle\langle l_1l_2\cdots l_N|.
\end{aligned}
\end{equation}
According to the definition given in Eq. \eqref{geffsub}, the effective Hamiltonian after eliminating the decoupled intermediate modes is reduced to
\begin{equation}\label{Heffsub} 
\begin{aligned}
H_{\rm eff}=&\sqrt{(n+1)(k+1)}g_{\rm eff}\\
&\times\left[|nk\rangle \langle (n+1)(k+1)|+{\rm h.c.}\right].
\end{aligned}
\end{equation}
Since this result holds for arbitrary $n$ and $k$ and the bosonic operators satisfy $a^\dag|n\rangle=\sqrt{n+1}|n+1\rangle$ and $c^\dag|k\rangle=\sqrt{k+1}|k+1\rangle$, the Hamiltonian in Eq. \eqref{Heffsub} defined on this subspace can be straightforwardly extended to the full Hilbert space of modes $a$ and $c$. Therefore, we eventually obtain 
\begin{equation}\label{Heffac1} 
H_{\rm eff}=g_{\rm eff}(a^\dag c^\dag+ac),
\end{equation}
where the analytical expression of $g_{\rm eff}$ has the form given in Eq. \eqref{gn}. The parameter $g_{\rm eff}$ denotes the effective coupling strength between microwave and optical modes, which depends on the coupling strengths, detunings, and transition frequencies of the intermediate modes. For the fixed $\{\Delta_a,\omega_s\}$, $g_{\rm eff}$ can be enhanced by increasing original coupling strengths. Taken together, Eqs.~\eqref{Heffac1} and \eqref{gn} provide explicit analytical expressions for the effective Hamiltonian and the corresponding coupling strength, respectively. It is worth noting that, under the conditions of large detuning perturbation and the degenerate-state approximation, the jointly tunable parameter space for the effective coupling strength $g_{\rm eff}$ is a limited but well-defined region, which delineates the regime of efficient control by filtering out parameter regions where the effective coupling is suppressed.

Our general theoretical framework can be applied to typical multipartite hybrid quantum systems, including the EOM and COMM systems with the effective coupling strengths given in Eqs. \eqref{geffeom} and \eqref{geffcomm}. Once the system parameters are given, the effective coupling strength $g_{\rm eff}$ between the two target modes can be directly obtained in terms of Eq. \eqref{gn}. The explicit construction of the effective MO Hamiltonians for the EOM and COMM systems, together with the corresponding path diagrams, are provided in Supplementary Notes IIA and IIIA, respectively.

This effective-Hamiltonian approach can be further extended to microwave-intermediate-optical hybrid systems with nonlinear interactions, enabling the generation of persistent non-Gaussian quantum resources. For the case of a single intermediate mode [$N=1$ in Eq. \eqref{generalH}], the corresponding Hamiltonian reads
\begin{equation}\label{EMgennlH}
\begin{aligned}
H=&\Delta_aa^\dag a+\omega_b b^\dag b+\Delta_c c^\dag c+g_a[\cos\theta(a^{\dag s} b+a^sb^\dag)\\&+\sin\theta(a^{\dag s} b^\dag+a^sb)]+g_c[\cos\phi(c^{\dag w} b+c^wb^\dag)\\&+\sin\phi(c^{\dag w} b^\dag+c^wb)],
\end{aligned}
\end{equation}
where the positive integers $s$ and $w$ specify the orders of the nonlinear interactions associated with modes $a$ and $c$, and the other parameters are defined as described in Eq. \eqref{generalH}.
When $g_a,g_c\ll\{|s\Delta_a-\omega_b|,|w\Delta_c-\omega_b|\}$ and $s\Delta_a\approx-w\Delta_c$, the effective Hamiltonian after neglecting the second-order energy shifts can be written as
\begin{equation}\label{EMgenHeff}
H_{\rm eff}=g_{\rm eff}(a^{\dag s}c^{\dag w}+a^sc^w),
\end{equation}
where the effective coupling strength is
\begin{equation}\label{EMgengeff}
g_{\rm eff}=g_ag_c\left(\cfrac{\cos\theta\sin\phi}{s\Delta_a-\omega_b}-\cfrac{\sin\theta\cos\phi}{s\Delta_a+\omega_b}\right).
\end{equation}
For the case of $s=2$, $w=1$, $\theta=0$, $\phi=\pi/4$, and $g_c\rightarrow\sqrt{2}g_c$, the general expression for the effective Hamiltonian in Eq. \eqref{EMgenHeff} reduces to the effective two-mode trisqueezing interaction given in Eq. \eqref{nlHeff}.

\section*{Data availability}
The data underlying the findings reported in this work are not publicly available. The data are available from the corresponding author upon reasonable request.

\section*{Code availability}
The codes used to generate the data presented in this work are available from the corresponding author upon reasonable request.

\section*{Acknowledgments}
This work was supported by NSFC (Grants No. 12404405, No. 12404330, and No. 11575051), the Guangdong Provincial Quantum Science Strategic Initiative (Grant No.GDZX2404001), Hebei NSF (Grant No. A2021205020 and No. A2025205030), Hebei 333 Talent Project (No. B20231005), and the funds of Hebei Normal University (Grants No. L2024B10 and No. L2026J02).

\section*{Author Contributions}
All authors discussed the work together and decided the structure and contents of the paper. F.L. wrote the paper with the help of comments from all authors. S.F.Q., Z.D.W. and Y.K.B. contributed the theoretical analysis. Z.D.W. and Y.K.B. conceived the project.

\section*{Competing interests}
The authors declare no competing interests.

\section*{Additional information}
\textbf{Supplementary information} This manuscript contains the supplementary material.

\textbf{Correspondence} and requests for materials should be addressed to Yan-Kui Bai.

%

\onecolumngrid
\clearpage

%

\setcounter{equation}{0}
\setcounter{figure}{0}
\setcounter{section}{0}
\setcounter{page}{1}
\setcounter{secnumdepth}{2}

\renewcommand{\theequation}{S\arabic{equation}}
\renewcommand{\thesection}{Supplementary Note \Roman{section}}

\renewcommand{\figurename}{Supplementary Figure}
\renewcommand{\thefigure}{\arabic{figure}}

\renewcommand{\theHfigure}{supp.figure.\arabic{figure}}
\renewcommand{\theHequation}{supp.equation.\arabic{equation}}
\renewcommand{\theHsection}{supp.section.\arabic{section}}

\begin{center}
\textbf{\large Supplementary Information for ``General theory of persistent microwave-optical quantum resources in hybrid-system dynamics''}
\end{center}

\begin{center}
Fan Li,$^1$ Shi-fan Qi,$^{1,*}$ Z. D. Wang,$^{2,3,4,\dag}$ and Yan-Kui Bai$^{1,4,\ddag}$\\
\vspace{0.3em}

\small {
\textit{$^1$College of Physics and Hebei Key Laboratory of Photophysics Research and Application,\\
Hebei Normal University, Shijiazhuang, Hebei 050024, China\\
$^2$Frontier Research Institute for Physics, South China Normal University, Guangzhou 510006, China\\
$^3$HK Institute of Quantum Science \& Technology and Department of Physics,\\
The University of Hong Kong, Pokfulam Road, Hong Kong, China\\
$^4$Hong Kong Branch for Quantum Science Center of Guangdong-Hong Kong-Macau Greater Bay Area, Shenzhen 518045, China}
}
\end{center}

\onecolumngrid
\begin{center}
\normalsize
\textbf{\normalsize CONTENTS}
\end{center}

\begin{tabbing}
\hspace{50.2em}\=\hspace{15em}\=\kill
{\color{blue}Supplementary Note I.~~}\hyperref[generala-c]{System dynamics and Gaussian quantum resources under the effective Hamiltonian} \>{2} \\
\quad~~{\color{blue}A.~~}\hyperref[SecCM]{Analytical solution for the covariance matrix} \>{2}\\
\quad~~{\color{blue}B.~~}\hyperref[SecMOen]{Microwave-optical entanglement} \>{3}\\
\quad~~{\color{blue}C.~~}\hyperref[SecMOst]{Microwave-optical quantum steering} \>{4}\\
\quad~~{\color{blue}D.~~}\hyperref[sectau]{The characteristic time window and its verification} \>{5}\\[2ex]

{\color{blue}Supplementary Note II.~~}\hyperref[SMEOM]{Application in electro-optomechanical systems} \>{7} \\
\quad~~{\color{blue}A.~~}\hyperref[SMEOMH]{Model and the effective Hamiltonian} \>{7}\\
\quad~~{\color{blue}B.~~}\hyperref[eomAmatrix]{System dynamics and quantum resources} \>{8}\\
\quad~~{\color{blue}C.~~}\hyperref[ValidityEOM]{Validity analysis and experimental feasibility} \>{9}
\\[2ex]

{\color{blue}Supplementary Note III.~~}\hyperref[SMCOMM]{Application in cavity optomagnomechanical systems} \>{11} \\
\quad~~{\color{blue}A.~~}\hyperref[SMCOMMH]{Model and the effective Hamiltonian} \>{11}\\
\quad~~{\color{blue}B.~~}\hyperref[commAmatrix]{Quantum resources and validity analysis} \>{13}\\[2ex]

{\color{blue}Supplementary Note IV.~~}\hyperref[MES]{The distributions of quantum resources in multipartite microwave-optical hybrid systems} \>{15}\\[2ex]

{\color{blue}Supplementary Note V.~~}\hyperref[SMCM]{Magnonic quantum resources in multipartite hybrid systems with linear interactions} \>{16} \\
\quad~~~{\color{blue}A.~~}\hyperref[SMCMH]{Model and the effective Hamiltonian} \>{16}\\
\quad~~~{\color{blue}B.~~}\hyperref[cmAmatrix]{System dynamics and quantum resources} \>{16}\\[2ex]

{\color{blue}Supplementary Note VI.~~}\hyperref[SMng]{Non-Gaussian microwave-optical quantum resources in multipartite nonlinear systems} \>{17} \\[2ex]

\quad~~~\hyperref[reference-sup]{References} \>{19}\\
\end{tabbing}
\newpage
\twocolumngrid

\section{System dynamics and Gaussian quantum resources under the effective Hamiltonian}\label{generala-c}
In this Supplementary Note, we investigate the system dynamics within the open-quantum-system framework governed by the effective Hamiltonian in Eq. (2) of the main text and analyze the generation of persistent entanglement and quantum steering between the target modes $a$ and $c$.

\subsection{Analytical solution for the covariance matrix}\label{SecCM}
Based on the effective Hamiltonian in Eq. (2) of the main manuscript, the dynamics of the target microwave-optical (MO) system under Markovian environments can be described by the following quantum Langevin equations (QLEs),
\begin{equation}\label{SMappaqleac}
\begin{aligned}
\dot a&=-ig_{\rm eff}c^\dag-\kappa_a a+\sqrt{2 \kappa_a}a_{in},\\
\dot c&=-ig_{\rm eff}a^\dag-\kappa_c c+\sqrt{2 \kappa_c}c_{in},	
\end{aligned}
\end{equation}
where $g_{\rm eff}$ denotes the effective coupling strength. $\kappa_a$ and $\kappa_c$ represent the decay rates of the modes $a$ and $c$, respectively. $a_{in}$ and $c_{in}$ are Gaussian noise operators and characterized by their covariance functions, $\langle o_{in}(t)o^\dag_{in}(t')\rangle=(N_o+1)\delta(t-t')$ and $\langle o^\dag_{in}(t)o_{in}(t')\rangle=N_o\delta(t-t')~ (o=a,c)$, where $N_o=[\exp(\hbar\omega_o/k_BT)-1]^{-1}$ is the mean population number of mode $o$ at the thermal equilibrium state, $k_B$ is the Boltzmann constant and $T$ is the environmental temperature.

The QLEs in Eq. \eqref{SMappaqleac} can be written in a matrix form
\begin{align}\label{SMueff}
\begin{aligned}
\dot u(t)=A_{\rm eff}u(t)+\xi_{\rm eff}(t),
\end{aligned}
\end{align}
where $u(t)=[X_{a}(t),Y_{a}(t),X_{c}(t),Y_{c}(t)]^T$ and $X_o(t)=(o+o^\dag)/\sqrt{2}, Y_o(t)=i(o^\dag-o)/\sqrt{2}~(o=a,c)$. The drift matrix $A_{\rm eff}$ is given by
\begin{align}
A_{\rm eff}=-\begin{pmatrix}
\kappa_a&0&0&g_{\rm eff}\\
0&\kappa_a&g_{\rm eff}&0\\
0&g_{\rm eff}&\kappa_c&0\\
g_{\rm eff}&0&0&\kappa_c\\
\end{pmatrix}.
\end{align}
$\xi_{\rm eff}(t)=[\sqrt{2\kappa_a}X_{a}^{in},\sqrt{2\kappa_a}Y_{a}^{in},\sqrt{2\kappa_c}X_{c}^{in},\sqrt{2\kappa_c}Y_{c}^{in}]^T$ is the vector of Gaussian noise operators, where $X^{in}_o=(o_{in}+o_{in}^\dag)/\sqrt{2},~ Y^{in}_o=i(o^\dag_{in}-o_{in})/\sqrt{2}$.

Given that the quantum state is Gaussian, the system dynamics can be fully characterized by a $4\times 4$ covariance matrix (CM). With the QLE in Eq. \eqref{SMueff}, the dynamics of the CM $v(t)$ satisfies
\begin{equation}\label{SMhatv}
\dot{v}(t)=A_{\rm eff}{v}(t)+{v}(t)A_{\rm eff}^T+D_{\rm eff}.
\end{equation}
The elements of $v(t)$ are defined as $v_{jl}(t)=\langle u_j(t)u_l(t)+ u_l(t)u_j(t)\rangle/2-\langle u_j(t)\rangle\langle u_l(t)\rangle~(j,l=1,2,3,4)$, where $u_j(t)$ is the $j$-term of $u(t)$ in Eq. \eqref{SMueff}. $D_{\rm eff}= {\rm diag}[\kappa_a(2N_a+1),\kappa_a(2N_a+1),\kappa_c(2N_c+1),\kappa_c(2N_c+1)]$ is the diffusion matrix and defined by $D_{{\rm eff}_{jl}} =\langle \xi_{{\rm eff}_{j}}(t)\xi_{{\rm eff}_{l}}(t) + \xi_{{\rm eff}_{l}}(t) \xi_{{\rm eff}_{j}}(t)\rangle/2$. 

At the initial time, two target modes are assumed to be in thermal states, i.e., the CM satisfies ${v}(0)={\rm diag}[N_a+1/2,N_a+1/2,N_c+1/2,N_c+1/2]$. Under this condition, the non-zero matrix elements in ${v}(t)$ can be analytically derived as
\begin{equation}
\begin{aligned}\label{SMv(t)}
v_{11}(t)=&(1+\sin\varphi)c_-e^{-(\Omega+\kappa_a+\kappa_c)t}-c_0e^{-(\kappa_a+\kappa_c)t}\\&+(1-\sin\varphi)c_+e^{(\Omega-\kappa_a-\kappa_c)t}+c_1,\\
v_{44}(t)=&(1-\sin\varphi)c_-e^{-(\Omega+\kappa_a+\kappa_c)t}+c_0e^{-(\kappa_a+\kappa_c)t}\\&+(1+\sin\varphi)c_+e^{(\Omega-\kappa_a-\kappa_c)t}+c_2,\\
v_{14}(t)=&\cos\varphi{c_-}e^{-(\Omega+\kappa_a+\kappa_c)t}+\tan\varphi c_0 e^{-(\kappa_a+\kappa_c)t}\\&-\cos\varphi c_+e^{(\Omega-\kappa_a-\kappa_c)t}+c_3,
\end{aligned}
\end{equation}
and $v_{22}(t)=v_{11}(t), v_{33}(t)=v_{44}(t), v_{23}(t)=v_{32}(t)=v_{41}(t)=v_{14}(t)$. The parameters are defined as
\begin{equation}\label{SMOmega}
\begin{aligned}
&\Omega=\sqrt{4g_{\rm eff}^2+(\kappa_a-\kappa_c)^2},~
\tan\varphi=\frac{\kappa_a-\kappa_c}{2g_{\rm eff}},\\
&c_{\pm}=\pm\frac{\kappa_+\mp\sin\varphi\kappa_-}{4[\Omega\mp(\kappa_a+\kappa_c)]}+\frac{N_+\mp\sin\varphi N_-+1}{4},\\
&c_0=\frac{\cos^2\varphi}{2}\left[\frac{\kappa_-}{\kappa_a+\kappa_c}-N_-\right],~N_{\pm}=N_a\pm N_c,\\
&\kappa_{\pm}=\kappa_a(2N_a+1)\pm\kappa_c(2N_c+1).\\
\end{aligned}
\end{equation}
The constants $c_1$, $c_2$, and $c_3$ are
\begin{equation}
\begin{aligned}\label{SMc1c2c3}
c_1&=N_a+\frac{1}{2}-\frac{g_{\rm eff}}{\kappa_a}c_3,\quad
c_2=N_c+\frac{1}{2}-\frac{g_{\rm eff}}{\kappa_c}c_3,\\
c_3&=\frac{g_{\rm eff}\kappa_a\kappa_c(N_a+N_c+1)}{(\kappa_a+\kappa_c)(g_{\rm eff}^2-\kappa_a\kappa_c)}.
\end{aligned}
\end{equation}

The steady-state regime requires the CM elements to be invariant, i.e., $\dot{v}(t)=0$. Under this condition, the elements are $v_{11}=c_1$, $v_{44}=c_2$ and $v_{14}=c_3$. Through the definitions in Eq. \eqref{SMc1c2c3}, a legitimate CM requires $g_{\rm eff}^2<\kappa_a\kappa_c$. Additionally, in the steady-state regime, one can easily demonstrate that the exponent factor $\Omega-\kappa_a-\kappa_c<0$ in Eq. \eqref{SMv(t)} via the definition of $\Omega$ in Eq. \eqref{SMOmega}. That leads to $v_{11}(\infty)=c_1$, $v_{44}(\infty)=c_2$, and $v_{14}(\infty)=c_3$. These elements under steady states are the asymptotic values as $t\to\infty$. By contrast, when $g_{\rm eff}^2>\kappa_a\kappa_c$, the CM elements are exponentially divergent due to the exponent factor $\Omega-\kappa_a-\kappa_c>0$, yielding unsteady CM dynamics of the system. At the critical point, $g_{\rm eff}^2=\kappa_a\kappa_c$, the exponent factor satisfies $\Omega-\kappa_a-\kappa_c=0$, and the system's CM exhibits an approximately linear divergence in this case. Accordingly, the system dynamics can be classified into two distinct regimes, the steady-state regime characterized by 
\begin{equation}\label{gss}
g_{\rm eff}^2<\kappa_a\kappa_c
\end{equation}
and the unsteady-state regime corresponding to
\begin{equation}\label{gus}
g_{\rm eff}^2\ge\kappa_a\kappa_c.
\end{equation}
The steady-state and unsteady-state regimes can also be identified from the Lyapunov stability analysis of the drift matrix~\cite{SMlyap87}, with the steady-state condition requiring all eigenvalues of the drift matrix $A_{\rm eff}$ to have negative real parts.

Moreover, when deriving the dynamics of the target MO system within the framework of open quantum systems, we retain only the environmental noise acting directly on the target modes and neglect the noise acting on the intermediate modes that mediate the effective coupling. This is justified because the Markovian environments associated with different modes are assumed to be local and independent, so that the corresponding input noises are uncorrelated. Consequently, the environmental noise of the intermediate modes has a negligible effect on the target modes.

\subsection{Microwave-optical entanglement}\label{SecMOen}
We now analyze the bipartite entanglement between the microwave mode $a$ and the optical mode $c$, based on the system dynamics presented in \ref{SecCM}. The resulting MO entanglement is quantified via the logarithmic negativity (LN)~\cite{SMentanglement07}, which is defined as
\begin{equation}\label{SMenac}
E_{ac}(t)=\max[0, E_{ac}^\prime(t)],
\end{equation}
where the quantity $E_{ac}^\prime(t)=-\ln(2\eta^-_{ac})$ and $\eta^-_{ac}=[\Gamma-(\Gamma^2-4\det v)^{1/2}]^{1/2}/\sqrt{2}$ is the minimum symplectic eigenvalue of the partial transpose of the CM $v(t)=[v_a,v_{ac};v_{ac}^T,v_c]$, with $v_a$, $v_c$, and $v_{ac}$ denoting the $2\times2$ subblocks of $v(t)$, and $\Gamma\equiv\det v_a+\det v_c-2\det v_{ac}$.
By the definition in Eq. \eqref{SMenac} and CM shown in Eq. \eqref{SMv(t)}, we can obtain the LN quantity
\begin{equation}\label{SMent}
\begin{aligned}
&E_{ac}^\prime(t)=-\ln[\zeta_{ac}(t)],\\
&\zeta_{ac}(t)=(v_{11}+v_{44})(1-\sqrt{1+x}),\\
&x=\frac{4(v_{14}^2-v_{11}v_{44})}{(v_{11}+v_{44})^2}.
\end{aligned}
\end{equation}

In the steady-state regime, $g^2_{\rm eff}<\kappa_a\kappa_c$, all of the CM elements given in Eq. \eqref{SMv(t)} approach steady values when $t\to\infty$, i.e., $v_{11}(\infty)=c_1$, $v_{44}(\infty)=c_2$, and $v_{14}(\infty)=c_3$. By substituting the values of $c_1$, $c_2$, and $c_3$ in Eq. \eqref{SMc1c2c3} into Eq. \eqref{SMent}, the finite asymptotic value LN quantity at $t\to\infty$ can be derived as
\begin{equation}\label{SMeninftynac}
\begin{aligned}
E_{ac}^\prime=-\ln(\zeta_{ac})=\ln\left[\cfrac{\kappa_a\kappa_c-g_{\rm eff}^2}{\tilde{N}(\kappa_a\kappa_c-g_{\rm eff}^2\chi)}\right],\\
\end{aligned}
\end{equation}
where $\tilde{N}=1+N_a+N_c$ and $\chi=\{4\kappa_a^2\kappa_c^2/[g_{\rm eff}^2(\kappa_a+\kappa_c)^2]+(\tilde{N}_--g_{\rm eff}^2\kappa_-)^2/[\tilde{N}^2g_{\rm eff}^4(\kappa_a+\kappa_c)^2]\}^{1/2}$ with $\tilde{N}_-=\kappa_a\kappa_c(\kappa_a+\kappa_c)(N_a-N_c)$ and $\kappa_-=\kappa_a(1+2N_a)-\kappa_c(1+2N_c)$. Here, the conventions $E_{ac}^\prime(\infty)\to E_{ac}^\prime$ and $\zeta_{ac}(\infty)\to\zeta_{ac}$ are applied for simplicity.

Next, we examine entanglement generation beyond the steady-state regime, i.e., for $g^2_{\rm eff}\ge\kappa_a\kappa_c$. In this unsteady-state regime, the CM elements in Eq. \eqref{SMv(t)} show that the parameter $x$ in Eq. \eqref{SMent} approaches zero as $t\to\infty$. Thus, expanding $\zeta_{ac}$ to first order in $x$ using the Taylor approximation $\sqrt{1+x}\approx 1+x/2$ gives $\zeta_{ac}=-(v_{11}+v_{44})x/2$. Consequently, the long-time entanglement quantity $E_{ac}^\prime$ is obtained as
\begin{equation}\label{SMeninftynac1}
E_{ac}^\prime=\ln\left(1+\cfrac{4g^2_{\rm eff}-\tilde{M}}{\tilde\chi+\tilde{M}}\right),\\
\end{equation}
where $\tilde\chi=\Omega(\kappa_a+\kappa_c)+(\kappa_a-\kappa_c)^2$, $\tilde{M}=2\kappa_aN_a(\Omega+\kappa_a-\kappa_c)+2\kappa_cN_c(\Omega-\kappa_a+\kappa_c)$, and $\Omega$ is given in Eq. \eqref{SMOmega}. 

By combining Eqs. \eqref{SMeninftynac} and \eqref{SMeninftynac1}, one can obtain
\begin{equation}\label{SMeninac}
\begin{aligned}
&E_{ac}^\prime=\begin{cases}
\ln\left[\cfrac{\kappa_a\kappa_c-g_{\rm eff}^2}{\tilde{N}(\kappa_a\kappa_c-g_{\rm eff}^2\chi)}\right],~~g_{\rm eff}^2<\kappa_a\kappa_c \\[1em]
\ln\left(1+\cfrac{4g^2_{\rm eff}-\tilde{M}}{\tilde\chi+\tilde{M}}\right),~~g_{\rm eff}^2\ge\kappa_a\kappa_c\\
\end{cases}\!\!\!\!.
\end{aligned}
\end{equation}
Moreover, in the limit $g^2_{\rm eff}\to (\kappa_a\kappa_c)^-$, Eq. \eqref{SMeninftynac} reduces to the steady-state entanglement expression, which coincides with the result obtained from Eq. \eqref{SMeninftynac1}, namely, 
\begin{equation}\label{SMeacboundnac}
\lim_{g^2_{\rm eff}\to(\kappa_a\kappa_c)^-\!} E_{ac}^\prime\!=\!\ln\left[\frac{(\kappa_a+\kappa_c)^2}{\kappa^2_a(1+2N_a)+\kappa^2_c(1+2N_c)}\right].\hspace{0cm}
\end{equation}
Therefore, the piecewise-defined LN $E_{ac}^\prime$ given in Eq. \eqref{SMeninac} is continuous with respect to the independent variable $g_{\rm eff}^2$. Furthermore, according to the definition of the LN in Eq. \eqref{SMenac}, $E_{ac}(\infty)={\rm max}[0, E_{ac}^\prime]$ converges to the fixed value $E_{ac}$.

Next, we analyze the monotonicity of the LN $E_{ac}$. For $g_{\rm eff}^2<\kappa_a\kappa_c$, the existence of entanglement between modes $a$ and $c$ requires $E_{ac}^\prime>0$, which is equivalent to the condition $\chi>1$ according to Eq. \eqref{SMeninac}. In this regime, $\chi$ decreases monotonically with increasing $g_{\rm eff}^2$, implying that $E_{ac}$ increases monotonically with $g_{\rm eff}^2$. Similarly, for $g_{\rm eff}^2\ge \kappa_a\kappa_c$, the condition $E_{ac}^\prime>0$ is equivalent to $4g_{\rm eff}^2>\tilde{M}$. Since $\Omega$ increases monotonically with $g_{\rm eff}^2$, $E_{ac}$ likewise increases monotonically with $g_{\rm eff}^2$. Therefore, the LN $E_{ac}$ defined in Eq. \eqref{SMeninac} is a continuous, monotonically increasing function of $g_{\rm eff}^2$. Consequently, operating in the unsteady-state regime allows stronger entanglement than in the steady-state regime. 

\begin{figure}[t] 
\centering
\includegraphics[width=0.35\textwidth]{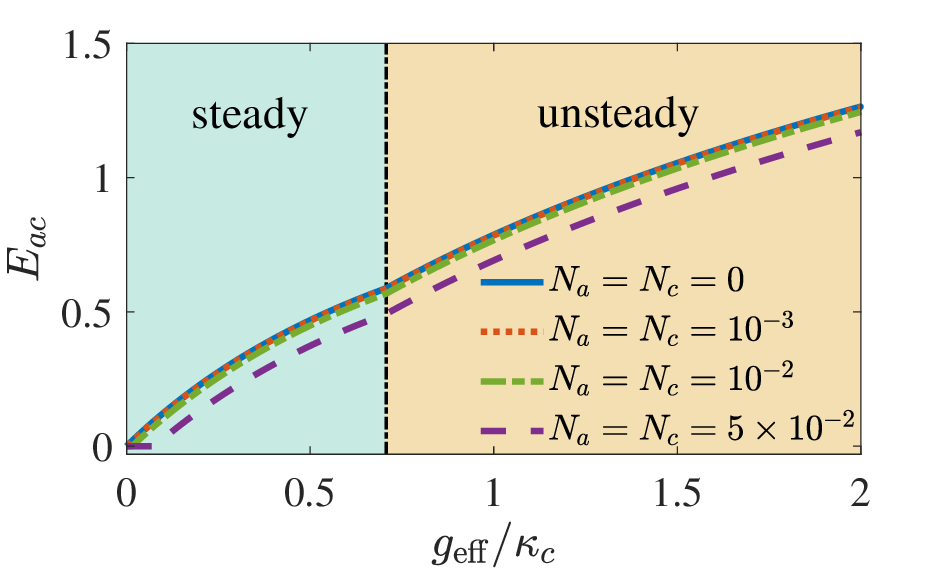}
\caption{Entanglement $E_{ac}$ versus the relative coupling strength $g_{\rm eff}/\kappa_c$ for decay rates $\kappa_c=2\kappa_a=1$, under different finite thermal occupation numbers. $N_a$ and $N_c$ are the thermal occupation number of mode $a$ and mode $c$, respectively. The black dotted-dashed line indicates the boundary between the steady-state and unsteady-state regimes.}
\label{ennac}
\end{figure}

In the cryogenic-temperature regime, where $N_a\approx N_c\approx0$, the parameters in Eq. \eqref{SMeninac} reduce to $\tilde{N}=1$ and $\tilde{M}=0$. Moreover, it follows from the CM elements in Eq. \eqref{SMv(t)} and the properties of the CM that $0<\zeta_{ac}<1$, implying $E_{ac}=E_{ac}^\prime$. Consequently, the LN $E_{ac}$ satisfies
\begin{equation}\label{SMeni}
\begin{aligned}
&E_{ac}=\begin{cases}
\ln\left(\cfrac{\kappa_a\kappa_c-g_{\rm eff}^2}{\kappa_a\kappa_c-g_{\rm eff}^2\chi}\right),~~g_{\rm eff}^2<\kappa_a\kappa_c \\[1em]
\ln\left(1+4\cfrac{g^2_{\rm eff}}{\tilde\chi}\right),~~g_{\rm eff}^2\ge\kappa_a\kappa_c\\
\end{cases}
\end{aligned}
\end{equation}
where $\chi=\{1+4\kappa_a\kappa_c(\kappa_a\kappa_c-g_{\rm eff}^2)/[g_{\rm eff}^2(\kappa_a+\kappa_c)^2]\}^{1/2}$ and $\tilde\chi$ is defined in Eq. \eqref{SMeninftynac1}. It is Eq. (7) in the main text. 

In addition, we investigate the behavior of MO entanglement for different finite thermal occupation numbers of modes $a$ and $c$, as shown in Supplementary Figure \ref{ennac}. In all cases, the entanglement increases monotonically with $g_{\rm eff}$. When $N_a=N_c=0$, entanglement exists for any nonzero coupling strength, i.e., $|g_{\rm eff}|>0$. As the total thermal occupation increases, a larger coupling strength is required for the onset of entanglement.

\subsection{Microwave-optical quantum steering}\label{SecMOst}
Similarly, we analyze the quantum steering between the microwave mode $a$ and the optical mode $c$ based on the system dynamics presented in \ref{SecCM}. The quantum steering can be measured by~\cite{SMLNsteeringK15}
\begin{equation}\label{SMstac1}
\begin{aligned}
S_{a\to c}(t)&={\rm max}[0,S_{ac}(t)],\\
S_{c\to a}(t)&={\rm max}[0,S_{ca}(t)],
\end{aligned}
\end{equation}
where the quantities $S_{ac}(t)=\ln [{\rm det}v_a/(4{\rm det} v)]/2$ and $S_{ca}(t)=\ln [{\rm det}v_c/(4{\rm det} v)]/2$. $S_{a\to c}>0$ $ (S_{c\to a}>0)$ indicates that the bipartite Gaussian state characterized by the CM is steerable from microwave (optical) mode to optical (microwave) mode. Based on the definition in Eq. \eqref{SMstac1} and the CM of Eq. \eqref{SMv(t)}, we derive the time-dependent quantum steering quantities
\begin{equation}\label{SMSact}
\begin{aligned}
S_{ac}(t)=\ln\left[\frac{v_{11}}{2(v_{11}v_{44}-v_{14}^2)}\right],\\
S_{ca}(t)=\ln\left[\frac{v_{44}}{2(v_{11}v_{44}-v_{14}^2)}\right].\\
\end{aligned}
\end{equation}

In the steady-state regime ($g_{\rm eff}^2<\kappa_a\kappa_c$), in the long-time limit, the values of $S_{ac}(\infty)$ and $S_{ca}(\infty)$ can be derived as 
\begin{equation}\label{SMstinftynac}
\begin{aligned}
S_{ac}=&\ln\left[\cfrac{\Xi_a-g_{\rm eff}^2\kappa_-(\kappa_a+\kappa_c)}{\Xi_a(1+2N_c)+g_{\rm eff}^2\kappa_-^2}\right],\\
S_{ca}=&\ln\left[\cfrac{\Xi_c+g_{\rm eff}^2\kappa_-(\kappa_a+\kappa_c)}{\Xi_c(1+2N_a)+g_{\rm eff}^2\kappa_-^2}\right],
\end{aligned}
\end{equation}
where $\Xi_o=\kappa_a\kappa_c(\kappa_a+\kappa_c)^2(1+2N_o)~(o=a,c)$ and $\kappa_-$ is defined in Eq. \eqref{SMOmega}. For simplicity, we apply the convention $S_{ac}(\infty)\to S_{ac}$ and $S_{ca}(\infty)\to S_{ca}$.

Moreover, by substituting the CM elements in Eq. \eqref{SMv(t)} into Eq. \eqref{SMSact}, the quantities $S_{ac}$ and $S_{ca}$ in the unsteady-state regime ($g_{\rm eff}^2\ge\kappa_a\kappa_c$) can be derived. In the long-time limit $t\to \infty$, they are given by 
\begin{equation}\label{SMstinftynac1}
\begin{aligned}
S_{ac}=&\ln\left(\cfrac{\Omega-\kappa_a+\kappa_c}{2\Omega}\right)+E_{ac}^\prime,\\
S_{ca}=&\ln\left(\cfrac{\Omega+\kappa_a-\kappa_c}{2\Omega}\right)+E_{ac}^\prime,
\end{aligned}
\end{equation}
where $\Omega$ is given in Eq. \eqref{SMOmega} and $E_{ac}^\prime$ is shown in Eq. \eqref{SMeninftynac1}.

\begin{figure}[t] 
\centering
\includegraphics[width=0.48\textwidth]{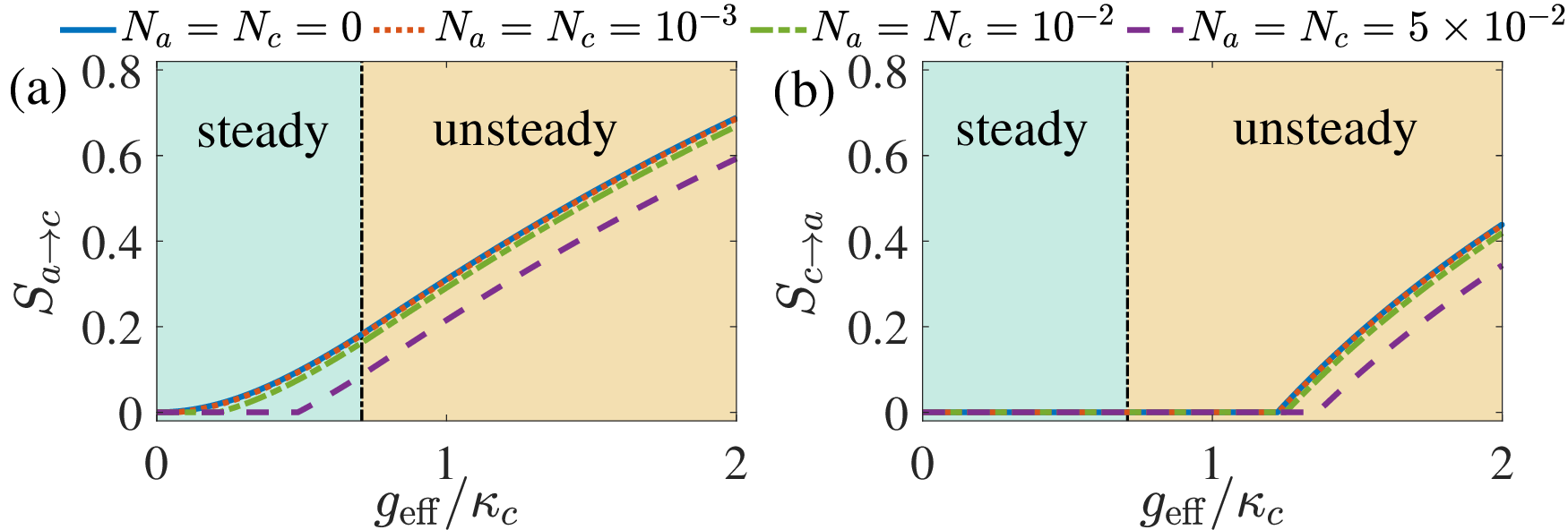}
\caption{Quantum steering (a) $S_{a\to c}$ and (b) $S_{c\to a}$  versus the relative coupling strength $g_{\rm eff}/\kappa_c$ under different finite thermal occupation numbers. The black dotted-dashed line indicates the boundary between the steady-state and unsteady-state regimes. The other parameters are the same as those in Supplementary Figure \ref{ennac}.}
\label{snac}
\end{figure}

By combining Eqs. \eqref{SMstinftynac} and \eqref{SMstinftynac1}, the quantity $S_{ac}$ can be expressed as
\begin{equation}\label{SMsactnac2}
\begin{aligned}
S_{ac}\!=\!\begin{cases}
\ln\left[\cfrac{\Xi_a-g_{\rm eff}^2\kappa_-(\kappa_a+\kappa_c)}{\Xi_a(1+2N_c)+g_{\rm eff}^2\kappa_-^2}\right],~g_{\rm eff}^2<\kappa_a\kappa_c\\[1em]
\ln\left(\cfrac{\Omega-\kappa_a+\kappa_c}{2\Omega}\right)+E_{ac}^\prime,~g_{\rm eff}^2\ge\kappa_a\kappa_c\\[1em]
\end{cases}\!\!\!\!.
\end{aligned}
\end{equation}
The steering value $S_{ca}$ is obtained by interchanging modes $a$ and $c$ in Eq. \eqref{SMsactnac2}.

Furthermore, when $g^2_{\rm eff}\to(\kappa_a\kappa_c)^-$, $S_{ac}$ and $S_{ca}$ in Eq. \eqref{SMstinftynac} under the steady-state regime correspond to the values of $S_{ac}$ and $S_{ca}$ in Eq. \eqref{SMstinftynac1}, i.e.,
\begin{equation}\label{SMsacboundnac}
\begin{aligned}
\lim_{g^2_{\rm eff}\to (\kappa_a\kappa_c)^-}S_{ac}=&\ln\left[\frac{\kappa_a\kappa_c+\kappa^2_c}{\kappa^2_a(1+2N_a)+\kappa^2_c(1+2N_c)}\right],\\
\lim_{g^2_{\rm eff}\to (\kappa_a\kappa_c)^-}S_{ca}=&\ln\left[\frac{\kappa^2_a+\kappa_a\kappa_c}{\kappa^2_a(1+2N_a)+\kappa^2_c(1+2N_c)}\right].
\end{aligned}
\end{equation}	
Both quantities are continuous functions of $g_{\rm eff}^2$. Furthermore, according to Eq. \eqref{SMstac1}, the asymptotic steering quantities are $S_{a\to c}(\infty)={\rm max}[0, S_{ac}]$ and $S_{c\to a}(\infty)={\rm max}[0, S_{ca}]$. For simplicity, we apply the convention $S_{a\to c}(\infty)\to S_{a\to c}$ and $S_{c\to a}(\infty)\to S_{c\to a}$ in the following content.

Then, we investigate the behavior of quantum steering $S_{a\to c}$ and $S_{c\to a}$ for different finite thermal occupation numbers of modes $a$ and $c$, as shown in Supplementary Figure \ref{snac}. In all cases, both steering quantities increase monotonically with the effective coupling strength $g_{\rm eff}$. As the total thermal occupation increases, a larger coupling strength is required for quantum steering to occur. Moreover, the threshold for two-way steering is also shifted to higher coupling strengths.

When $N_a\approx N_c\approx 0$, the expression for the quantity $S_{ac}$ in Eq. \eqref{SMsactnac2} can be simplified to 
\begin{equation}\label{SMsact2}
\begin{aligned}
S_{ac}=\begin{cases}
\ln\left[\cfrac{g_{\rm eff}^2(\kappa_c^2-\kappa_a^2)+\Xi}{g_{\rm eff}^2(\kappa_a-\kappa_c)^2+\Xi}\right],~g_{\rm eff}^2<\kappa_a\kappa_c\\[1em]
\ln\left(\cfrac{\Omega-\kappa_a+\kappa_c}{2\Omega}\right)+E_{ac}^\prime,~g_{\rm eff}^2\ge\kappa_a\kappa_c\\[1em]
\end{cases},
\end{aligned}
\end{equation}
where $\Xi=\kappa_a\kappa_c(\kappa_a+\kappa_c)^2$ and $E_{ac}^\prime$ is given by in Eq. \eqref{SMeni}. It corresponds to Eq. (9) in the main text.

Next, we identify the parameter regimes for realizing quantum steering by analyzing Eq. \eqref{SMsact2}. In the steady-state regime, for $\kappa_a<\kappa_c$, one finds $S_{ac}>0$ and $S_{ca}<0$, indicating that the bipartite entangled state is steerable from mode $a$ to mode $c$, whereas steering from $c$ to $a$ is forbidden. Conversely, when $\kappa_a>\kappa_c$, $S_{ca}>0$ while $S_{ac}<0$, implying that mode $c$ can steer mode $a$. The special case $\kappa_a=\kappa_c$ corresponds to a critical point at which quantum steering vanishes in both directions, i.e., $S_{ac}=S_{ca}=0$. Accordingly, the parameter ranges required to realize one-way steering in the steady-state regime are given by
\begin{equation}\label{SMstable}
\begin{aligned}
S_{a\to c}:~&\kappa_a<\kappa_c,0<g_{\rm eff}^2<\kappa_a\kappa_c,\\
S_{c\to a}:~&\kappa_a>\kappa_c,0<g_{\rm eff}^2<\kappa_a\kappa_c.
\end{aligned}
\end{equation}
Similarly, the ranges required to realize persistent quantum steering in the unsteady-state regime can be derived as
\begin{equation}\label{SMunstable}
\begin{aligned}
&S_{a\to c}
\begin{cases}
\kappa_a\le\kappa_c, &g_{\rm eff}^2>\kappa_a\kappa_c\\
\kappa_a>\kappa_c, &g_{\rm eff}^2>\kappa_a(2\kappa_a-\kappa_c)
\end{cases},\\
&S_{c\to a}
\begin{cases}
\kappa_a<\kappa_c, &g_{\rm eff}^2>\kappa_c(2\kappa_c-\kappa_a)\\
\kappa_a\ge\kappa_c, &g_{\rm eff}^2>\kappa_a\kappa_c
\end{cases}.
\end{aligned}
\end{equation}
Notably, two-way quantum steering between modes $a$ and $c$ emerges when the coupling strength $g_{\rm eff}$ is increased such that $g^2_{\rm eff}+\kappa_a\kappa_c>2\kappa^2_a,2\kappa^2_c$.

Then, we analyze the monotonicity of the steering quantities. In the steady-state regime ($g_{\rm eff}^2<\kappa_a\kappa_c$), the monotonic behavior of the steering quantities can be demonstrated by analyzing the derivatives $\partial S_{ac}/\partial g_{\rm eff}^2$ and $\partial S_{ca}/\partial g_{\rm eff}^2$, which are given by
\begin{equation}
\begin{aligned}
\frac{\partial S_{ac}}{\partial g_{\rm eff}^2}=&\frac{2\kappa_a\Xi(\kappa_c-\kappa_a)}{[\Xi+g_{\rm eff}^2(\kappa_a-\kappa_c)^2][\Xi+g_{\rm eff}^2(\kappa_c^2-\kappa_a^2)]},\\
\frac{\partial S_{ca}}{\partial g_{\rm eff}^2}=&\frac{2\kappa_c\Xi(\kappa_a-\kappa_c)}{[\Xi+g_{\rm eff}^2(\kappa_a-\kappa_c)^2][\Xi+g_{\rm eff}^2(\kappa_a^2-\kappa_c^2)]},\\
\end{aligned}
\end{equation}
where $\Xi>0$ is defined in Eq. \eqref{SMsact2}. Under the condition $S_{ac}>0$ for $\kappa_a<\kappa_c$, one directly obtains $\partial S_{ac}/\partial g_{\rm eff}^2>0$. Similarly, when $S_{ca}>0$ for $\kappa_a>\kappa_c$, it follows that $\partial S_{ca}/\partial g_{\rm eff}^2>0$. 

Moreover, in the unsteady-state regime ($g_{\rm eff}^2\ge\kappa_a\kappa_c$), the derivatives $\partial S_{ac}/\partial g_{\rm eff}^2$ and $\partial S_{ca}/\partial g_{\rm eff}^2$ can be derived as
\begin{equation}\label{SMscaf}
\begin{aligned}
\frac{\partial S_{ac}}{\partial g_{\rm eff}^2}
=&\frac{4\left[2\kappa_+g_{\rm eff}^2+\Omega\kappa_-^2+\kappa_-(\kappa_+\kappa_-+2\kappa_a\kappa_c)\right]}{\Omega(\Omega-\kappa_-)(\Omega+\kappa_+)(\Omega\kappa_++\kappa_-^2)},\\
\frac{\partial S_{ca}}{\partial g_{\rm eff}^2}=&\frac{4\left[2\kappa_+g_{\rm eff}^2+\Omega\kappa_-^2+\kappa_-(\kappa_+\kappa_--2\kappa_a\kappa_c)\right]}{\Omega(\Omega+\kappa_-)(\Omega+\kappa_+)(\Omega\kappa_++\kappa_-^2)},
\end{aligned}
\end{equation}
respectively, where $\kappa_\pm=\kappa_a\pm\kappa_c$. One can show that $\partial S_{ac}/\partial g_{\rm eff}^2>0$ for $\kappa_a\le\kappa_c$ with $g_{\rm eff}^2>\kappa_a\kappa_c$, and for $\kappa_a>\kappa_c$ with $g_{\rm eff}^2>\kappa_a(2\kappa_a-\kappa_c)$. Similarly, one finds that $\partial S_{ca}/\partial g_{\rm eff}^2>0$ under the conditions given in Eq. \eqref{SMunstable}. Therefore, both $S_{a\to c}$ and $S_{c\to a}$ increase monotonically with the magnitude of $g_{\rm eff}$.

Furthermore, we clarify the relation between MO entanglement [Eq. \eqref{SMeni}] and quantum steering [Eq. \eqref{SMsact2}]. In the steady-state regime, we define the ratios $\mathcal{R}_{ac}=\exp(E_{ac})/\exp(S_{ac})$ and $\mathcal{R}_{ca}=\exp(E_{ac})/\exp(S_{ca})$, which can be derived as 
\begin{equation}
\begin{aligned}
\mathcal{R}_{ac}=&1+\frac{\sqrt{4\kappa_a^2\kappa_c^2g_{\rm eff}^2+\mathcal{K}_{ac}^2}-\mathcal{K}_{ac}}{\kappa_a\kappa_c(\kappa_a+\kappa_c)+\mathcal{K}_{ac}},\\
\mathcal{R}_{ca}=&1+\frac{\sqrt{4\kappa_a^2\kappa_c^2g_{\rm eff}^2+\mathcal{K}_{ca}^2}-\mathcal{K}_{ca}}{\kappa_a\kappa_c(\kappa_a+\kappa_c)+\mathcal{K}_{ca}},
\end{aligned}
\end{equation}
where $\mathcal{K}_{ac}=g_{\rm eff}^2(\kappa_c-\kappa_a)$ and $\mathcal{K}_{ca}=g_{\rm eff}^2(\kappa_a-\kappa_c)$. Under the condition $\kappa_a<\kappa_c$, one has $S_{ac}>0$ and $\mathcal{K}_{ac}>0$, leading to a ratio $\mathcal{R}_{ac}>1$. Similarly, for $\kappa_a>\kappa_c$, $S_{ca}>0$ and $\mathcal{K}_{ca}>0$, yielding $\mathcal{R}_{ca}>1$. These results provide an analytical demonstration that, in the steady-state regime, quantum steering constitutes a strict subset of quantum entanglement. In the unsteady-state regime, according to the definition of $\Omega$ in Eq. \eqref{SMOmega}, $\Omega-\kappa_a+\kappa_c<2\Omega$ and $\Omega+\kappa_a-\kappa_c<2\Omega$. Consequently, the first logarithmic terms in Eqs. \eqref{SMsact2} are negative, implying $S_{ac},S_{ca}<E_{ac}$. This shows that quantum steering is still strictly bounded by the corresponding quantum entanglement in the unsteady-state condition.

\subsection{The characteristic time window and its verification}\label{sectau}
In practical numerical calculations, the evolution time cannot be taken to infinity. For the steady-state regime, the asymptotic solution can be obtained by setting $\dot{v}(t)=0$ in Eq. \eqref{SMhatv}. However, this approach does not apply to the unsteady-state regime. We therefore introduce a characteristic time $\tau$ to approximate the asymptotic long-time behavior in the unsteady-state regime. Furthermore, for realistic MO multipartite hybrid systems, we identify the characteristic time window
\begin{equation}
t\in[\tau,2\tau],
\end{equation}
within which the generated quantum resources remain essentially unchanged.

We now turn to the determination of the characteristic time $\tau$. Under the transformation
\begin{equation}
X=\cos\tilde\varphi X_a+\sin\tilde\varphi Y_c,~Y=\sin\tilde\varphi X_a- \cos\tilde\varphi Y_c,
\end{equation}
where $\tilde\varphi=\pi/4-\varphi/2$ and $\varphi$ is given by Eq. \eqref{SMOmega}. $\zeta_{ac}(t)$ in Eq. \eqref{SMent} turns into
\begin{equation}\label{SMzac}
\zeta_{ac}(t)=\Delta X+\Delta Y-\sqrt{(\Delta X-\Delta Y)^2+4\langle X Y\rangle^2}.
\end{equation}
Using the CM elements given in Eq. \eqref{SMv(t)}, in the cryogenic-temperature regime ($N_a\approx N_c\approx0$), the variance $\Delta X=\langle X^2\rangle-\langle X\rangle^2$ can be derived as
\begin{equation}\label{SMdX}
\begin{aligned}
\Delta X(t)=\frac{1}{2}+2c_{-}e^{-(\Omega+\kappa_a+\kappa_c)t}-2c_-,
\end{aligned}
\end{equation}
the variance $\Delta Y=\langle Y^2\rangle-\langle Y\rangle^2$ is
\begin{equation}\label{dY}
\begin{aligned}
\Delta Y(t)=\frac{1}{2}+2c_{+}e^{(\Omega-\kappa_a-\kappa_c)t}-2c_{+},
\end{aligned}
\end{equation}
and the correlation term $\langle X(t)Y(t)\rangle$ is obtained as
\begin{equation}\label{XX}
\langle X(t)Y(t)\rangle=\frac{c_0}{\cos\varphi}\left[1-e^{-(\kappa_a+\kappa_c)t}\right].
\end{equation}
The definitions of $c_\pm$, $c_0$, $\varphi$, and $\Omega$ are given in Eq. \eqref{SMOmega}.

\begin{figure}[t] 
\centering
\includegraphics[width=0.47\textwidth]{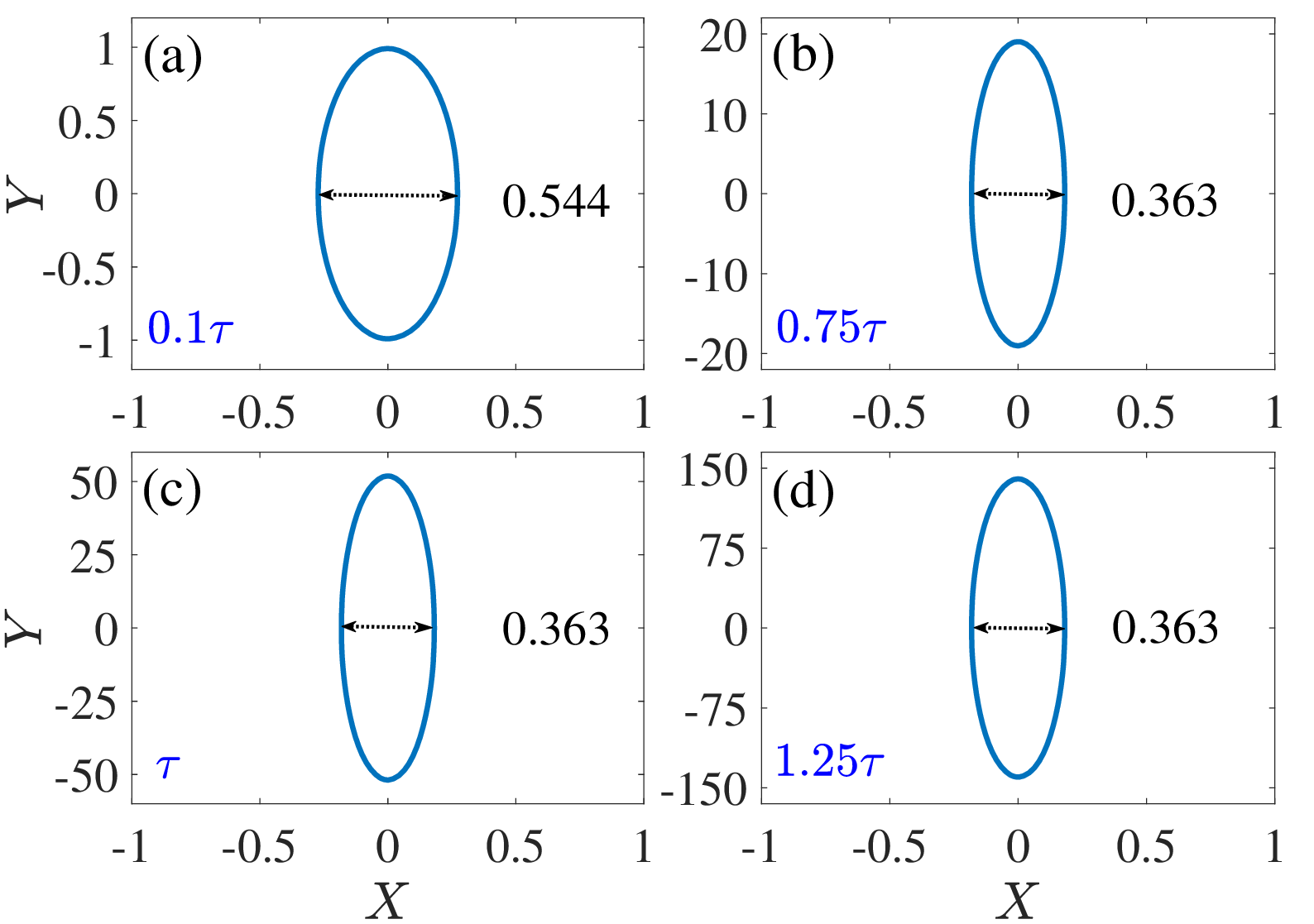}
\caption{Variances in the $X$ and $Y$ quadratures at different evolution times, $0.1\tau$, $0.75\tau$, $\tau$, and $1.25\tau$. We set $g_{\rm eff}^2/\kappa_a\kappa_c=4$ and $\kappa_c=2\kappa_a$ for all panels.}
\label{XY}
\end{figure}

In the unsteady-state regime, the exponential term in Eq. \eqref{dY} is positive, i.e., $\Omega-\kappa_a-\kappa_c>0$. Consequently, after sufficiently long-time evolution, one finds $\Delta X(t), \langle X(t)Y(t)\rangle\ll \Delta Y(t)$. Under this condition, $\zeta_{ac}(t)$ can be approximated as
\begin{equation}
\zeta_{ac}(t)\approx2\Delta X(t)=1 + 4c_- e^{-(\Omega+\kappa_a+\kappa_c)t} - 4c_-.
\end{equation}
We can then define a characteristic time
\begin{equation}\label{tau}
\tau = \frac{4\pi}{\Omega+\kappa_a+\kappa_c},
\end{equation}
which approximately represents the timescale for the system to establish persistent quantum resources. At $t = \tau$, $e^{-4\pi}\ll 1$ (numerically $\sim 10^{-6}$), so that $\zeta_{ac}(\tau)\approx 2\Delta X(\tau) \approx 2\Delta X(2\tau) \approx 2\Delta X(\infty) \approx \zeta_{ac}(\infty)$. Consequently, in the unsteady-state regime, the entanglement in Eq. \eqref{SMent} can be reduced to
\begin{equation}\label{SMEacDx}
E_{ac}=-\ln(\zeta_{ac})
\approx-\ln(2\Delta X).
\end{equation}
For simplicity and without loss of generality, we adopt the conventions $\zeta_{ac}(\infty)\to\zeta_{ac}$ and $\Delta X(\infty)\to\Delta X$. Similarly, substituting Eq. \eqref{SMEacDx} into Eq. \eqref{SMsact2}, the steering quantity $S_{ac}$ can also be expressed in terms of the variance $\Delta X$, namely,
\begin{equation}\label{SMSacDx}
S_{ac}\approx\ln\left(\cfrac{\Omega-\kappa_a+\kappa_c}{4\Omega\Delta X}\right).
\end{equation}
From Eqs. \eqref{SMdX}, \eqref{SMEacDx}, and \eqref{SMSacDx}, increasing $|g_{\rm eff}|$ reduces $\tau$, thereby accelerating the convergence of the quantum resources to their asymptotic values.

\begin{figure}[t] 
\centering
\includegraphics[width=0.47\textwidth]{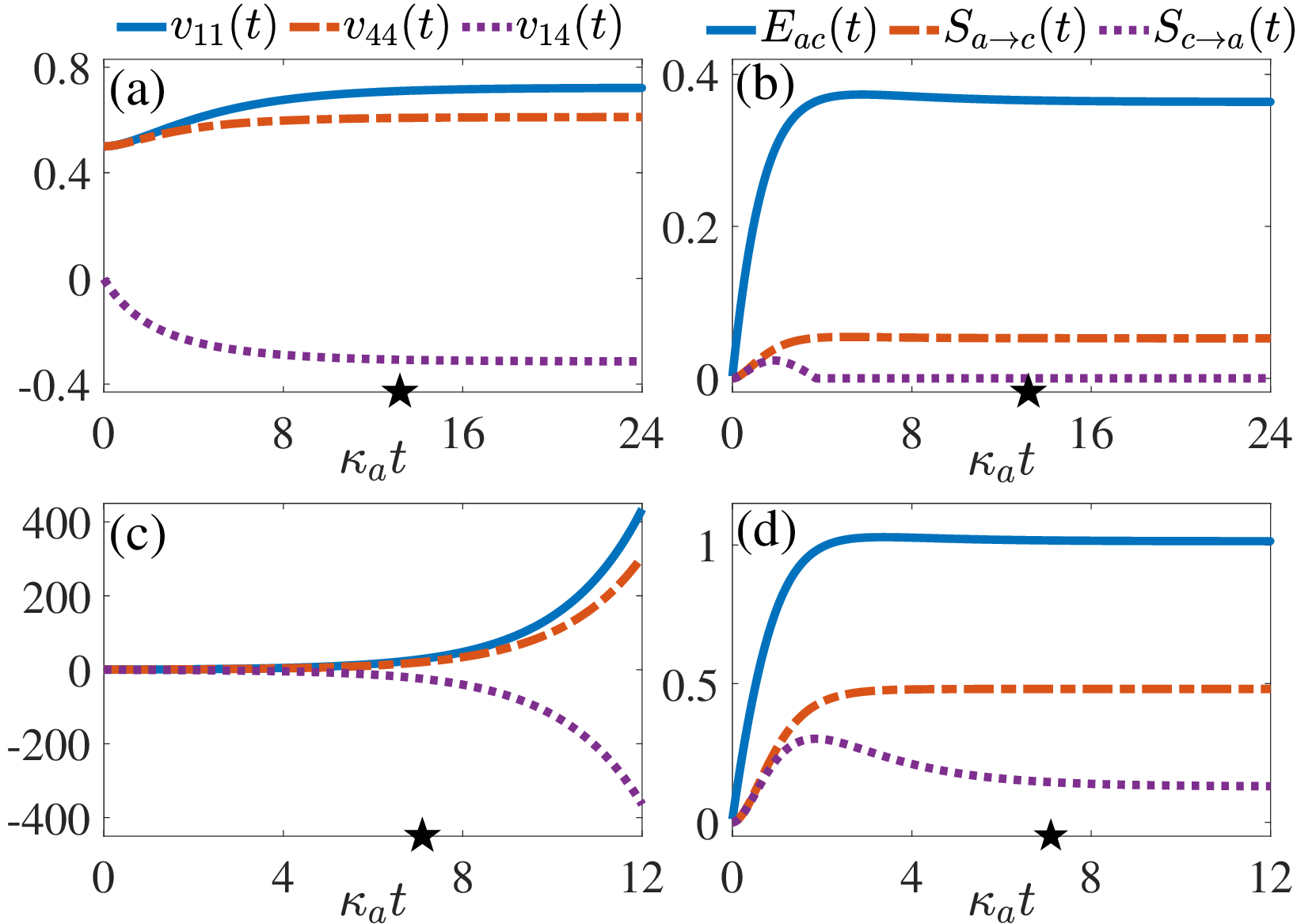}
\caption{Numerical verification of theoretical results. [(a), (c)] The dynamical evolution of the CM elements $v_{11}(t)$, $v_{44}(t)$, and $v_{14}(t)$ under steady-state and unsteady-state conditions, respectively. [(b), (d)] Time evolution of the entanglement $E_{ac}$, steering $S_{a\to c}$, and steering $S_{c\to a}$ under steady-state and unsteady-state conditions, respectively. For (a) and (b), $g_{\rm eff}^2/\kappa_a\kappa_c=0.25$, and for (c) and (d), $g_{\rm eff}^2/\kappa_a\kappa_c=4$. $\kappa_c=2\kappa_a$ is fixed for all panels. The star symbols mark the point corresponding to the characteristic time $\tau$.}
\label{v-ac}
\end{figure}

In Supplementary Figure \ref{XY}, we investigate the evolution of the variance of the generalized quadrature operator $\hat{O}=\cos\psi X+\sin\psi Y$, where $\psi\in[0,2\pi)$. In this plot, the ellipse’s semi-minor axis along $x$ represents $\Delta X$, while the semi-major axis along $y$ represents $\Delta Y$. As time evolves, $\Delta Y$ increases rapidly, whereas $2\Delta X$ initially increases slightly before approaching its asymptotic value of $0.363$ over the characteristic timescale $\tau$.

Our theoretical results from the preceding \ref{SecCM}--\ref{sectau} can be numerically verified through the CM elements and quantum resources. In Supplementary Figure \ref{v-ac}(a), the CM elements $v_{11}(t)$, $v_{44}(t)$, and $v_{14}(t)$ approach constant values after extended evolution, reflecting the steady-state regime. In contrast, Supplementary Figure \ref{v-ac}(c) shows these elements diverging exponentially over time, indicating unsteady system dynamics. Unlike the CM elements, which exhibit qualitatively distinct behaviors in the steady-state and unsteady-state regimes, quantum entanglement and steering display similar dynamical features in both regimes, approaching finite asymptotic values at long times. Panels (b) and (d) correspond to the CM dynamics shown in panels (a) and (c), respectively. In both cases, $E_{ac}(t)$ and $S_{a\to c}(t)$ increase rapidly at early times, become nearly constant before the characteristic time $\tau$, and subsequently approach their asymptotic plateaus. By contrast, $S_{c\to a}(t)$ initially increases and then gradually decays toward its asymptotic value. Notably, the values of $E_{ac}(t)$, $S_{a\to c}(t)$, and $S_{c\to a}(t)$ in Supplementary Figure \ref{v-ac}(d) are larger than the corresponding values in Supplementary Figure \ref{v-ac}(b). Furthermore, Supplementary Figure \ref{v-ac}(d) clearly shows two-way steering, with $S_{a\to c}(t)>S_{c\to a}(t)>0$. These observations indicate that the unsteady-state regime can exhibit stronger quantum resources.

From a physical perspective, within the framework of open quantum systems, the effective two-mode squeezing interaction and environmental decoherence jointly affect the system dynamics. The former continuously generates and enhances quantum entanglement and steering, whereas the latter suppresses these quantum correlations. Their joint effect causes the MO entanglement $E_{ac}$ and the steering quantities $S_{a\to c}$ and $S_{c\to a}$ to approach finite asymptotic values.

\section{Application in electro-optomechanical systems}\label{SMEOM}
\subsection{Model and the effective Hamiltonian}\label{SMEOMH}

\begin{figure}[htbp] 
\centering
\includegraphics[width=0.33\textwidth]{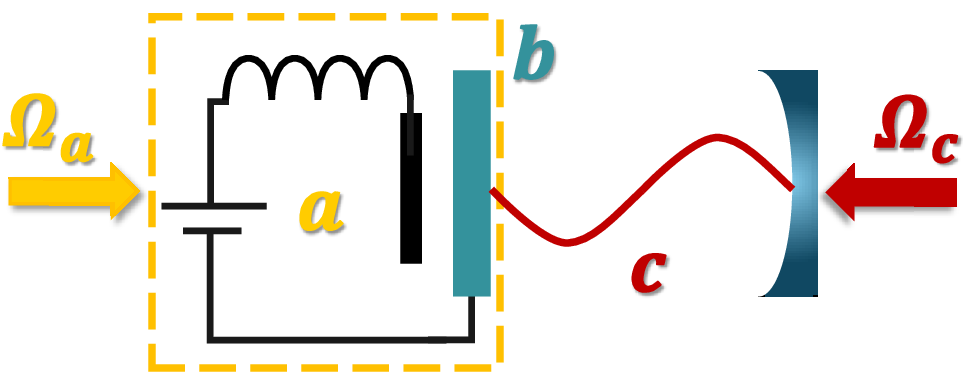}
\caption{Schematic of the electro-optomechanical system. The mechanical oscillator acts as an intermediate mode $b$, coupling with the microwave resonator $a$ and optical cavity $c$. The microwave and optical modes are driven by a microwave drive field $\Omega_a$ and an optical driving laser $\Omega_c$, respectively.}
\label{eomdiagram}
\end{figure}

A hybrid electro-optomechanical (EOM) system is considered for generating MO quantum resources. As shown in Supplementary Figure \ref{eomdiagram}, the system consists of a microwave inductor-capacitor resonator, a mechanical oscillator, and an optical cavity, which has been realized in recent works~\cite{SMEOM12,SMEOM15}. The mechanical oscillator is capacitively coupled to the microwave resonator on one side, while on the other side, it is coupled to the optical cavity via the radiation pressure. A microwave driving field and an optical laser are simultaneously applied to the microwave resonator and optical cavity, respectively. In the rotating frame with respect to the driving frequencies, the Hamiltonian of the EOM system~\cite{SMEOM12,SMEOM15} is ($\hbar\equiv1$)
\begin{equation}\label{SMeomH1}
\begin{aligned}
H_{S_1}=\sum_{o=a,c}&\Delta_o o^\dag o+\omega_b b^\dag b+g_{ob}o^\dag o(b+b^\dag)\\
&+i\Omega_o(o^\dag-o),
\end{aligned}
\end{equation}		
where $a$ ($a^\dag$), $b$ ($b^\dag$), and $c$ ($c^\dag$) are the annihilation (creation) operators of the microwave, mechanical, and optical modes, respectively. $\Delta_o=\omega_o-\epsilon_o~(o=a,c)$ denotes the detuning of mode $o$, where $\omega_o$ is the transition frequency and $\epsilon_o$ is the corresponding driving-field frequency. $\omega_b$ is the transition frequency of the mechanical mode. The coupling strengths $g_{ab}$ and $g_{cb}$ describe the microwave-mechanical and optomechanical interactions, respectively, which can be enhanced by strongly driving the microwave and optical modes with Rabi frequencies $\Omega_a$ and $\Omega_c$. 

The microwave and optical modes under strong driving are assumed to have large expectation amplitudes $|\langle a\rangle|\gg1$ and $|\langle c\rangle|\gg1$, respectively.  This allows us to linearize the system dynamics by writing the operators $o=\delta o+\langle o\rangle~(o=a,c)$, with $\delta o$ describing the fluctuation of the mode $o$. Neglecting the second-order fluctuation terms~\cite{SMEOM12,SMEOM15}, the Hamiltonian in Eq. \eqref{SMeomH1} becomes
\begin{equation}\label{SMeomH20}
\begin{aligned}
H_{S_1}&=H_0+V,~H_0=\sum_{o=a,c}\Delta_o\delta o^\dag \delta o+\omega_b b^\dag b,\\
V&=\sum_{o=a,c}g_o(\delta o+\delta o^\dag)(b+b^\dag),
\end{aligned}
\end{equation}	
where $g_a=g_{ab}\langle a\rangle$ and $g_c=g_{cb}\langle c\rangle$ are the enhanced microwave-mechanical and optomechanical coupling strengths, respectively. For simplicity and without loss of generality, we assume that the values $\langle o\rangle~(o=a,c)$ are real numbers and make $\delta o\to o$ in the following content. The linearized Hamiltonian in Eq. \eqref{SMeomH20} is obtained by setting $\theta=\phi=\pi/4$ and $N=1$ in the general Hamiltonian in the main text, along with the substitutions $\omega_1\to\omega_b$, $b_1\to b$, and $g_o\to\sqrt{2}g_o$. Then, we can use the general effective Hamiltonian construction approach presented in the Methods section of the main text to derive the effective coupling of the MO subsystem in the specific EOM system.

\begin{figure}[t] 
\centering
\includegraphics[width=0.48\textwidth]{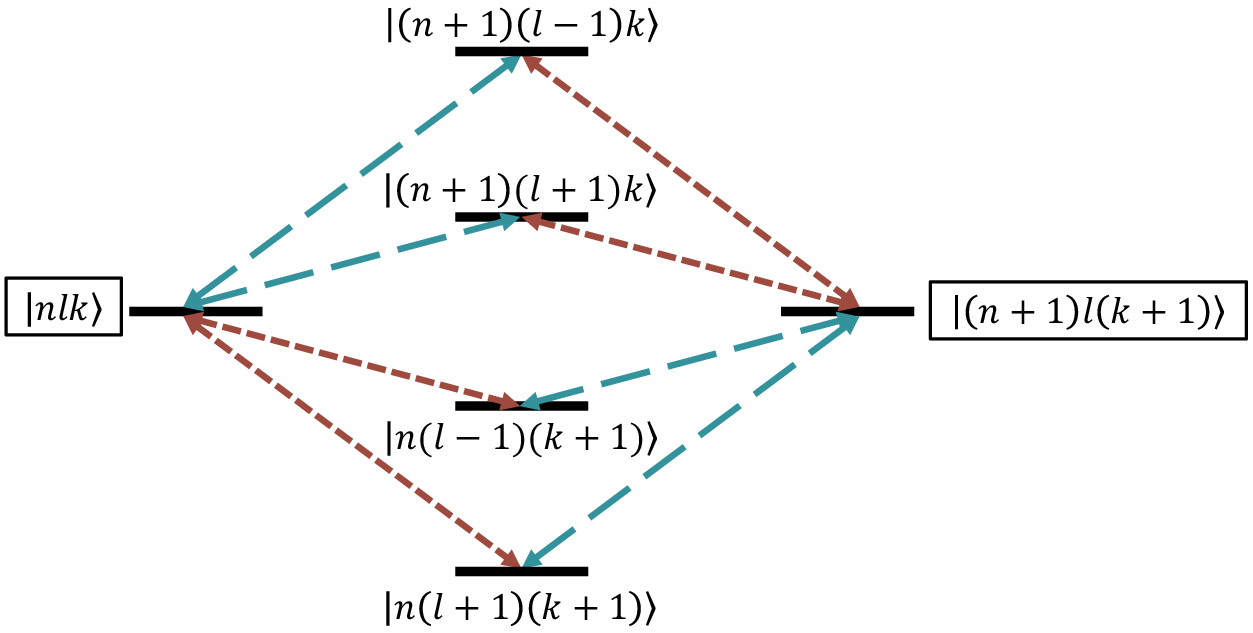}
\caption{All four second-order paths connecting $|nlk\rangle$ and $|(n+1)l(k+1)\rangle$ are depicted. The dashed lines represent the coupling between the microwave and mechanical modes, while the dotted lines indicate the coupling between the optical and mechanical modes.}
\label{eomacpath}
\end{figure}

Specifically, when the microwave detuning $\Delta_a$ is nearly opposite to the optical detuning $\Delta_c$, and both of them are far resonant from mechanical frequency $\omega_b$, i.e., $\Delta_a+\Delta_c\approx0$ and $g_a,g_c\ll\{|\Delta_a-\omega_b|, |\Delta_c-\omega_b|\}$, the tensor product state $|nlk\rangle\equiv|n\rangle_a|l\rangle_{b}|k\rangle_c$ approximately degenerates with $|(n+1)l(k+1)\rangle$. Here, the subscripts $a$, $b$, and $c$ represent the microwave, mechanical, and optical modes, respectively, and $n$, $l$, and $k$ denote the corresponding Fock states. Under these cases, the interaction term $V$ in Eq. \eqref{SMeomH20} can effectively couple the arbitrary base pairs $|nlk\rangle$ and $|(n+1)l(k+1)\rangle$, yielding an effective Hamiltonian expressed in the following form
\begin{equation}\label{SMeomHeac}
\begin{aligned}
H_{\rm eff}\!=&\epsilon_1|nlk\rangle\langle nlk|+(\Delta_{a}+\Delta_{c}+\epsilon_2)\\&\times|(n+1)l(k+1)\rangle\langle(n+1)l(k+1)|\\
&+\tilde g_{\rm eff}(|nlk\rangle\langle (n+1)l(k+1)|+{\rm h.c.}).
\end{aligned}
\end{equation}
Here, $\epsilon_1$ and $\epsilon_2$ represent the energy shifts caused by the coupling between the states $|nlk\rangle$ and $|(n+1)l(k+1)\rangle$, respectively, and $\tilde g_{\rm eff}$ is the effective coupling strength. These are three coefficients to be determined. It should be noted that the common unperturbed eigenenergies of the two bases, $n\Delta_a+l\omega_b+k\Delta_c$, have been omitted.

We first consider the effective coupling strength $\tilde{g}_{\rm eff}$ between states $|nlk\rangle$ and $|(n+1)l(k+1)\rangle$. As shown in Supplementary Figure \ref{eomacpath}, we summarize all four paths connecting these states and their corresponding contributions to the effective coupling. According to Eq. (18) in the Methods section of the main text, we obtain
\begin{equation}
\tilde{g}_{\rm eff}=\frac{2g_ag_c\omega_b}{\Delta_a^2-\omega^2_b}\sqrt{(n+1)(k+1)}
\end{equation}
up to the second-order perturbation of coupling strengths $g_{a}$ and $g_c$. Similarly, according to Eq. (22) in the Methods section of the main text, the difference $\delta$ can be derived as
\begin{equation}\label{SMeomdelta}
\delta\equiv\epsilon_1-\epsilon_2=\frac{2(g_a^2+g_c^2)\omega_b}{\omega_b^2-\Delta_a^2}.
\end{equation}

Under the condition $\Delta_a=-\Delta_c+\delta$, the effective Hamiltonian in Eq. \eqref{SMeomHeac} can be expressed as
\begin{equation}\label{SMeomHeac1}
\begin{aligned}
H_{\rm eff}=&\tilde{g}_{\rm eff}(|nlk\rangle\langle (n+1)l(k+1)|+{\rm h.c.})\\
=& g_{\rm eff}\sqrt{(n+1)(k+1)}(|nk\rangle\langle (n+1)(k+1)|\\&+{\rm h.c.})\otimes|l\rangle\langle l|.
\end{aligned}
\end{equation}
By eliminating the mechanical mode, the Hamiltonian in Eq. \eqref{SMeomHeac1} can be promoted to the full Hilbert space of the microwave and optical modes, yielding
\begin{equation}\label{SMeomHefac}
\begin{aligned}
H_{\rm eff}=g_{\rm eff}(a^\dag c^\dag+ac),
\end{aligned}
\end{equation}
and the coupling strength can be derived as
\begin{equation}\label{SMeomgeff}
g_{\rm eff}=\frac{2\omega_bg_{a}g_{c}}{\Delta_a^2-\omega^2_b}.
\end{equation}
This is exactly the effective coupling strength of the EOM system reported in the manuscript.

\subsection{System dynamics and quantum resources} \label{eomAmatrix}
{This Supplementary Note} is devoted to calculating the CM of the full hybrid EOM system. Within the open-quantum-system framework and under standard Markovian environments, one can derive the QLEs for the full Hamiltonian in Eq. \eqref{SMeomH20}
\begin{equation}\label{SMeom-line}
\begin{aligned}
\dot a=&-(i\Delta_a+\kappa_a)a-ig_{a}(b+b^\dag)+\sqrt{2\kappa_a}a_{in},\\	
\dot c=&-(i\Delta_c+\kappa_c)c-ig_{c}(b+b^\dag)+\sqrt{2\kappa_c}c_{in},\\	
\dot b=&-(i\omega_b+\kappa_{b})b-i\sum_{o=a,c}g_{o}(o+o^\dag)+\sqrt{2\kappa_{b}}b_{in},\\
\end{aligned}
\end{equation}
where $\kappa_a$, $\kappa_b$, and $\kappa_c$ are the decay rates of the microwave, mechanical, and optical modes, respectively. The input noise operator $o_{in}~(o=a,c)$ has zero mean and satisfies the correlation functions, $\langle o_{in}(t)o^\dag_{in}(t') \rangle=[N_o(\omega_o)+1]\delta(t-t')$ and $\langle o^\dag _{in}(t)o_{in}(t')\rangle=N_o(\omega_o)\delta(t-t')$. Here, $N_o(\omega_o)=[\exp(\hbar\omega_o/k_BT)-1]^{-1}$ is the average thermal excitation number of mode-$o$, with the Boltzmann constant $k_B$ and the environmental temperature $T$.

By introducing the quadrature operators $X_o(t)=(o+o^\dag)/\sqrt{2}$ and $Y_o(t)=i(o^\dag-o)/\sqrt{2}$, the above Eq. \eqref{SMeom-line} can be written in the matrix form
\begin{equation}\label{SMuteom}
\dot{\tilde{u}}(t)=A\tilde{u}(t)+\xi(t)
\end{equation}
where $\tilde{u}(t)=[u(t)^T,X_{b}(t),Y_{b}(t)]^T$ and $u(t)$ is shown in Eq. \eqref{SMueff}. $\xi(t)=[\xi_{\rm eff}^T(t),\sqrt{2\kappa_{b}}X_{b}^{in}(t),\sqrt{2\kappa_{b}}Y_{b}^{in}(t)]^T$, $\xi_{\rm eff}(t)$ is the noise vector in Eq. \eqref{SMueff}, $X_{b}^{in}(t)=(b_{in}+b_{in}^\dag)/\sqrt{2}$ and $Y_{b}^{in}(t)=i(b_{in}^\dag-b_{in})/\sqrt{2}$ are the quadratures of the input noise operators. The drift matrix $A$ is given by
\begin{align}\label{SMeomA}
A=\begin{pmatrix}
-\kappa_a&\Delta_a&0&0&0&0\\
-\Delta_a&-\kappa_a&0&0&-2g_a&0\\
0&0&-\kappa_c&\Delta_c&0&0\\
0&0&-\Delta_c&-\kappa_c&-2g_c&0\\
0&0&0&0&-\kappa_{b}&\omega_b\\
-2g_a&0&-2g_c&0&-\omega_b&-\kappa_{b}
\end{pmatrix}.
\end{align}

Owing to the linear dynamics in Eq. \eqref{SMuteom} and Gaussian input noises, the system state remains Gaussian. Accordingly, the EOM system can be fully characterized by the $6\times6$ CM $\tilde{v}_{jl}(t)=\langle \tilde{u}_j(t)\tilde{u}_l(t)+ \tilde{u}_l(t)\tilde{u}_j(t)\rangle/2-\langle\tilde u_j(t)\rangle\langle \tilde u_l(t)\rangle~(j,l=1,2,...,6)$, which satisfies
\begin{equation}\label{SMeomV}
\dot{\tilde{v}}(t)=A\tilde{v}(t)+\tilde{v}(t)A^T+D,
\end{equation}
where $D= {\rm diag}[\kappa_a(2N_a+1),\kappa_a(2N_a+1),\kappa_c(2N_c+1),\kappa_c(2N_c+1),\kappa_{b}(2N_{b}+1),\kappa_{b}(2N_{b}+1)]$ is the diffusion matrix and defined by $D_{jl}=\langle\xi_j(t)\xi_l(t)+\xi_l(t)\xi_j(t)\rangle/2$. The CM of the MO subsystem is given by $v=\tilde{v}(1:4;1:4)$. The corresponding numerical results for the linearized Hamiltonian in Eq. \eqref{SMeomH20} can then be obtained by solving the above differential equation.

\begin{figure}[t] 
\centering
\includegraphics[width=0.48\textwidth]{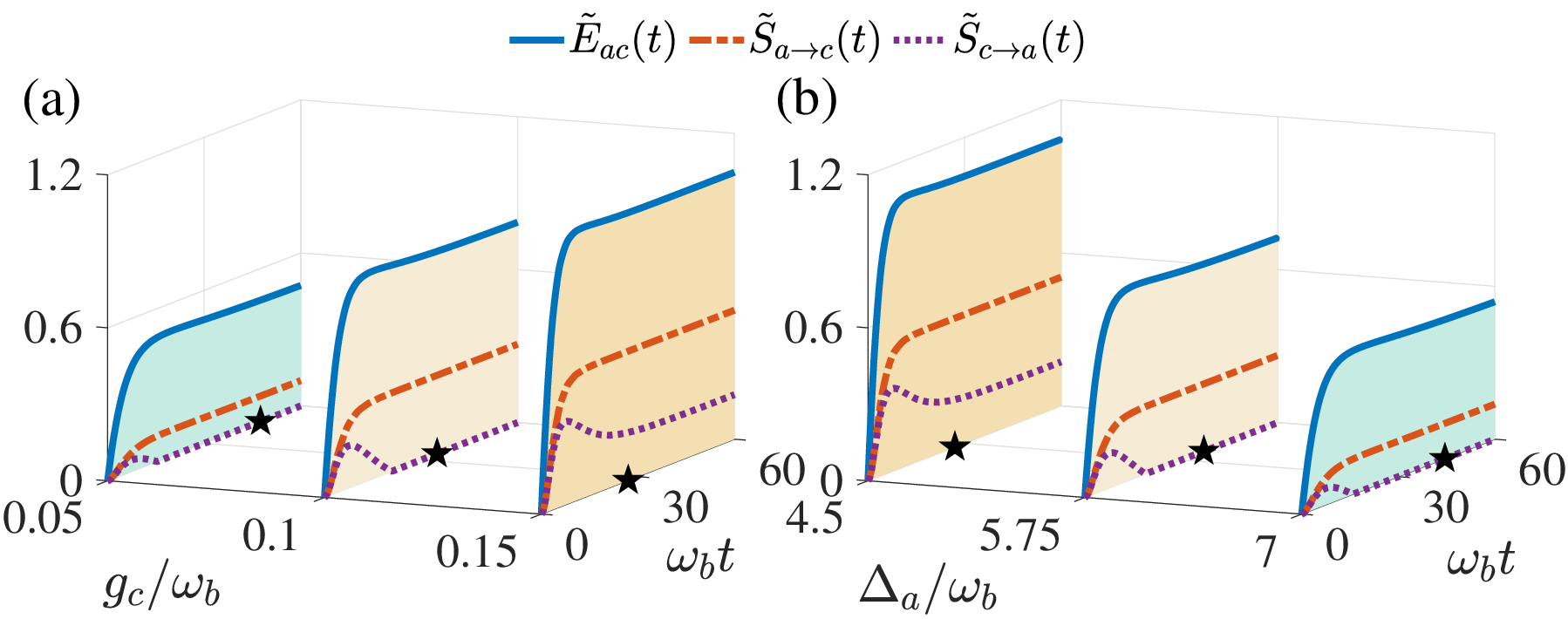}
\includegraphics[width=0.48\textwidth]{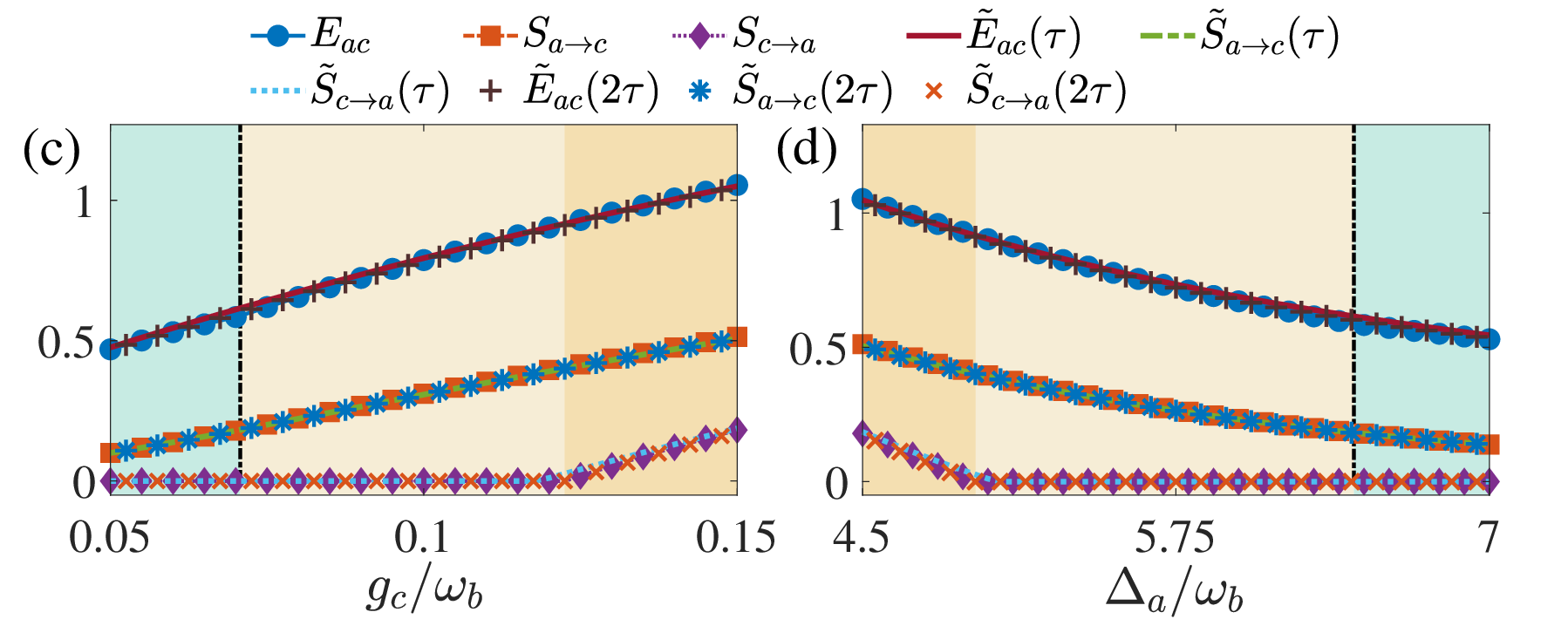}
\caption{The persistent MO entanglement and quantum steering in the EOM system. [(a), (b)] Dynamics of MO quantum resources for different coupling strengths $g_c$ and detunings $\Delta_a$, respectively. Star symbols indicate the values of $\tau$. [(c), (d)] Theoretical predictions (lines and marks) of MO quantum resources, together with numerical results at times $\tau$ (lines) and $2\tau$ (marks), shown versus the coupling strength $g_c$ and detuning $\Delta_a$, respectively. The black dot-dashed line denotes the boundary between steady-state (the teal region) and unsteady-state (the beige and golden-yellow regions) regimes in (c) and (d). $E_{ac}$, $S_{a\to c}$, and $S_{c\to a}$ are evaluated from the theoretical predictions ($t\to\infty$), while $\tilde E_{ac}$, $\tilde S_{a\to c}$, and $\tilde S_{c\to a}$ are obtained from the full Hamiltonian. $\Delta_a=5\omega_b$ for (a) and (c). $g_c=0.12\omega_b$ for (b) and (d). Other parameters are set as $g_a=0.12\omega_b$, $\kappa_c=2\kappa_a=10^{-3}\omega_b$, $\kappa_{b} = 10^{-6}\omega_b$, $N_a=N_c=0$, and $N_{b}=10$.}
\label{eom1}
\end{figure} 

For the specific EOM system, the main manuscript presents numerical results for various quantum resources under different coupling strengths $g_a$, demonstrating that these resources can be precisely controlled via  $g_a$. In the following, we provide additional results to elucidate the dependence of quantum resources on other physical parameters, including the coupling strength $g_c$ and the detuning $\Delta_a$.

As shown in Supplementary Figures \ref{eom1}(a) and (b), both the MO entanglement $\tilde E_{ac}$ and quantum steering $\tilde S_{a\to c}$ initially increase rapidly and converge to their asymptotic values before the characteristic time $\tau$, after which they remain on a plateau. In contrast, $\tilde S_{c\to a}$ exhibits distinct dynamics, rising to a peak before decaying toward its asymptotic value and eventually reaching a plateau.
In Supplementary Figure \ref{eom1}(a), quantum resources increase with $g_c$. At $g_c=0.05\omega_b$, $\tilde E_{ac}\approx0.47$ and $\tilde S_{a\to c}\approx0.1$ are relatively weak; at $g_c=0.1\omega_b$, they rise to $\tilde E_{ac}\approx0.79$ and $\tilde S_{a\to c}\approx0.31$; and further enhancement to $g_c=0.15\omega_b$ yields asymmetric two-way steering ($\tilde E_{ac}\approx1.05$, $\tilde S_{a\to c}\approx0.51$, and $\tilde S_{c\to a}\approx0.17$). By contrast, in Supplementary Figure \ref{eom1}(b), the quantum resources decrease with increasing detuning $\Delta_a$. At $\Delta_a=4.5\omega_b$, asymmetric two-way steering is observed ($\tilde S_{a\to c}>\tilde S_{c\to a}>0$). Dynamically persistent quantum resources can thus be controlled via $g_c$ and $\Delta_a$. A quantitative analysis of the MO entanglement and quantum steering at times $\tau$ and $2\tau$ is presented in Supplementary Figures \ref{eom1}(c) and (d). The numerical results for $\tilde E_{ac}$, $\tilde S_{a\to c}$, and $\tilde S_{c\to a}$ at times $\tau$ and $2\tau$ agree well with the discrete markers representing theoretical predictions from Eqs. \eqref{SMeni} and \eqref{SMsact2}.

\begin{figure}[t] 
\centering
\includegraphics[width=0.48\textwidth]{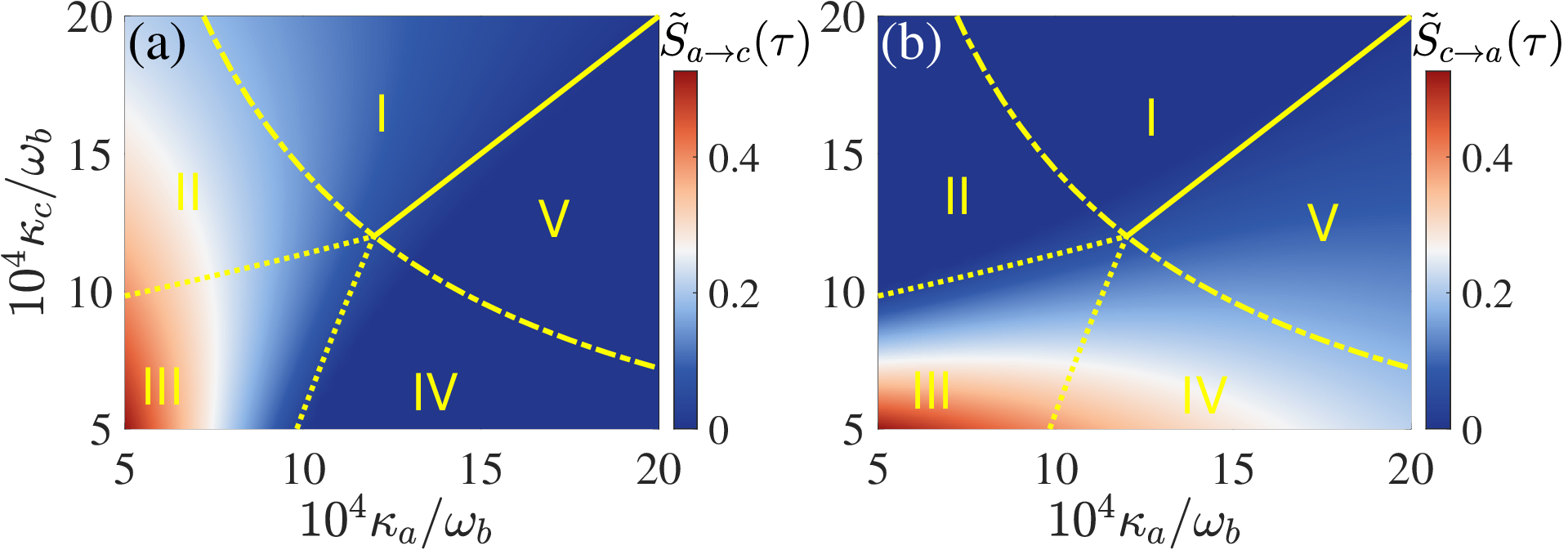}
\caption{The performance of quantum steering in the EOM system, including (a) $\tilde{S}_{a\to c}(\tau)$ and (b) $\tilde{S}_{c\to a}(\tau)$, with varying decay rates $\kappa_a$ and $\kappa_c$, respectively. The yellow dot-dashed line delineates the theoretically predicted boundary between the unsteady-state (left) and steady-state (right) regions. The yellow solid line marks the theoretically predicted disappearance of quantum steering. The yellow dotted line represents the theoretical regime of two-way steering. $g_c=0.12\omega_b$ and other parameters are the same as Supplementary Figure \ref{eom1}(a).}
\label{eom2}
\end{figure}

Quantum steering $\tilde S_{a\to c}(\tau)$ and $\tilde S_{c\to a}(\tau)$ are shown in Supplementary Figure \ref{eom2} over the parameter space of microwave and optical decay rates, respectively. $\tilde S_{a\to c}>0$ arises in regions I and II for $\kappa_a<\kappa_c$, while $\tilde S_{c\to a}>0$ occurs in regions IV and V for $\kappa_a>\kappa_c$, independent of whether the system dynamics is steady or unsteady. Quantum steering vanishes in the steady state near $\kappa_a\simeq\kappa_c$, whereas two-way steering emerges ($S_{a\leftrightarrow c}>0$) in the unsteady-state regime at lower decay rates in region III. Moreover, the smaller the decay rate $\kappa_a$, the broader the range of $\kappa_c$ that sustains high-quality $\tilde S_{a\to c}$, and vice versa. For instance, at $\kappa_a=5\times10^{-4}\omega_b$, a wide range $5\le10^4\kappa_c/\omega_b\le20$ yields $\tilde S_{a\to c}>0.2$, as shown in Supplementary Figure \ref{eom2}(a). All numerical results are in excellent agreement with the theoretical predictions of Eqs. \eqref{SMstable} and \eqref{SMunstable}.

\subsection{Validity analysis and experimental feasibility}\label{ValidityEOM}

\begin{figure*}[t] 
\centering
\includegraphics[width=0.85\textwidth]{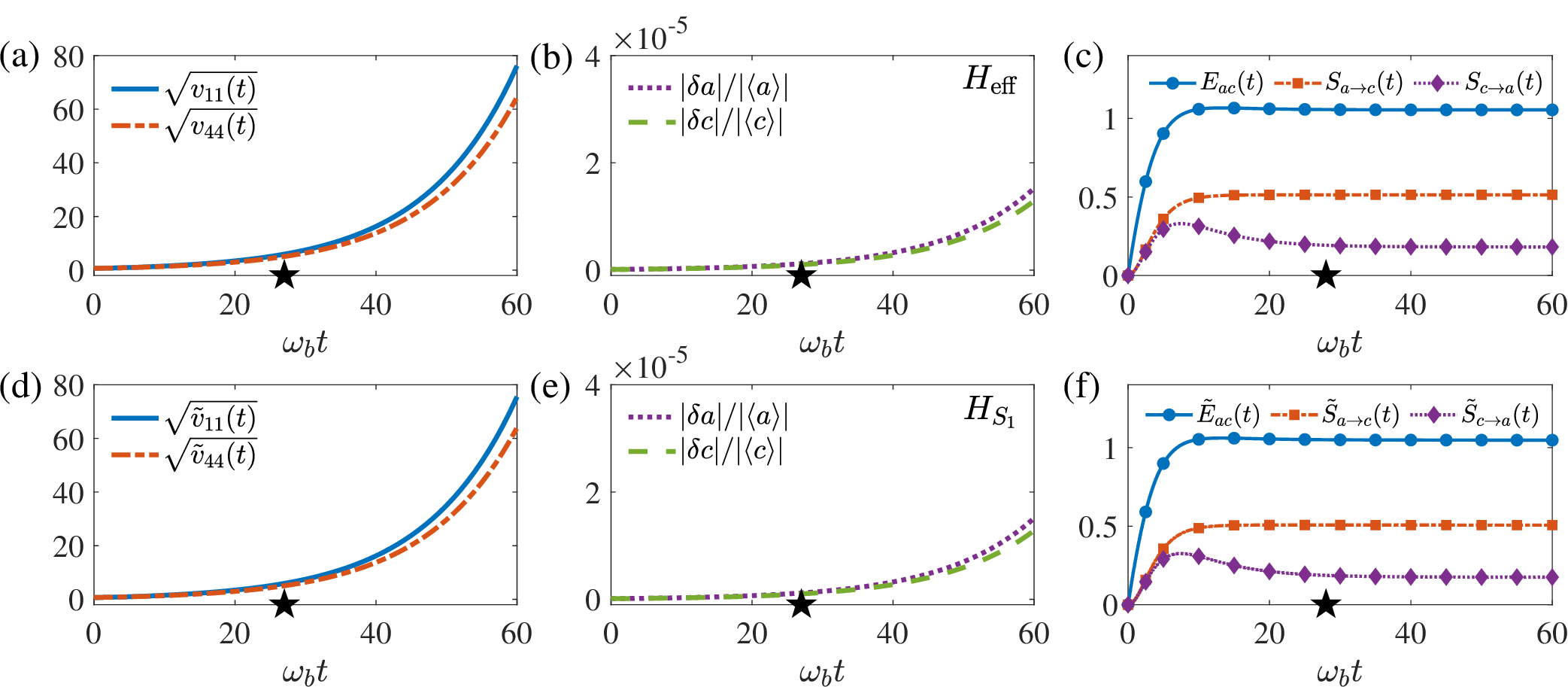}
\caption{Comparison of the dynamical evolution of the square roots of the diagonal CM elements [(a) and (d)], relative quantum fluctuations [(b) and (e)], and quantum resources [(c) and (f)] in the EOM system, calculated using the effective Hamiltonian [(a)--(c)] and the full Hamiltonian [(d)--(f)]. The star symbols mark the characteristic time $\tau$. Here, $g_a=0.15\omega_b$, $g_c=0.12\omega_b$, the mean-field amplitudes are set to $5\times10^{6}$, and the other parameters are the same as those in Supplementary Figure \ref{eom1}(a).}
\label{eomv}
\end{figure*}

In this Supplementary Note, we examine the validity and experimental feasibility of the protocol presented in \ref{SMEOMH} and \ref{eomAmatrix}. We first verify the self-consistency of the linearized treatment within the characteristic time window $t\in[\tau,2\tau]$, during which the generated quantum resources remain essentially unchanged. The validity of the linearized description requires that the quantum fluctuations remain much smaller than the corresponding mean-field amplitudes throughout the evolution, namely,
\begin{equation}
|\delta a|\ll|\langle a\rangle|,\quad|\delta c|\ll|\langle c\rangle|.
\end{equation}
Here, $\delta a$ and $\delta c$ denote the quantum fluctuations of the microwave and optical modes, respectively. Their evolution can be predicted from the CM associated with the effective Hamiltonian in Eq. \eqref{SMeomHefac}. Specifically, the corresponding fluctuation occupations are related to the diagonal elements of the CM [Eq. \eqref{SMv(t)}] as
\begin{equation}
\begin{aligned}
\langle\delta a^\dagger\delta a\rangle
&=\frac{1}{2}\left[v_{11}(t)+v_{22}(t)-1\right],\\
\langle\delta c^\dagger\delta c\rangle
&=\frac{1}{2}\left[v_{33}(t)+v_{44}(t)-1\right].\\
\end{aligned}
\end{equation}
Since $v_{11}(t)=v_{22}(t)$ and $v_{33}(t)=v_{44}(t)$ throughout the evolution, and the relevant CM elements become large in the long-time limit in the unsteady-state regime, one obtains
\begin{equation}\label{SMdelac}
\begin{aligned}
\langle\delta a^\dagger\delta a\rangle
&\approx v_{11}(t),&|\delta a|&\approx\sqrt{v_{11}(t)},\\
\langle\delta c^\dagger\delta c\rangle&\approx v_{44}(t),&|\delta c|&\approx\sqrt{v_{44}(t)},
\end{aligned}
\end{equation}
for sufficiently large $t$. The analytical solutions show that $v_{11}(t)$ and $v_{44}(t)$ are dominated by an exponentially growing term with exponent $\Omega-\kappa_a-\kappa_c>0$. Accordingly,
\begin{equation}
v_{11}(2\tau),\,v_{44}(2\tau)\propto e^{2\tau(\Omega-\kappa_a-\kappa_c)}.
\end{equation}
In addition, the mean-field amplitudes obtained from the classical equations of motion can be approximately estimated as $|\langle o\rangle|\approx|{\Omega_o}/{\Delta_o}|~(o=a,c)$~\cite{SMEOM12,SMCOMMLJ23}.

The predictions based on the CM dynamics governed by the effective Hamiltonian are shown in Supplementary Figures \ref{eomv}(a)--\ref{eomv}(c), while the corresponding numerical simulations of the full EOM dynamics are presented in Supplementary Figures \ref{eomv}(d)--\ref{eomv}(f). The two sets of results are in excellent agreement. They show that the square roots of the diagonal CM elements satisfy $\sqrt{v_{11}(t)},\sqrt{v_{44}(t)}\lesssim80$ throughout the considered time interval, implying $|\delta a|,|\delta c|\lesssim80$ for $t\in[\tau,2\tau]$. Since the corresponding mean-field amplitudes are of order $10^{6}$, the relative fluctuations satisfy ${|\delta o|}/{|\langle o\rangle|}\lesssim 10^{-5}$ ($o=a,c$) over the relevant timescale $t\lesssim2\tau$. Therefore, the relative quantum fluctuations remain sufficiently small throughout the relevant evolution, while the MO entanglement and quantum steering remain essentially unchanged within the characteristic time window $t\in[\tau,2\tau]$. These results verify the self-consistency of the linearized treatment and confirm the validity of the characteristic time window in which persistent quantum resources can be generated in the unsteady-state regime.

\begin{figure}[t] 
\centering
\includegraphics[width=0.49\textwidth]{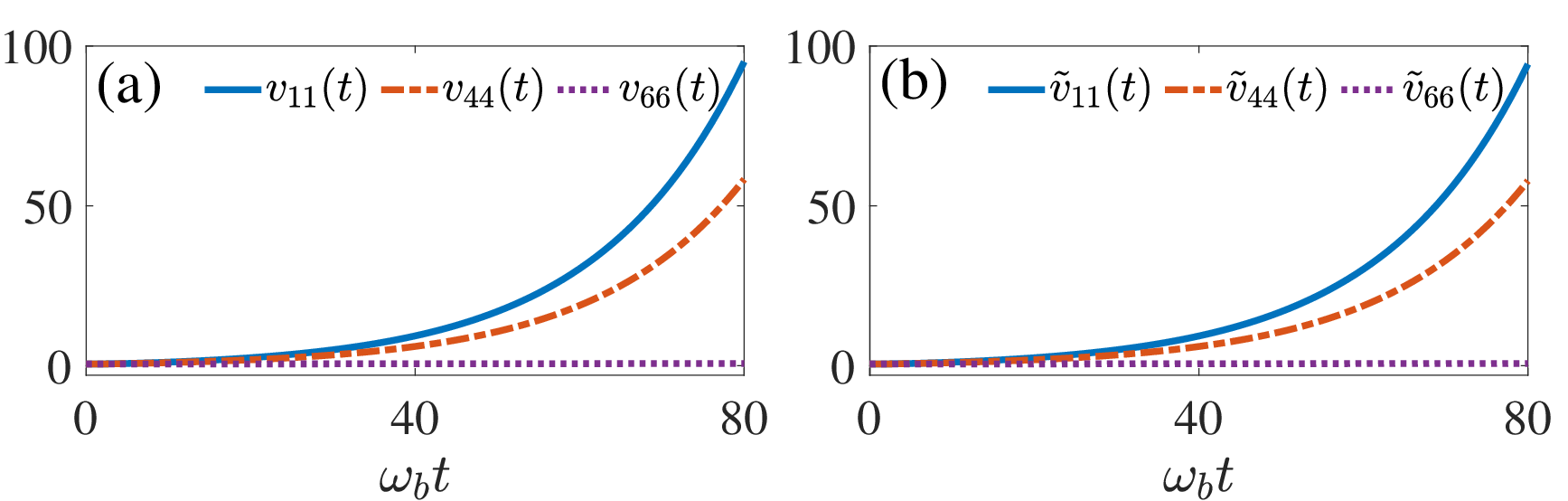}
\caption{Dynamical evolution of the CM elements in the EOM system. (a) $v_{11}(t)$ and $v_{44}(t)$ obtained from the effective Hamiltonian, together with $v_{66}(t)$ obtained from the free Hamiltonian of the mechanical mode. (b) $\tilde{v}_{11}(t)$, $\tilde{v}_{44}(t)$ and $\tilde{v}_{66}(t)$ obtained from the full Hamiltonian. Here, $g_c=0.1\omega_b$, and the other parameters are the same as those in Supplementary Figure \ref{eom1}(a).}
\label{eomef}
\end{figure}

We next justify neglecting the environmental effects associated with the intermediate mode, thereby allowing the effective description to be formulated as a reduced open quantum system involving only the environments of the target modes. Under the effective-Hamiltonian description, the EOM Hamiltonian can be decomposed as $H=H_{\rm eff}\oplus H_{FB}$, where $H_{\rm eff}$ is given by Eq. \eqref{SMeomHefac}, and $H_{FB}$ is the free Hamiltonian of the mechanical mode. The interactions between the target and intermediate modes are incorporated into the effective coupling $g_{\rm eff}$ in $H_{\rm eff}$, whereas $H_{FB}$ describes only the free evolution of the mechanical mode. Within the open-quantum-system framework, each mode is coupled to an independent Markovian environment. Consequently, the environmental noise acting on different modes is uncorrelated, and no bath-induced correlations are generated. In particular, the environment coupled to mode $b$ affects only its local dynamics, which are governed by the effective QLE
\begin{equation}
\dot{b}=-(i\omega_b+\kappa_b)b+\sqrt{2\kappa_b}b_{in}.
\end{equation}
The dissipation term $\kappa_b$ and the corresponding input noise $\sqrt{2\kappa_b}b_{in}$ influence only the mechanical mode $b$ and do not introduce any additional coupling or feedback between the target microwave mode $a$ and the optical mode $c$.

Supplementary Figure \ref{eomef} compares the evolution of the CM elements obtained from the reduced open-system description (supplemented by the free evolution of the intermediate mode) with those obtained from the full-system dynamics. The excellent agreement between the two approaches validates the reduced open-system description and confirms that the environmental effects associated with the mechanical mode have a negligible influence on the target modes. Moreover, the close agreement between $v_{66}(t)$ and $\tilde{v}_{66}(t)$ shows that the intermediate mode evolves approximately as a free mode, indicating that the environmental noise acting on this mode is confined to its local dynamics and does not significantly affect the target modes.

Finally, we discuss the experimental feasibility of the parameters used in our numerical simulations. Throughout the manuscript, all parameters are expressed in dimensionless units normalized to the mechanical frequency. Typical mechanical frequencies and decay rates are $\omega_b/2\pi\sim10-100~\mathrm{MHz}$ and $\kappa_b\sim10^{-6}\omega_b$, respectively~\cite{SMCOMMLJ23,SMcavityoptomechanics14}. The cavity decay rates are typically in the range $\kappa_o\sim10^{-4}\omega_b-10^{-2}\omega_b$ ($o=a,c$)~\cite{SMCMM22,SMEOM12,SMEOM15,SMEOM25,SMCOMMLJ23,SMcavityoptomechanics14}. Meanwhile, the driving strengths can reach $\Omega_o\sim10^{13}-10^{14}~\mathrm{Hz}$, yielding enhanced cavity-mechanical coupling strengths of $g_o\sim0.1\omega_b$~\cite{SMEOM15,SMCOMMLJ23}. Such coupling strengths readily access the unsteady-state regime considered in this work. Moreover, at cryogenic temperatures ($T\sim10~\mathrm{mK}$), the thermal occupation numbers of the microwave and optical reservoirs are effectively zero ($N_a\approx N_c\approx0$), whereas the mechanical reservoir typically has a thermal occupation of $N_b\sim1-10$~\cite{SMCOMMLJ23}. However, for mechanical frequencies in the range $\omega_b/2\pi\sim10-100~\mathrm{MHz}$, achieving the relatively small optical cavity decay rate $\kappa_c=10^{-3}\omega_b$ remains experimentally challenging. Nevertheless, continued advances in microcavity engineering are expected to overcome this limitation. Moreover, electromechanical frequencies in the GHz regime have already been demonstrated in EOM hybrid systems~\cite{SMEOM13}. In this case, the corresponding optical cavity decay rate falls within the MHz regime, which is readily accessible with current experimental techniques. 

Furthermore, the effective coupling strength $g_{\rm eff}$ and the optical cavity decay rate $\kappa_c$ are connected through a multi-parameter trade-off. As shown in Eq. \eqref{SMeomgeff}, $g_{\rm eff}\propto g_c$, where the enhanced coupling satisfies $g_c\propto\sqrt{{2P_c\kappa_c}/(\hbar\omega_L)}$, where $P_c$ is the driving power and $\omega_L$ is the driving frequency. Consequently, ${g_{\rm eff}}\propto\sqrt{{2P_c\kappa_c}/{(\hbar\omega_L)}}$. This relation indicates that, for a fixed optical cavity decay rate $\kappa_c$, the effective coupling strength $g_{\rm eff}$ can be further enhanced by increasing the driving power $P_c$, making the unsteady-state condition $g_{\rm eff}^2\ge\kappa_a\kappa_c$ experimentally achievable with continued technological advances. Finally, the generated MO quantum resources can be experimentally characterized via homodyne detection~\cite{SMEOM12,SMEOM13}.

\section{Application in cavity optomagnomechanical systems}\label{SMCOMM}
\subsection{Model and the effective Hamiltonian}\label{SMCOMMH}
\begin{figure}[htbp] 
\centering
\includegraphics[width=0.33\textwidth]{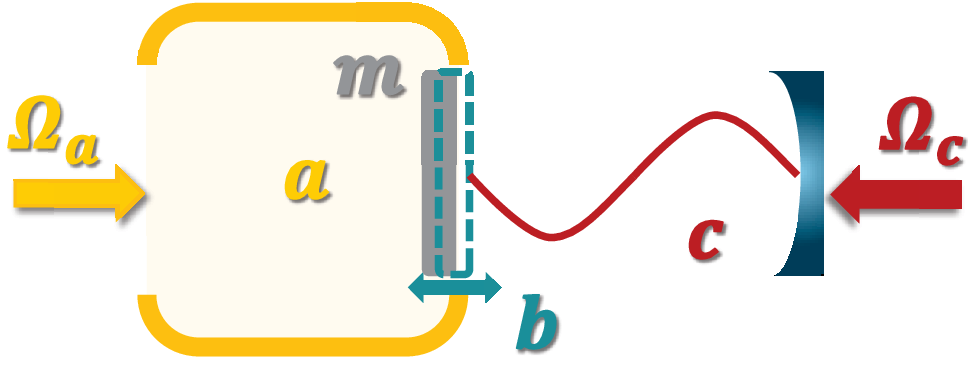}
\caption{Schematic of the cavity optomagnomechanical system. A yttrium iron garnet crystal is placed inside a microwave cavity near the maximum magnetic field in microwave mode $a$, which provides the magnon mode $m$. It simultaneously serves as a vibrating end mirror (mechanical mode $b$) of the optical cavity $c$. The microwave and optical photon modes are driven by a microwave drive field $\Omega_a$ and an optical driving laser $\Omega_c$, respectively.}
\label{commdiagram}
\end{figure}

To demonstrate the generality of our approach, we further analyze the generation of MO resources in a cavity optomagnomechanical (COMM) system~\cite{SMCOMMLJ23}, as illustrated in Supplementary Figure \ref{commdiagram}. In the rotating frame with respect to the driving frequencies, the Hamiltonian of the COMM system is ($\hbar\equiv1$)
\begin{equation}\label{SMcommH1}
\begin{aligned}
H_{S_2}=&\Delta_aa^\dag a+\Delta_m{m}^\dag {m}+\omega_b{b}^\dag {b}+\Delta_cc^\dag c\\&+g_a(a^\dag m+am^\dag)+g_{mb}{m}^\dag m(b+b^\dag)\\
&+g_{bc}c^\dag c(b+b^\dag)+i\Omega_a(a^\dag-a)\\&+i\Omega_c(c^\dag-c),
\end{aligned}
\end{equation}	
where $a$ ($a^\dag$), $m$ ($m^\dag$), $b$ ($b^\dag$), and $c$ ($c^\dag$) are the annihilation (creation) operators of the microwave, magnon, mechanical, and optical modes, respectively. $\Delta_o=\omega_o-\epsilon_o~(o=a,m,c)$ denotes the detuning of mode $o$, with $\omega_o$ and $\epsilon_o$ being the transition and driving-field frequencies, respectively, and $\epsilon_m=\epsilon_a$. $\omega_b$ is the transition frequency of the mechanical mode. $g_a$ is the microwave-magnon coupling strength. The magnomechanical (optomechanical) coupling strength $g_{mb}$ ($g_{bc}$) is typically small, considering the large frequency mismatch between the magnon (optical) and the mechanical modes, yet it can be significantly enhanced by driving the microwave (optical) mode with a strong field with Rabi frequency $\Omega_a$ ($\Omega_c$).

All microwave, magnon, and optical modes have considerable expectation values under strong driving fields. This allows us to linearize~\cite{SMCOMMLJ23} the system dynamics by writing the operators $o=\delta o+\langle o\rangle~(o=a,m,c)$, where $\langle o\rangle$ is the steady-state value and $\delta o$ represents the quantum fluctuation of the mode $o$. The Hamiltonian in Eq. \eqref{SMcommH1} turns to
\begin{equation}\label{SMcommH2}
\begin{aligned}
H_{S_2}=~& H_0+V,\quad H_0=\omega_bb^\dag b+\sum_{o=a,m,c}\Delta_o\delta o^\dag \delta o,\\
V=~&g_a(\delta a^\dag \delta m+\delta a\delta m^\dag)+g_m(\delta m+\delta m^\dag)\\&\times(b+b^\dag)+g_c(\delta c+\delta c^\dag)(b+b^\dag),\\
\end{aligned}
\end{equation}	
where $g_m=g_{mb}\langle m\rangle$ and $g_{c}=g_{bc}\langle c\rangle$ are the effective magnomechanical and optomechanical coupling strengths, respectively. For simplicity and without loss of generality, we assume $\langle o\rangle$ is a real number and make $\delta o\to o$ in the following content. It corresponds to the linearized Hamiltonian given in Eq. (1) of the main text for the specific case $\theta=0$, $\phi=\pi/4$, and $N=2$, with the substitutions $\omega_1\to\Delta_m$, $\omega_2\to\omega_b$, $g_1\to g_m$, $g_c\to\sqrt{2}g_c$, $b_1\to m$, and $b_2\to b$. Then, we apply the general effective Hamiltonian construction approach introduced in the Methods section of the main text to derive the effective coupling of the MO target subsystem in the COMM system.

When the microwave detuning $\Delta_a$ is near opposite the optical detuning $\Delta_c$, and both of them are far resonant from the magnon detuning $\Delta_m$ and the mechanical frequency $\omega_b$, i.e., $\Delta_a\approx -\Delta_c$ and $g_a,g_m,g_c\ll\{|\Delta_a-\Delta_m|, |\Delta_c-\omega_b|\}$, it is found that the tensor-product state $|nljk\rangle\equiv|n\rangle_a|l\rangle_{m}|j\rangle_{b}|k\rangle_c$ is near-degenerate with $|(n+1)lj(k+1)\rangle$. Using perturbation theory, the effective Hamiltonian for transitions between any base-pair $|nljk\rangle$ and $|(n+1)lj(k+1)\rangle$ can be analytically derived. It can be written in the following form
\begin{equation}\label{SMcommHeac}
\begin{aligned}
H_{\rm eff}=&\epsilon_1|nljk\rangle\langle nljk|+(\Delta_{a}+\Delta_{c}+\epsilon_2)\\&\times|(n+1)lj(k+1)\rangle\langle(n+1)lj(k+1)|\\&+\tilde g_{\rm eff}(|nljk\rangle\langle (n+1)lj(k+1)|+{\rm h.c.}),
\end{aligned}
\end{equation}
where $\epsilon_1$ and $\epsilon_2$ are the energy shifts induced by the coupling of the states $|nljk\rangle$ and $|(n+1)lj(k+1)\rangle$, respectively, and $\tilde g_{\rm eff}$ is the effective coupling strength. Here, we omit the common unperturbed eigenenergy of two bases.

\begin{figure}[t] 
\centering
\includegraphics[width=0.48\textwidth]{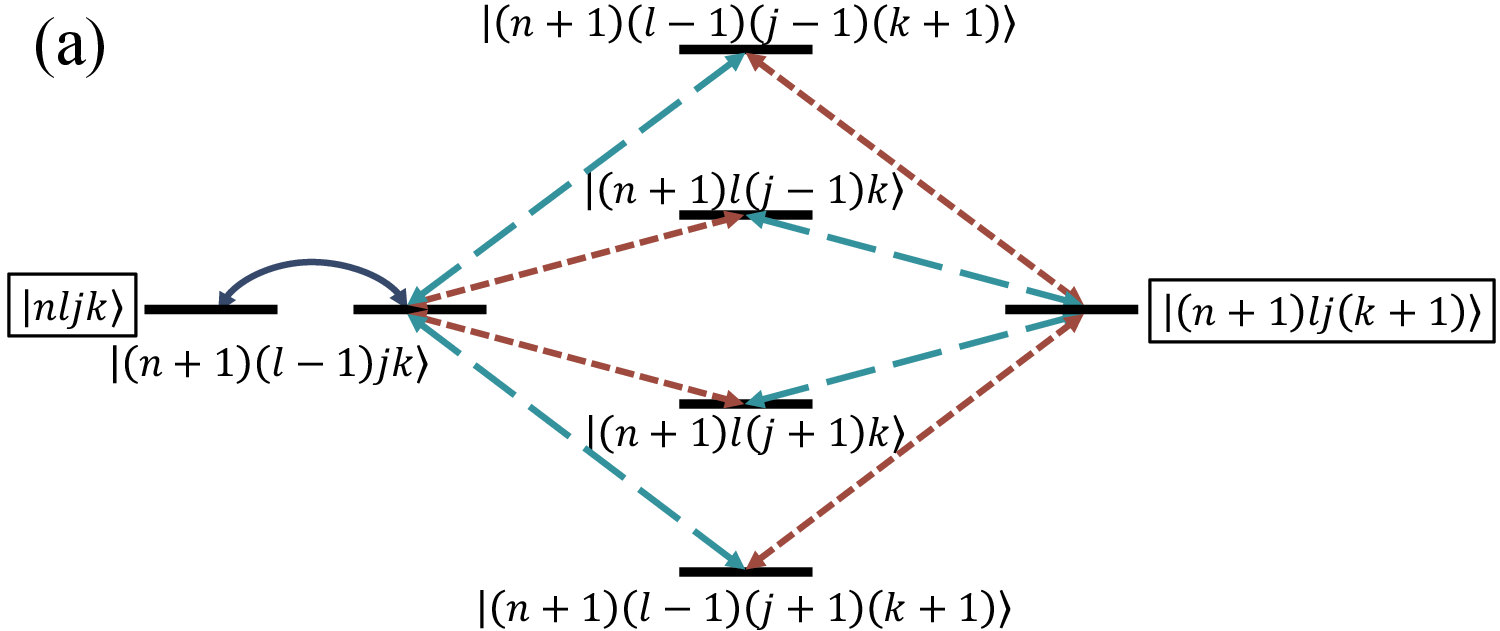}
\includegraphics[width=0.48\textwidth]{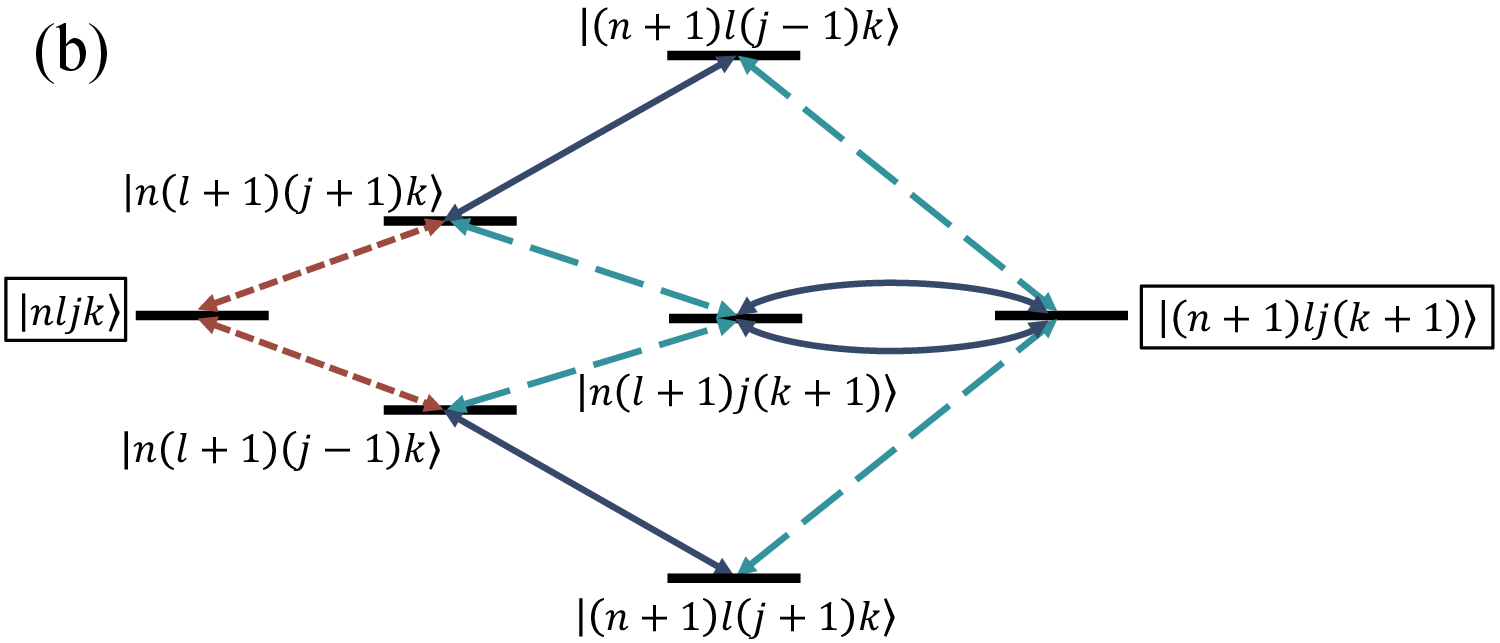}
\includegraphics[width=0.48\textwidth]{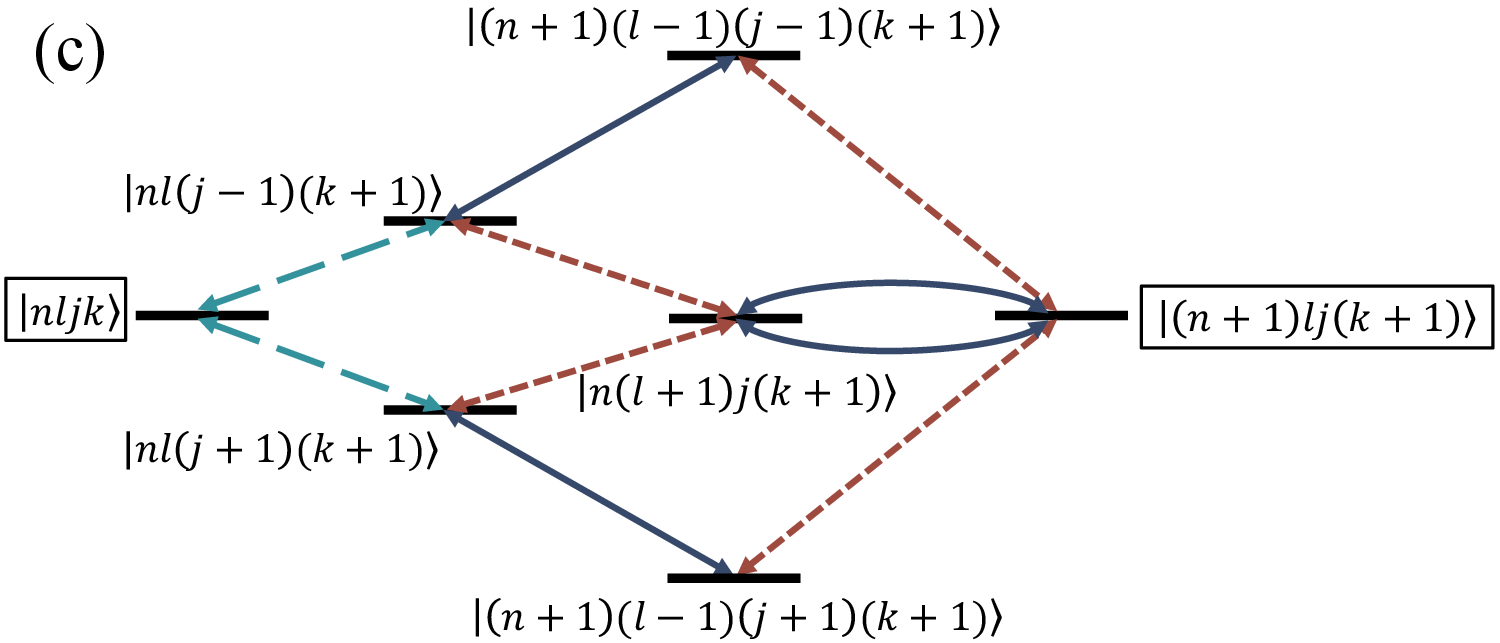}
\caption{All twelve third-order paths connecting $|nljk\rangle$ and $|(n+1)lj(k+1)\rangle$. Solid, dotted, and dashed lines represent the couplings between the microwave and magnon modes, the magnon and mechanical modes, and the optical and mechanical modes, respectively.}
\label{commacpath}
\end{figure}

Using Eq. (18) in the Methods section of the main text and summing over all twelve paths in Supplementary Figure \ref{commacpath}, we obtain the effective coupling strength
\begin{equation}
\tilde{g}_{\rm eff}=\frac{2\omega_bg_ag_mg_c}{(\Delta_m-\Delta_a)(\omega_b^2-\Delta_a^2)}\sqrt{(n+1)(k+1)}
\end{equation}
up to the third-order perturbation of coupling strengths. Similarly, via Eq. (22) in the Methods section of the main text, the second-order energy shift $\delta$ can be derived as
\begin{equation}	
\delta\equiv\epsilon_1-\epsilon_2=\frac{g_a^2}{\Delta_m-\Delta_a}+\frac{g_c^2}{\omega_b-\Delta_a}+\frac{g_c^2}{\omega_b+\Delta_a}.
\end{equation}

\begin{figure*}[t] 
\centering
\includegraphics[width=0.96\textwidth]{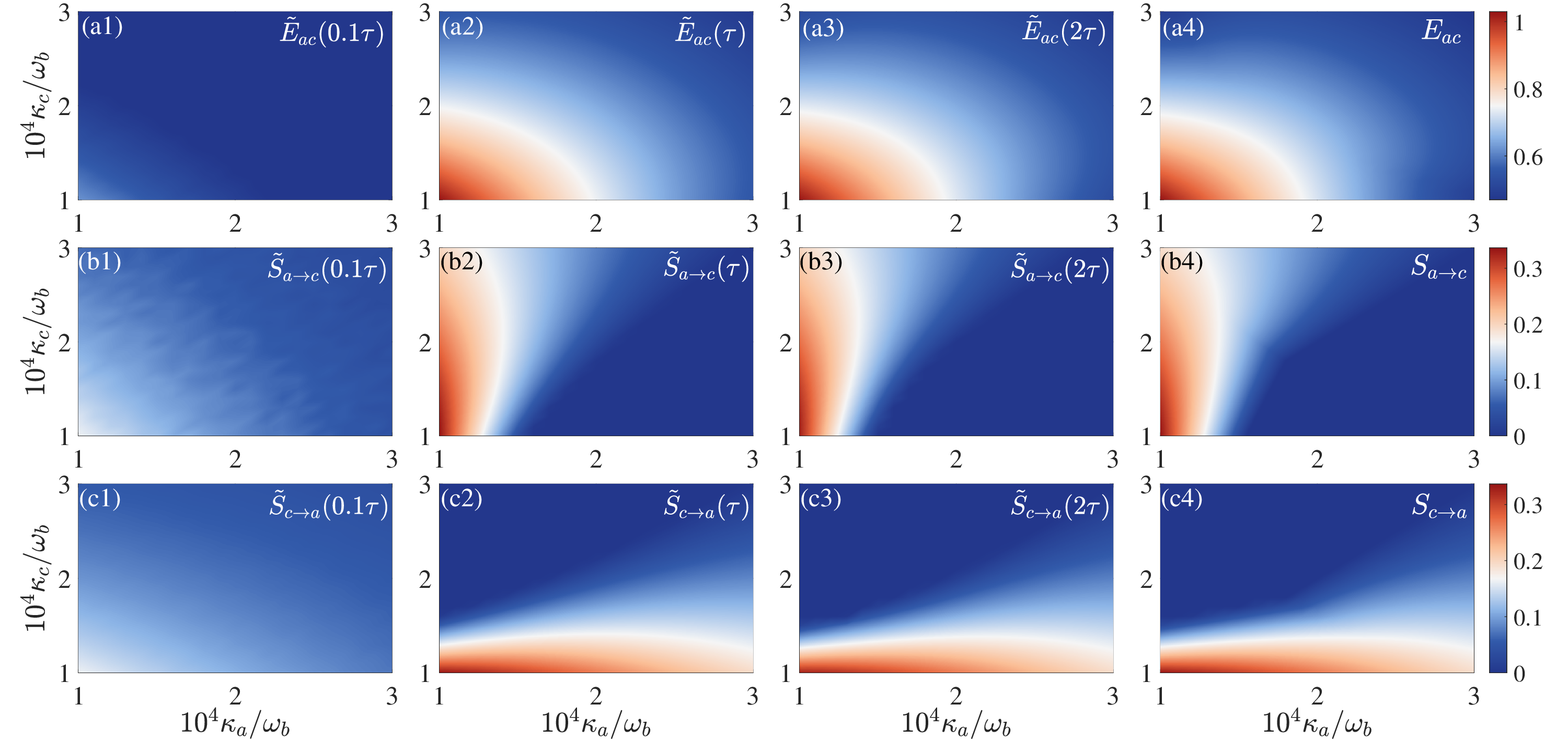}
\caption{The performance of (a) MO entanglement $\tilde E_{ac}$ and quantum steering (b) $\tilde S_{a\to c}$  and (c) $\tilde S_{c\to a}$  in the decay rates $\kappa_a$ and $\kappa_c$ space at times $0.1\tau$, $\tau$, and $2\tau$, respectively, in the COMM system, in agreement with the corresponding theoretical predictions at $t\to\infty$. The parameters are set as $g_a=g_c=0.12\omega_b$, $g_m=0.1\omega_b$, $\Delta_m=\omega_b$, $\Delta_a=3\omega_b$, $\kappa_m=10^{-3}\omega_b$, $\kappa_b=10^{-6}\omega_b$, $N_a=N_m=N_c=0$, and $N_{b}=10$.}
\label{comm1}
\end{figure*}

Therefore, under the condition $\Delta_a=-\Delta_c+\delta$, the effective Hamiltonian in Eq. \eqref{SMcommHeac} can be derived as
\begin{equation}\label{SMcommHeac1}
\begin{aligned}
H_{\rm eff}=&\tilde{g}_{\rm eff}(|nljk\rangle\langle (n+1)lj(k+1)|+{\rm h.c.})\\
=&g_{\rm eff}\sqrt{(n+1)(k+1)}(|nk\rangle\langle (n+1)(k+1)|\\
&+{\rm h.c.})\otimes|lj\rangle\langle lj|.
\end{aligned}
\end{equation}
The magnon and mechanical modes can be eliminated. Expanding the effective Hamiltonian in Eq. \eqref{SMcommHeac1} from the subspace to the full Hilbert space of microwave and optical modes, the effective Hamiltonian in Eq. \eqref{SMcommHeac1} eventually becomes
\begin{equation}\label{SMHefac}
\begin{aligned}
H_{\rm eff}=g_{\rm eff}(a^\dag c^\dag+ac),
\end{aligned}
\end{equation}
and the coupling strength can be written as
\begin{equation}\label{SMcommeff}
g_{\rm eff}=\frac{2g_ag_mg_c\omega_b}{(\Delta_m-\Delta_a)(\omega_b^2-\Delta_a^2)}.
\end{equation}
This is precisely the effective coupling strength in the COMM system given in the manuscript.

\subsection{Quantum resources and validity analysis}\label{commAmatrix}

This Supplementary Note contributes to the calculation of CM for full hybrid COMM system. In the open-quantum-system framework, under the standard Markovian environments, one can derive the QLEs for the full linearized Hamiltonian in Eq. \eqref{SMcommH2}
\begin{equation}\label{SMlineLangevin}
\begin{aligned}
&\dot a=-(i\Delta_a+\kappa_a)a-ig_am+\sqrt{2\kappa_a}a_{in},\\
&\dot c=-(i\Delta_c+\kappa_c)c-ig_{c}(b+b^\dag)+\sqrt{2\kappa_c}c_{in},\\	
&\dot m=-(i\Delta_m+\kappa_{m}){m}-ig_aa-ig_m(b+b^\dag)\\&\quad\quad+\sqrt{2\kappa_{m}}m_{in},\\
&\dot b=-(i\omega_b+\kappa_{b})b-i\sum_{o=m,c}g_o(o+o^\dag)+\sqrt{2\kappa_{b}}b_{in},\\
\end{aligned}
\end{equation}
where $\kappa_o (o=a,c,m,b)$ are the decay rates of the microwave, optical, magnon, and mechanical modes, respectively. The input noise operator $o_{in}$ is zero mean and satisfies the correlation functions, $\langle o_{in}(t)o^\dag_{in}(t') \rangle=[N_o(\omega_o)+1]\delta(t-t')$ and $\langle o^\dag _{in}(t)o_{in}(t')\rangle=N_o(\omega_o)\delta(t-t')$, where $N_o(\omega_o)$ is the corresponding average thermal excitation number. 

By introducing the quadrature operators $X_o(t)=(o+o^\dag)/\sqrt{2}$ and $Y_o(t)=i(o^\dag-o)/\sqrt{2}$, the above Eqs. \eqref{SMlineLangevin} can be written in the matrix form
\begin{equation}\label{SMutcomm}
\dot{\tilde{u}}(t)=A\tilde{u}(t)+\xi(t)
\end{equation}
where $\tilde{u}(t)=[u(t)^T,X_{m}(t),Y_{m}(t),X_{b}(t),Y_{b}(t)]^T$ is the operator vector, and $u(t)$ is shown in Eq. \eqref{SMueff}. $\linebreak$ $\xi(t)=[\xi^T_{\rm eff}(t),\sqrt{2\kappa_{m}}X_{m}^{in}(t),\sqrt{2\kappa_{m}}Y_{m}^{in}(t),\sqrt{2\kappa_{b}}X_{b}^{in}(t), \linebreak\sqrt{2\kappa_{b}}Y_{b}^{in}(t)]^T$, $\xi_{\rm eff}(t)$ is the noise vector in Eq. \eqref{SMueff}, $X_o^{in}(t)=(o_{in}+o_{in}^\dag)/\sqrt{2}$ and $Y_o^{in}(t)=i(o^\dag_{in}-o_{in})/\sqrt{2}~(o=m,b)$ are the quadratures of the input noise operators. The drift matrix $A$ is given by
\begin{align}\label{SMcommA}
A \!\!=\!\begin{pmatrix}
-\kappa_a&\Delta_a&0&0&0&g_a&0&0\\
-\Delta_a&-\kappa_a&0&0&-g_a&0&0&0\\
0&0&-\kappa_c&\Delta_c&0&0&0&0\\
0&0&-\Delta_c&-\kappa_c&0&0&-2g_c&0\\
0&g_a&0&0&-\kappa_{m}&\Delta_m&0&0\\
-g_a&0&0&0&-\Delta_m&-\kappa_{m}&-2g_m&0\\
0&0&0&0&0&0&-\kappa_{b}&\omega_b\\
0&0&-2g_c&0&-2g_m&0&-\omega_b&-\kappa_{b}
\end{pmatrix}\!.
\end{align}
Using the above linear dynamics described in Eq. \eqref{SMutcomm}, the full system can be characterized by a time-dependent CM, whose elements are defined as $\tilde{v}_{jl}(t)=\langle \tilde{u}_j(t)\tilde{u}_l(t)+ \tilde{u}_l(t)\tilde{u}_j(t)\rangle/2-\langle \tilde u_j(t)\rangle\langle \tilde u_l(t)\rangle~(j,l=1,2,...,8)$, and which satisfies
\begin{equation}\label{commvt}
\dot{\tilde v}(t)=A\tilde{v}(t)+\tilde{v}(t)A^T+D,
\end{equation}
where $D= {\rm diag}[\kappa_a(2N_a+1),\kappa_a(2N_a+1),\kappa_c(2N_c+1),\kappa_c(2N_c+1),\kappa_{m}(2N_{m}+1),\kappa_{m}(2N_{m}+1),\kappa_{b}(2N_{b}+1),\kappa_{b}(2N_{b}+1)]$ is the diffusion matrix and defined by $D_{jl}=\langle\xi_j(t)\xi_l(t)+\xi_l(t)\xi_j(t)\rangle/2$. The CM of the MO subsystem is $v=\tilde{v}(1:4;1:4)$. Then, the numerical results via the full Hamiltonian [Eq. \eqref{SMcommH2}] can be obtained by calculating the above differential equation.

\begin{figure}[t] 
\centering
\includegraphics[width=0.89\linewidth]{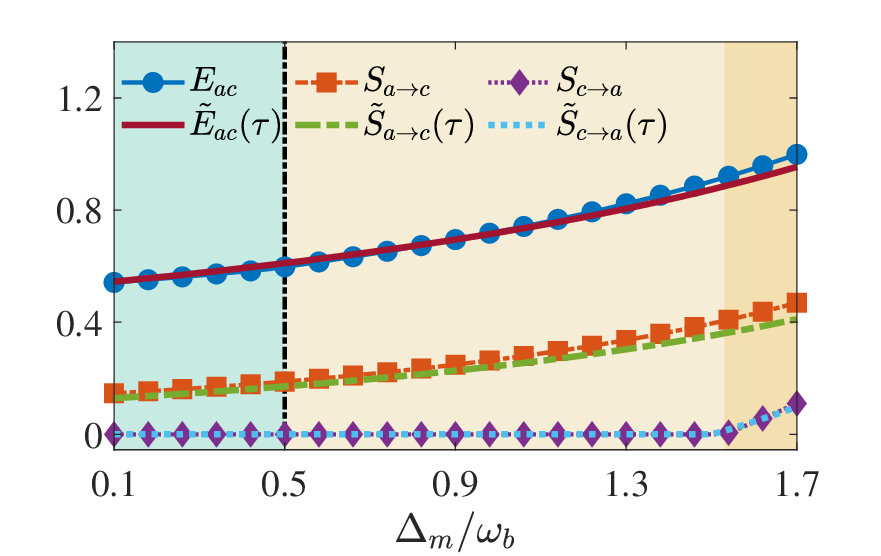}
\caption{Theoretical predictions and numerical results for the MO entanglement and quantum steering versus the detuning $\Delta_m$ in the COMM system. The black dot-dashed line indicates the boundary between the steady-state (left) and unsteady-state (right) regimes. $\kappa_a=0.5\kappa_c=10^{-4}\omega_b$ and the other parameters are the same as Supplementary Figure \ref{comm1}.}
\label{comm2}
\end{figure}

In Supplementary Figure \ref{comm1}, we present the numerical MO quantum resources at specific times $0.1\tau$, $\tau$, and $2\tau$, based on the full system dynamics described by Eq. \eqref{commvt}, along with the corresponding theoretical predictions under identical parameters. At $0.1\tau$, the values $\tilde E_{ac}$, $\tilde S_{a\to c}$, and $\tilde S_{c\to a}$ remain relatively weak compared to those at $\tau$. By $\tau$, the system develops quantum resources that remain nearly unchanged, as evidenced by the nearly identical values at $\tau$ and $2\tau$, consistent with theoretical predictions, $\tilde E_{ac}(\tau)\approx\tilde E_{ac}(2\tau)\approx E_{ac}$, $\tilde S_{a\to c}(\tau)\approx\tilde S_{a\to c}(2\tau)\approx S_{a\to c}$, and $\tilde S_{c\to a}(\tau)\approx\tilde S_{c\to a}(2\tau)\approx S_{c\to a}$. This demonstrates that persistent MO entanglement and quantum steering are established after a sufficient evolution time $\tau$, in agreement with theory. 

\begin{figure}[t] 
\centering
\includegraphics[width=0.38\textwidth]{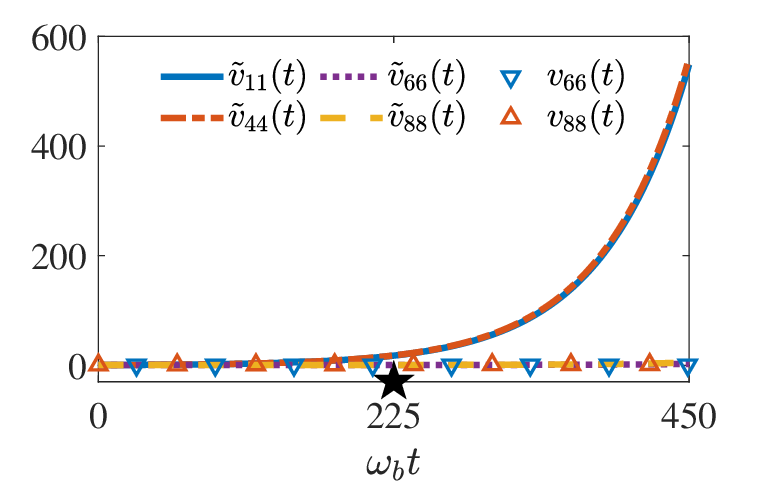}
\caption{Dynamical evolution of the CM elements in the COMM system. $\tilde{v}_{11}(t)$, $\tilde{v}_{44}(t)$, $\tilde{v}_{66}(t)$, and $\tilde{v}_{88}(t)$ are obtained from the full-system Hamiltonian, while $v_{66}(t)$ and $v_{88}(t)$ are obtained from the free Hamiltonians of the magnon and mechanical modes, respectively. Here, $\kappa_a=\kappa_c=10^{-4}\omega_b$, and all other parameters are the same as in Supplementary Figure \ref{comm1}.}
\label{commv}
\end{figure}

Supplementary Figure \ref{comm2} presents theoretical and numerical results at the characteristic time $\tau$ versus the detuning $\Delta_m$, showing good agreement in both steady-state and most unsteady-state regimes, consistent with Eqs. \eqref{SMeni} and \eqref{SMsact2}. As $\Delta_m$ increases, the quantum resources are enhanced, and the system undergoes a transition from one-way to two-way steering. Moreover, $\tilde E_{ac}(\tau)$  [$\tilde S_{a\to c}(\tau)$] exhibits a slight deviation from the corresponding theoretical prediction $E_{ac}$ ($S_{a\to c}$) when $\Delta_m\gtrsim 1.5\omega_b$, since the perturbative validity condition $g_a,g_c,g_m\ll |\Delta_a-\Delta_m|$ is no longer satisfied for $\Delta_a=3\omega_b$.

We calculate the CM elements of the microwave and optical modes from the full-system COMM Hamiltonian to verify the self-consistency of the linearization procedure. Supplementary Figure \ref{commv} shows the dynamical evolution of the corresponding CM elements $\tilde{v}_{11}(t)$ and $\tilde{v}_{44}(t)$. The associated quantum fluctuation amplitudes are estimated as $|\delta a|\approx\sqrt{\tilde{v}_{11}(t)}\lesssim25$ for the microwave mode and $|\delta c|\approx\sqrt{\tilde{v}_{44}(t)}\lesssim25$ for the optical mode. Since the corresponding classical amplitudes satisfy $|\langle o\rangle|\sim10^6$ ($o=a,c$), the relative fluctuations remain bounded by $|\delta o|/|\langle o\rangle|\lesssim2.5\times10^{-5}$ throughout the evolution. Therefore, the condition required for linearization is well satisfied over the entire time interval $t\lesssim2\tau$, confirming the self-consistency of the linearized treatment even in the unsteady-state regime. In particular, the linearization remains valid within the characteristic time window $t\in[\tau,2\tau]$, during which the generated quantum resources remain nearly invariant, as shown in Supplementary Figure \ref{comm1}.

The validity of the reduced open-system description is further verified in Supplementary Figure \ref{commv}. The CM elements of the magnon and mechanical modes, $v_{66}(t)$ and $v_{88}(t)$, obtained from their free evolution agree excellently with the corresponding elements, $\tilde{v}_{66}(t)$ and $\tilde{v}_{88}(t)$, calculated from the full COMM Hamiltonian. This agreement indicates that the environmental noise acting on the intermediate modes is confined to their local dynamics and has a negligible influence on the target modes, thereby confirming the validity of the reduced open-system description.

In the COMM system~\cite{SMCOMMLJ23}, the enhanced coupling strengths $g_m$ and $g_c$, as well as the detunings $\Delta_a$, $\Delta_c$, and $\Delta_m$, can be precisely tuned by the microwave and optical driving fields. The driving strengths can reach $\Omega_o\sim10^{13}-10^{14}~\mathrm{Hz}$ ($o=a,c$), yielding steady-state amplitudes of $|\langle o\rangle|\approx|\Omega_o/\Delta_o|\sim10^6$, which are sufficiently large to justify the linearization procedure. The effective coupling strength depends on both the couplings and detunings, as given by Eq. \eqref{SMcommeff}. The mechanical frequency $\omega_b$ is typically in the range of $10-100$ MHz, with a decay rate of $\kappa_b\sim 10^{-6}\omega_b$~\cite{SMcavityoptomechanics14,SMCOMMLJ23}. The decay rates of the magnon, microwave, and optical modes are $\kappa_m,\kappa_a,\kappa_c \sim 10^{-4}\omega_b-10^{-3}\omega_b$~\cite{SMcavitymagnonics22,SMcavityoptomechanics14,SMCOMMLJ23,SMEOM25,SMCMM22}, respectively. The microwave-magnon, magnomechanical, and optomechanical coupling strengths are $g_a, g_m, g_c \sim 0.1\omega_b$~\cite{SMCOMMLJ23}, allowing the unsteady-state condition $g_{\rm eff}^2\ge\kappa_a\kappa_c$ to be readily achieved. At cryogenic temperatures ($T\sim10~\mathrm{mK}$), the thermal occupation numbers of the microwave, magnon, and optical reservoirs are negligible, whereas the mechanical reservoir typically exhibits a thermal occupation of $N_b\sim 10$~\cite{SMCOMMLJ23}. 

\section{The distributions of quantum resources in multipartite microwave-optical hybrid systems}\label{MES}
In multipartite quantum systems, quantum resources such as entanglement and quantum steering are constrained by monogamy relations and cannot be freely shared~\cite{SMmonogamy07,SMmonogamy16}. In this Supplementary Note, we analyze the distribution of quantum resources in hybrid quantum systems within the validity regime of the effective Hamiltonian.

\begin{figure}[t] 
\centering
\includegraphics[width=0.48\textwidth]{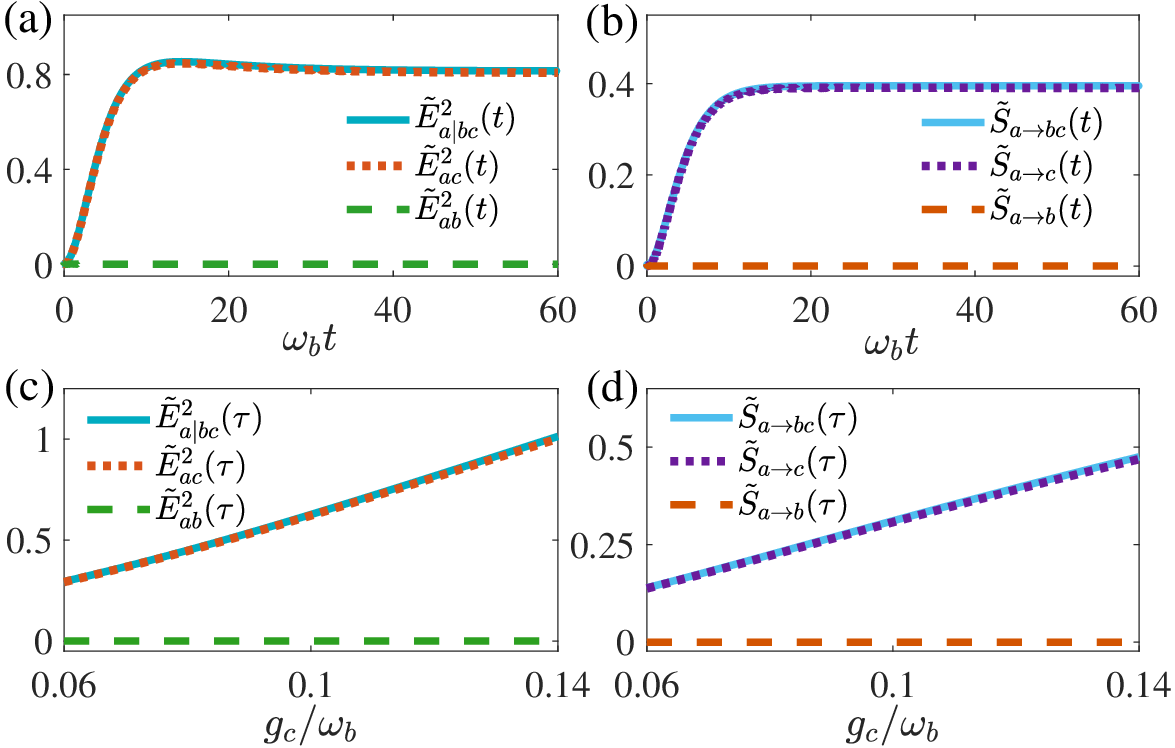}
\caption{Time evolution of the entanglement (a) and quantum steering (b) for $g_c = 0.12\omega_b$, and efficient quantum control of the MO entanglement (c) and quantum steering (d) versus the coupling strength $g_c$, in the EOM system. Other parameters are the same as those in Supplementary Figure \ref{eom1}(a).}
\label{mo-eom}
\end{figure}

At first, we quantify the quantum entanglement between a single mode $a$ and the remaining modes $B$ in a hybrid quantum system using the LN~\cite{SMentanglement07}. The definition of single-vs-multi-mode LN $E_{a|B}(t)$ is given by
\begin{equation}\label{en}
E_{a|B}(t)=\max\big[0,-\ln(2\eta^-_{a|B})\big],
\end{equation}
where $B\equiv b_1\dots b_s\dots b_Nc$, $\eta^-_{a|B}=\min({\rm eig}|i\sigma\mathcal{V}|)$ is the minimum symplectic eigenvalue. Here, $\sigma=\bigoplus^{N+2}_{1}i\sigma_y$ with $\sigma_y$ denoting the $y$-Pauli matrix. $\mathcal{V}=P_{a|B}\tilde{v}P_{a|B}$, where $\tilde{v}$ represents the CM of whole system and $P_{a|B}={\rm{diag}}(1,-1,1,1,\dots,1,1)$. 

Similarly, the criterion~\cite{SMLNsteeringK15} to measure the Gaussian quantum steering between single-mode and multi-mode is given by
\begin{equation}\label{SMscma}
S_{a\to B}(t)={\rm max}\left[0,-\sum_{j:\bar{\nu}_j^B<1/2}\ln(2\bar{\nu}_j^B)\right],
\end{equation}
The quantities $\bar{\nu}_j^B$s are the symplectic eigenvalues of the Schur complement matrix $\nu=v_B-v_{aB}^Tv_a^{-1}v_{aB}$ in which $v_a$, $v_B$, and $v_{aB}$ are the subblocks of the CM $\tilde{v}=[v_a, v_{aB}; v_{aB}^T, v_B]$. When subsystem $B$ is a single mode, Eq. \eqref{SMscma} reduces to the same result as Eq. \eqref{SMstac1}~\cite{SMLNsteeringK15}. 

Both entanglement and quantum steering are constrained by monogamy relations. Specifically, the entanglement satisfies the Coffman-Kundu-Wootters-type monogamy inequality~\cite{SMmonogamy07,SMmonogamy16}
\begin{equation}\label{Ec-bma}
E_{a|B}^2-\sum_{s=1}^{N}E_{ab_s}^2-E_{ac}^2\ge 0,
\end{equation}
where $E_{ap}$ is the entanglement between mode $a$ and mode $p$ ($p=b_s,c$). Similarly, the quantum steering obeys
\begin{equation}\label{G}
S_{a\to B}-\sum_{s=1}^{N}S_{a\to b_s}-S_{a\to c}\ge 0,
\end{equation}
where $S_{a\to B}$ denotes the multi-mode steering from mode $a$ to the set of other modes, and $S_{a\to b_s} (S_{a\to c})$ represent the steering from mode $a$ to mode $b_s (c)$.

We analyze the distributions of entanglement and quantum steering among different bipartitions of the EOM system. In Supplementary Figures \ref{mo-eom}(a) and (b), the time evolution of these distributions is shown for $g_c/\omega_b=0.12$. The results indicate that dynamically persistent bipartite entanglement and quantum steering can be established, with $\tilde{E}_{ac}^2(t)\approx \tilde{E}_{a|bc}^2(t)$ and $\tilde{S}_{a\to c}(t)\approx \tilde{S}_{a\to bc}(t)$ throughout the entire time evolution, while all other bipartite resources remain negligible. In Supplementary Figures \ref{mo-eom}(c) and (d), the entanglement and quantum steering distributions at the characteristic time $\tau$ are plotted versus the relative coupling strength $g_c/\omega_b$. The values of $\tilde{E}^2_{ac}(\tau)$ [$\tilde{S}_{a\to c}(\tau)$] closely approach those of $\tilde{E}^2_{a|bc}(\tau)$ [$\tilde{S}_{a\to bc}(\tau)$], whereas the other resources remain negligibly small.

\begin{figure}[t]  
\centering
\includegraphics[width=0.48\textwidth]{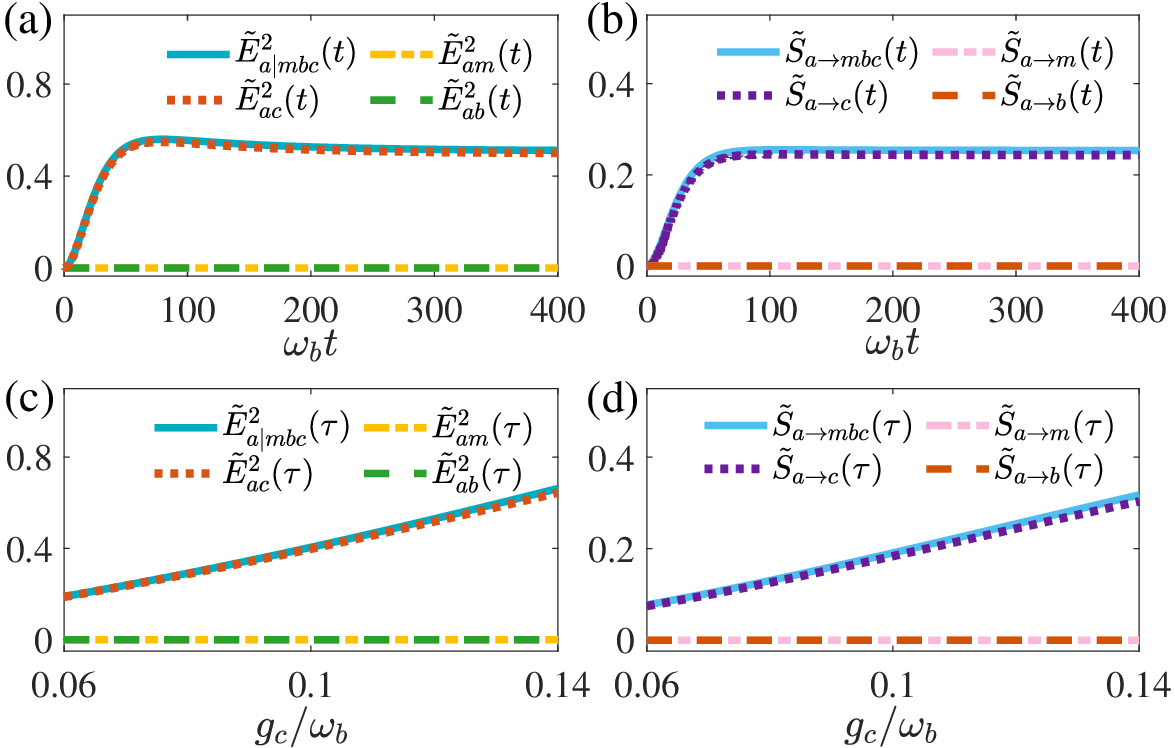}
\caption{Time evolution of the entanglement (a) and quantum steering (b) for $g_c=0.12\omega_b$, and efficient quantum control of the MO entanglement (c) and quantum steering (d) versus the coupling strength $g_c$, in the COMM system, where the decay rates are $\kappa_a = 0.5\kappa_c = 10^{-4}\omega_b$, and other parameters are the same as those in Supplementary Figure \ref{comm1}.}
\label{mo-comm}
\end{figure}

We also investigate the distribution of entanglement and quantum steering among different bipartitions of the COMM system. Supplementary Figures \ref{mo-comm}(a) and (b) show the dynamical evolution of these distributions. The results indicate that dynamically persistent bipartite entanglement and quantum steering can be established, satisfying $\tilde{E}_{ac}^2(t)\approx \tilde{E}_{a|mbc}^2(t)$ and $\tilde{S}_{a\to c}(t)\approx \tilde{S}_{a\to mbc}(t)$, whereas the remaining bipartite contributions are consistently negligible throughout the evolution. Supplementary Figures \ref{mo-comm}(c) and (d) further show the dependence of the entanglement and quantum steering distributions on the relative coupling strength $g_c/\omega_b$ at the characteristic time $\tau$. The values of $\tilde{E}^2_{ac}(\tau)$ [$\tilde{S}_{a\to c}(\tau)$] closely approach those of $\tilde{E}^2_{a|mbc}(\tau)$ [$\tilde{S}_{a\to mbc}(\tau)$], while all the other quantum resources remain negligible.

\section{Magnonic quantum resources in multipartite hybrid systems with linear interactions}\label{SMCM}
Our general theoretical framework is applicable to arbitrary two-mode subsystems coupled via linearly interacting intermediate modes. As an example, we apply it to a cavity magnonic system~\cite{SMCM22} to study the generation of quantum resources between two magnon modes.

\subsection{Model and the effective Hamiltonian}\label{SMCMH}
The hybrid cavity magnonic system~\cite{SMCM22} consists of a microwave cavity coupled to two magnon modes. Under a two-photon parametric drive and in the rotating frame at the pump frequency, the total Hamiltonian reads
\begin{equation}\label{SMCMH1}
\begin{aligned}
H_{S_{3}}=&\tilde\Delta_aa^\dag a+\Omega_a(a^{\dag2}+a^{2})\\&+\sum_{j=1,2}\Delta_jm_j^\dag m_j+g_j(a^\dag m_j+am_j^\dag)
\end{aligned}
\end{equation}
where $a$ ($a^\dag$) and  $m_j$ ($m_j^\dag$, $j=1,2$) are the annihilation (creation) operators of the microwave and magnon modes, respectively. $\tilde\Delta_a=\omega_a-\omega_d$ and $\Delta_j=\omega_j-\omega_d$ are the detunings of the microwave and magnon modes, respectively. $\omega_a$ and $\omega_j$ are the corresponding transition frequencies, and $\omega_d$ is the driving frequency. The coupling strength $g_{j}$ describes the microwave-magnon interaction. The parameter $\Omega_a$ denotes the Rabi frequency, and $\omega_d$ is the driving frequency.

In the rotating frame with the unitary transformation $U=\exp\left[(a^2-a^{\dag2})r/2\right]$, the Hamiltonian in Eq. \eqref{SMCMH1} turns into
\begin{equation}\label{SMCMH3}
\begin{aligned}
H_{S_{3}}=&\sum_{j=1,2}\Delta_jm_j^\dag m_j+g_j\cosh r(am_j^\dag+a^\dag m_j)\\
&+\Delta_aa^\dag a-g_j\sinh r(am_j+a^\dag m_j^\dag),
\end{aligned}
\end{equation}
where $\Delta_a=\tilde\Delta_a/\cosh(2r)$ and $\tanh(2r)=2\Omega_a/\tilde\Delta_a$. The Hamiltonian in Eq. \eqref{SMCMH3} can be mapped onto the general Hamiltonian in Eq. (1) considered in the main text by setting $\theta=\phi$ and $N=1$, together with the substitutions $\Delta_{a(c)}\rightarrow\Delta_{1(2)}$, $\omega_1\rightarrow\Delta_a$, $g_{a(c)}\rightarrow g_{1(2)}$, $\cos\theta\rightarrow\cosh r$, $\sin\theta\rightarrow-\sinh r$, $a\to m_1$, $b_1\to a$, and $c\to m_2$. Consequently, the general procedure for constructing the effective Hamiltonian, presented in the Methods section of the main text, can be directly applied to derive the effective magnon-magnon coupling in the cavity magnonic system.

When the magnon-1 detuning $\Delta_1$ is nearly opposite to the magnon-2 detuning $\Delta_2$, while both detunings are far off-resonant from the cavity detuning $\Delta_a$, i.e., $\Delta_1=-\Delta_2+\delta$ and $g_1,g_2\ll\{|\Delta_1-\Delta_a|, |\Delta_2-\Delta_a|\}$. According to the general effective Hamiltonian construction approach presented in the Methods section of the main text, the effective Hamiltonian between the two magnon modes can be obtained as
\begin{equation}\label{SMCMHefac}
\begin{aligned}
H_{\rm eff}=g_{\rm eff}(m_1^\dag m_2^\dag+m_1m_2).
\end{aligned}
\end{equation}
Here, the corresponding effective coupling strength $g_{\rm eff}$ is derived from Eq. (18) in the Methods section of the main text as
\begin{equation}\label{SMCMgeff}
g_{\rm eff}=\frac{g_1g_2\Delta_a\sinh(2r)}{\Delta_a^2-\Delta_1^2}
\end{equation}
and the difference $\delta$ is similarly obtained as
\begin{equation}\label{SMCMdelta}
\begin{aligned}
\delta=&\frac{(g_1^2+g_2^2)\Delta_a\cosh(2r)+(g_1^2-g_2^2)\Delta_1}{\Delta_a^2-\Delta_1^2}.
\end{aligned}
\end{equation}

Moreover, to obtain the analytical expressions for the quantum resources between the two magnon modes, we replace the decay rate $\kappa_a$ ($\kappa_c$) of the microwave (optical) mode in Eq. \eqref{SMeni} and Eq. \eqref{SMsact2} with the decay rate $\kappa_1$ ($\kappa_2$) of magnon-1 (magnon-2) mode. Meanwhile, the effective coupling strength obtained in Eq. \eqref{SMCMgeff} is substituted into Eq. \eqref{SMeni} and Eq. \eqref{SMsact2}, respectively. With these replacements and substitutions, the analytical expressions for the quantum resources between the two magnon modes can be obtained. 

\subsection{System dynamics and quantum resources} \label{cmAmatrix}

\begin{figure}[t] 
\centering
\includegraphics[width=0.4\textwidth]{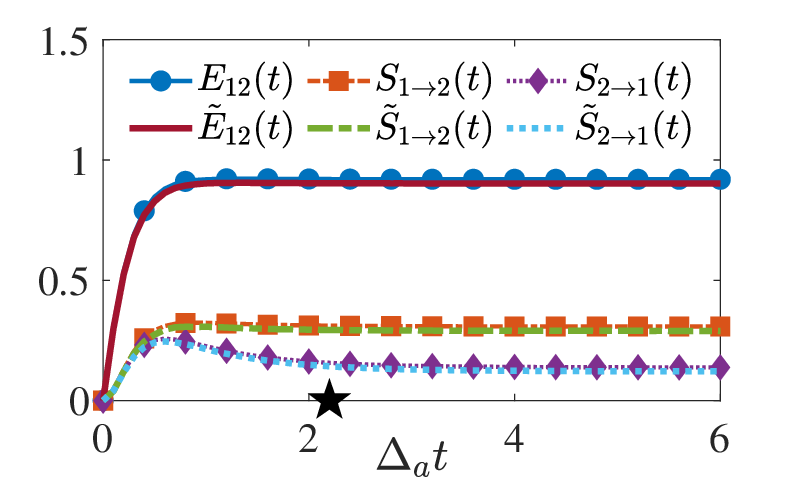}
\caption{The unsteady-state dynamics of magnon-magnon quantum resources with the star symbol indicating the characteristic time $\tau$. The system parameters are $g_1=g_2=0.15\Delta_a$, $\Delta_1=2.5\Delta_a$, $\kappa_a=10^{-4}\Delta_a$, $\kappa_1=10^{-3}\Delta_a$, $\kappa_2=1.3\times10^{-3}\Delta_a$, and the thermal occupation numbers are $N_{m_1}=N_{m_2}=N_{a}=0$.}
\label{cmdy}
\end{figure}

According to the Hamiltonian in Eq. \eqref{SMCMH3}, the dynamics of the system within the open-quantum-system framework can be analytically described by QLEs. Given that the initial state is Gaussian, the system dynamics in a Markovian environment can be fully characterized by a CM, which satisfies
\begin{equation}\label{SMCMV}
\dot{\tilde{v}}(t)=A\tilde{v}(t)+\tilde{v}(t)A^T+D,
\end{equation}
where $\dot{\tilde{v}}(t)$ is the derivation of $\tilde{v}(t)$. The elements of $\tilde{v}(t)$ are defined as $\tilde{v}_{jl}(t)=\langle \tilde{u}_j(t)\tilde{u}_l(t)+ \tilde{u}_l(t)\tilde{u}_j(t)\rangle/2$, with $j,l=1,2,\cdots,6$, where $\tilde{u}_j(t)$ is the $j$-term of $\tilde{u}(t)=[X_{m_1}(t),Y_{m_1}(t),X_{m_2}(t),Y_{m_2}(t),X_{a}(t),Y_{a}(t)]^T$. Here, the quadrature operators are defined as $X_o(t)=(o+o^\dag)/\sqrt{2}$ and $Y_o(t)=i(o^\dag-o)/\sqrt{2}$, with $o=m_1,m_2,a$. Moreover, the drift matrix $A$ in Eq. \eqref{SMCMV} has the form
\begin{align}\label{SMCMA}
A \!\!=\!\begin{pmatrix}
\!	-\kappa_1&\!\Delta_1 &\! 0 &\! 0& \!0&\! g_1^+\\
\!	-\Delta_1&\!-\kappa_1&\!0&\!0&\!g_1^-&\!0\\
\!	0&\!0&\!-\kappa_2&\!\Delta_2&\!0&\! g_2^+\\
\!	0&\!0&\!-\Delta_2&\!-\kappa_2&\!g_2^-&\!0\\
\!	0&\! g_1^+&\!0&\!g_2^+&\!-e^{2r}\kappa_a&\!\Delta_a\\
\!	g_1^-&\!0&\!g_2^-&\!0&\!-\Delta_a&\!-e^{2r}\kappa_a
\end{pmatrix},                                            
\end{align}
where $g_j^\pm=g_j(\sinh r\pm\cosh r)$ with $j=1,2$. $\kappa_1$, $\kappa_2$, and $\kappa_a$ represent the decay rates of the magnon-1, magnon-2, and cavity modes, respectively, and $D= {\rm diag}[\kappa_1(2N_{m_1}+1),~\kappa_1(2N_{m_1}+1),~\kappa_2(2N_{m_2}+1),~\kappa_2(2N_{m_2}+1),~e^{2r}\kappa_{a}(2N_{a}+1),~e^{2r}\kappa_{a}(2N_{a}+1)]$ is the diffusion matrix. Here, $N_o(\omega_o)=[\exp(\hbar\omega_o/k_BT)-1]^{-1}$ is the mean population of mode-$o$ ($o=m_1,m_2,a$ and  $\omega_{m_1}\equiv\omega_1,~\omega_{m_2}\equiv\omega_2$), with the Boltzmann constant $k_B$ and the environmental temperature $T$.

As shown in Supplementary Figure \ref{cmdy}, both the entanglement between the two magnon modes, $E_{12}(t)$ [$\tilde E_{12}(t)$], and the steering from magnon-1 to magnon-2, $S_{1\to2}(t)$ [$\tilde S_{1\to2}(t)$], increase rapidly at early times and subsequently approach their asymptotic values. In contrast, the steering $S_{2\to1}(t)$ [$\tilde S_{2\to1}(t)$] exhibits a distinct dynamical behavior: it first increases to a maximum and then gradually decreases toward its asymptotic value. In the long-time limit, asymmetric two-way steering emerges, satisfying $\tilde S_{1\to2}>\tilde S_{2\to1}>0$. Moreover, these dynamical quantum resources reach their asymptotic values before the characteristic time $\tau$ (indicated by the star symbols), enabling persistent quantum resources to be obtained throughout the time window $t\in[\tau,2\tau]$. Furthermore, the numerical results for $\tilde E_{12}$, $\tilde S_{1\to2}$, and $\tilde S_{2\to1}$ agree well with the analytical predictions represented by the lines with markers.

The parameter regime considered here is experimentally accessible in current cavity-magnonic platforms~\cite{SMCM22}. The decay rates of the two magnon modes are typically $\kappa_1,\kappa_2\sim1~\mathrm{MHz}$~\cite{SMCM22}, while the microwave cavity decay rate is $\kappa_a\sim0.1~\mathrm{MHz}$~\cite{SMcavitymagnonics22,SMcavityoptomechanics14}. In cavity magnonic systems, the cavity detuning can reach $\Delta_a\sim1~\mathrm{GHz}$~\cite{SMcavitymagnonics22}. Under two-photon parametric driving, the effective cavity detuning is given by $\Delta_a=\tilde{\Delta}_a/\cosh(2r)$, with the squeezing parameter satisfying $\tanh(2r)=2\Omega_a/\tilde{\Delta}_a$. For the parameter regime considered in this work ($r=0.2$), the corresponding parametric driving strength is $\Omega_a\approx0.19\tilde{\Delta}_a$. In addition, the microwave-magnon coupling strengths typically lie in the range $g_1,g_2\sim50-150~\mathrm{MHz}$~\cite{SMCM22}. Under these conditions, the cooperativity can exceed unity, $C\equiv g_{\rm eff}^2/(\kappa_1\kappa_2)>1$, enabling access to the unsteady-state dynamical regime. Furthermore, at cryogenic temperatures of $T\sim10~\mathrm{mK}$, the thermal occupations of the magnon and microwave modes are strongly suppressed, yielding $N_{m_1}\approx N_{m_2}\approx N_a\approx0$~\cite{SMCOMMLJ23}. Moreover, we apply the same validation procedure as that used for the EOM and COMM systems to the cavity magnonic system, and the consistent results further support the validity of our reduced open-system description.

\section{Non-Gaussian microwave-optical quantum resources in multipartite nonlinear systems}\label{SMng}
Non-Gaussian quantum resources are essential for various applications, particularly in universal quantum computing. Although the experimental realization of effective nonlinear MO interactions remains highly challenging, theoretical investigations of MO nonlinear effects and non-Gaussian quantum resources are both timely and important. In this Supplementary Note, based on the effective Hamiltonian approach, we further investigate dynamical generation of persistent non-Gaussian MO quantum resources in a three-mode microwave-intermediate-optical hybrid system with nonlinear interactions. Specifically, we consider a hybrid quantum system described by the following Hamiltonian ($\hbar\equiv1$)
\begin{equation}\label{SMnlH}
\begin{aligned}
H_{\rm nl}=~&\Delta_aa^\dag a+\omega_b b^\dag b+\Delta_c c^\dag c\\
&+g_a(a^{\dag 2} b+a^2b^\dag)+g_c(c+c^\dag)(b+b^\dag).
\end{aligned}
\end{equation} 
Here, $a$ ($a^\dag$), $b$ ($b^\dag$), and $c$ ($c^\dag$) are the annihilation (creation) operators of the microwave, intermediate, and optical modes, respectively. $\Delta_o$ ($o=a,c$) denotes the detuning of mode $o$. $\omega_b$ is the transition frequency of the intermediate mode. The coupling strengths $g_{a}$ and $g_{c}$ characterize the nonlinear interactions between modes $a$ and $b$ and the linear interactions between modes $c$ and $b$, respectively. The Hamiltonian in Eq. \eqref{SMnlH} follows from the general Hamiltonian given by Eq. (26) in the Methods section of the main text by setting $s=2$, $w=1$, $\theta=0$, and $\phi=\pi/4$, together with the substitution $g_c\to\sqrt{2}g_c$.

\begin{figure*}[t] 
\centering
\includegraphics[width=0.9\textwidth]{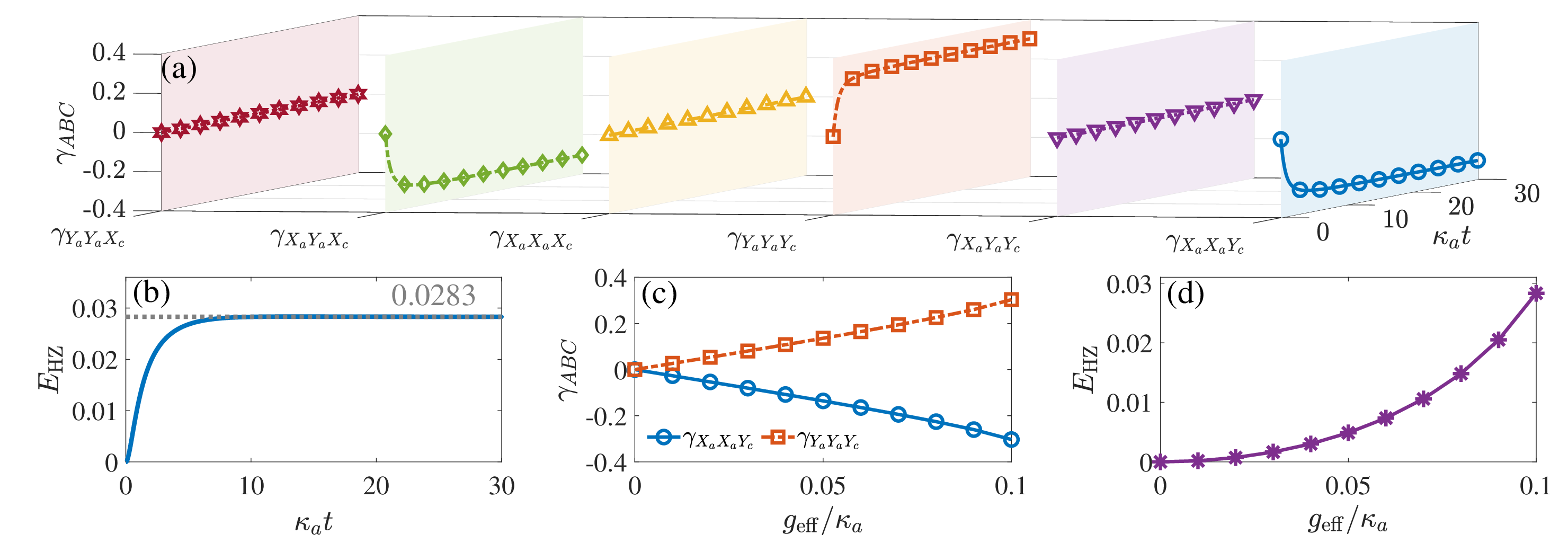}\\
\caption{(a) Time evolution of various coskewnesses $\gamma_{ABC}$. (b) Dynamically generated persistent non-Gaussian MO entanglement quantified by $E_{\rm HZ}$ based on the Hillery-Zubairy criterion. (c) Coskewness $\gamma_{ABC}$ and (d) entanglement quantity $E_{\rm HZ}$ as functions of the effective coupling strength $g_{\rm eff}$. $g_{\rm eff}/\kappa_a=0.1$ for panels (a) and (b). The system parameters are $\kappa_a=\kappa_c$ and the thermal occupation numbers are $N_a=N_c=0$.}
\label{gcen40}
\end{figure*}

Then, by applying the general effective-Hamiltonian construction approach presented in the Methods section of the main text to the Hamiltonian in Eq. \eqref{SMnlH}, we derive the corresponding effective coupling. Specifically, when twice the microwave detuning is nearly opposite to the optical detuning, i.e., $2\Delta_a+\Delta_c\approx0$, and both detunings are far off-resonant from the intermediate frequency $\omega_b$, satisfying $g_a,g_c\ll\{|2\Delta_a-\omega_b|,|\Delta_c-\omega_b|\}$ the tensor product state $|nlk\rangle\equiv|n\rangle_a|l\rangle_{b}|k\rangle_c$ approximately degenerates with $|(n+2)l(k+1)\rangle$. Here, the subscripts $a$, $b$, and $c$ represent the microwave, intermediate, and optical modes, respectively. Under these cases, the interaction term in Eq. \eqref{SMnlH} can effectively couple the arbitrary base pairs $|nlk\rangle$ and $|(n+2)l(k+1)\rangle$, yielding an effective Hamiltonian expressed in the following form
\begin{equation}\label{SMnlHef}
\begin{aligned}
H_{\rm eff}\!=&\epsilon_1|nlk\rangle\langle nlk|+(2\Delta_{a}+\Delta_{c}+\epsilon_2)\\&\times|(n+2)l(k+1)\rangle\langle(n+2)l(k+1)|\\
&+\tilde{g}_{\rm eff}(|nlk\rangle\langle (n+2)l(k+1)|+{\rm h.c.}).
\end{aligned}
\end{equation}
Here, $\epsilon_1$ and $\epsilon_2$ represent the energy shifts caused by the coupling between the states $|nlk\rangle$ and $|(n+2)l(k+1)\rangle$, respectively, and $\tilde{g}_{\rm eff}$ is the effective coupling strength. These are three coefficients to be determined. It should be noted that the common unperturbed eigenenergy $n\Delta_a+l\omega_b+k\Delta_c$ shared by the two basis states has been omitted.

We first consider the effective coupling strength $\tilde{g}_{\rm eff}$ between states $|nlk\rangle$ and $|(n+2)l(k+1)\rangle$. According to Eq. (18) in the Methods section of the main text, we obtain
\begin{equation}
\tilde{g}_{\rm eff}=\frac{g_ag_c}{2\Delta_a-\omega_b}\sqrt{(n+2)(n+1)(k+1)}.
\end{equation}
The effective Hamiltonian in Eq. \eqref{SMnlHef} after neglecting the second-order energy shifts can be expressed as
\begin{equation}\label{SMnlHef1}
\begin{aligned}
H_{\rm eff}=&\tilde{g}_{\rm eff}(|nlk\rangle\langle (n+2)l(k+1)|+{\rm h.c.})\\
=&g_{\rm eff}\sqrt{(n+2)(n+1)(k+1)}(|nk\rangle\\&\times\langle (n+2)(k+1)|+{\rm h.c.})\otimes|l\rangle\langle l|.
\end{aligned}
\end{equation}
By eliminating the intermediate mode, the Hamiltonian in Eq. \eqref{SMnlHef1} can be promoted to the full Hilbert space of the microwave and optical modes, yielding
\begin{equation}\label{SMnlHeff}
\begin{aligned}
H_{\rm eff}=g_{\rm eff}(a^{\dag2}c^\dag+a^2c),
\end{aligned}
\end{equation}
and the coupling strength is
\begin{equation}\label{SMnlgeff}
g_{\rm eff}=\frac{g_ag_c}{2\Delta_a-\omega_b}.
\end{equation}
The effective coupling strength $g_{\rm eff}$ determines the strength of this nonlinear interaction and is therefore associated with the dynamical generation of non-Gaussian correlations.

With the system initially prepared in the vacuum state, the nonlinear interaction described by the effective Hamiltonian in Eq. \eqref{SMnlHeff} drives the system into a non-Gaussian state. Within the open-quantum-system framework, the corresponding density-matrix evolution is governed by the Lindblad master equation
\begin{equation}\label{rhot}
\begin{aligned}
\dot{\rho}(t)=-i[H_{\rm eff},\rho(t)]+\sum_{o=a,c}\kappa_o\mathcal{L}[o],
\end{aligned}
\end{equation}
where $\mathcal{L}[o]$ is the Lindblad superoperator defined as $\mathcal{L}[o]\equiv\left\{o\rho(t) o^\dag-\frac{1}{2}[o^\dag o\rho(t)+\rho(t) o^\dag o]\right\}$. Here, $\kappa_a$ and $\kappa_c$ denote the decay rates of the microwave and optical modes, respectively. The nonlinear interaction naturally gives rise to third-order correlations during the system evolution. To quantify these non-Gaussian features, we employ the normalized third-order statistical moment, known as the coskewness~\cite{SMcoskewness},
\begin{equation}\label{rABC}
\gamma_{ABC}=\frac{\langle ABC\rangle}{\sigma_A\sigma_B\sigma_C}.
\end{equation}
Here, $A$, $B$, and $C$ denote quadrature operators $X_o(t)=(o+o^\dag)/\sqrt{2}$ and $Y_o(t)=i(o^\dag-o)/\sqrt{2}$ with $o=a,c$. The quantities $\sigma_A$, $\sigma_B$, and $\sigma_C$ are the corresponding standard deviations, defined as $\sigma_O=[\langle O^2\rangle-\langle O\rangle^2]^{1/2}$ ($O=A,B,C$). Nonzero $\gamma_{ABC}$ indicates the existence of third-order correlations. We further explored their connection to quantum entanglement. Specifically, we considered the Hillery-Zubairy entanglement criterion~\cite{SMHZ06}, given by
\begin{equation}
|\langle a^2c\rangle|^2>\langle a^{\dag2}a^2\rangle\langle c^{\dag}c\rangle.
\end{equation}
We define the quantity $E_{\rm HZ}=|\langle a^2c\rangle|^2-\langle a^{\dag2}a^2\rangle\langle c^{\dag}c\rangle$ to characterize the non-Gaussian MO entanglement. A positive value $E_{\rm HZ}>0$ certifies the presence of entanglement. Furthermore, by expressing $|\langle a^2c\rangle|^2$ in terms of the quadrature operators of modes $a$ and $c$, one can establish a analytical connection between entanglement and coskewnesses
\begin{equation}
\begin{aligned}
E_{\rm HZ}=&2\sigma_{X_a}^4\sigma_{X_c}^2(\gamma_{X_aX_aX_c}^2+\gamma_{X_aX_aY_c}^2)\\&-\langle a^{\dag2}a^2\rangle\langle c^{\dag}c\rangle.
\end{aligned}
\end{equation}

Within this dynamical framework, we verify that all first-order moments and second-order quadrature covariances between the microwave and optical modes remain zero throughout the evolution. In the numerical simulations, the density matrix $\rho$ is evolved in a truncated Hilbert space with a sufficiently large cutoff to ensure validity~\cite{SMdim98}. We then investigate the normalized third-order statistical moments, namely the coskewnesses defined in Eq. \eqref{rABC}. As shown in Supplementary Figure \ref{gcen40}(a), the coskewnesses $\gamma_{X_aY_aX_c}$, $\gamma_{Y_aY_aY_c}$, and $\gamma_{X_aX_aY_c}$ increase rapidly at early times and subsequently approach approximately constant values over the time window considered, whereas $\gamma_{Y_aY_aX_c}$, $\gamma_{X_aX_aX_c}$, and $\gamma_{X_aY_aY_c}$ remain identically zero throughout the evolution. These third-order correlations arise from the joint action of nonlinear interactions and decoherence in the open-system dynamics. Moreover, as shown in Supplementary Figure \ref{gcen40}(b), a nonzero entanglement value $E_{\rm HZ}\simeq0.0283>0$ is achieved during the dynamical evolution. Therefore, these results demonstrate the generation of dynamically persistent non-Gaussian MO entanglement induced by nonlinear interactions. We further examine the dependence of the coskewnesses $\gamma_{ABC}$ and entanglement quantity $E_{\rm HZ}$ on the effective coupling strength $g_{\rm eff}$. As shown in Supplementary Figures \ref{gcen40}(c) and (d), within the parameter regime considered, increasing $g_{\rm eff}$ enhances both the magnitudes of the coskewnesses $\gamma_{ABC}$ and the entanglement quantity $E_{\rm HZ}$, indicating enhanced non-Gaussian quantum resources. Since $g_{\rm eff}$ is determined by the coupling strengths $g_a$ and $g_c$ and the detuning $\Delta_a$, these quantum resources can be indirectly controlled and enhanced by tuning the system parameters.

\begin{figure}[t] 
\centering
\includegraphics[width=0.48\textwidth]{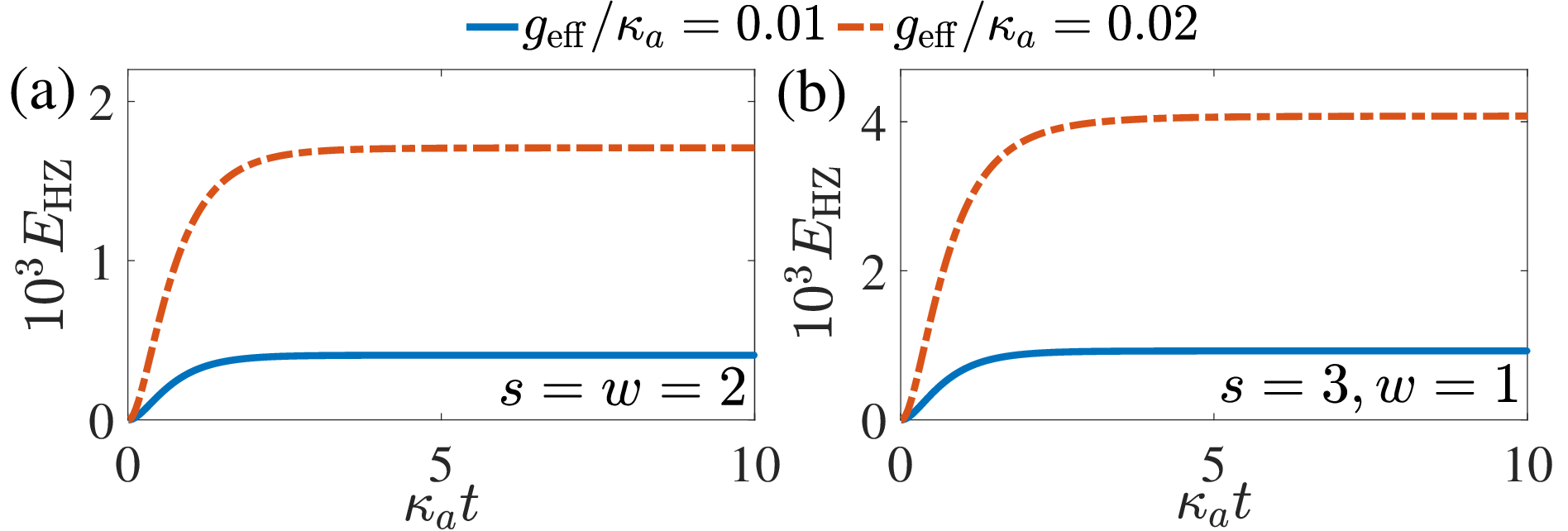}\\
\caption{Dynamically generated persistent entanglement quantity $E_{\rm HZ}$ under different coupling strength $g_{\rm eff}$. The nonlinear interaction integers are $s=w=2$ for (a) and $s=3, w=1$ for (b), respectively. Other parameters are the same as those in Supplementary Figure \ref{gcen40}.}
\label{ascw}
\end{figure}

To further investigate the non-Gaussian entanglement generated by nonlinear MO interactions, we consider the general nonlinear effective Hamiltonian derived for the microwave-intermediate-optical hybrid system in Eq. (27) of the Methods section of the main text,
\begin{equation}\label{SMgenHeff}
H_{\rm eff}=g_{\rm eff}(a^{\dag s}c^{\dag w}+a^sc^w),
\end{equation}
where the positive integers $s$ and $w$ specify the orders of the nonlinear interactions associated with the $a$ and $c$ modes, respectively. The two-mode trisqueezing effective Hamiltonian considered in Eq. \eqref{SMnlHeff} corresponds to the special case of $s=2$ and $w=1$ within this general nonlinear interaction. Taking the two-mode vacuum state as the initial state, we characterize the generated entanglement using a quantity associated with the Hillery-Zubairy criterion~\cite{SMHZ06},
\begin{equation}
E_{\rm HZ}= \left|\left\langle a^{s}c^{w}\right\rangle\right|^{2}-\left\langle a^{\dagger s}a^{s}\right\rangle\left\langle c^{\dagger w}c^{w}\right\rangle .
\end{equation}
This quantity allows us to characterize the MO entanglement generated by different nonlinear interactions. A positive value, $E_{\rm HZ}>0$, is sufficient to certify the presence of entanglement. As shown in Supplementary Figure \ref{ascw}, our numerical simulations of the dynamical evolution demonstrate that different nonlinear effective Hamiltonians can generate persistent non-Gaussian quantum resources over different coupling-strength regimes.

\end{document}